\documentclass[prd,reprint,10pt,twoside,tightenlines,nofootinbib,showpacs,notitlepage,superscriptaddress,aps]{revtex4-1}
\pdfoutput=1
\usepackage{dcolumn}
\usepackage{amsmath,amsfonts,amssymb,latexsym}
\usepackage[sort&compress]{natbib}
\usepackage{grffile,epstopdf}
\usepackage{graphicx}
\usepackage{hhline}
\usepackage{xcolor}
\usepackage{hhline}
\usepackage{physics}

\newcommand{\beq}{\begin{equation}}
\newcommand{\eeq}{\end{equation}}
\newcommand{\bea}{\begin{eqnarray}}
\newcommand{\eea}{\end{eqnarray}}
\newcommand{\vp}{\vspace{0.4cm}}
\newcommand{\hp}{\hspace{1.0cm}}
\newcommand{\bfr}{\begin{flushright}}
\newcommand{\efr}{\end{flushright}}

\begin{document}

\title{Quantum correlation dynamics and in-medium 3$\leftrightarrow$3 collisions of fermions}
\author{Wolfgang Cassing}
\affiliation{Institute for Theoretical Physics, University of Gie\ss en,
Heinrich-Buff Ring 16, Gie\ss en, 35392, Germany}

\begin{abstract}
In this study we aim for quantifying the role of in-medium 3$\leftrightarrow$3 collisions for systems of fermions which initially are out-off equilibrium.  The formulation of the 3-body dynamics is based on the equations of motion method for identical fermions -- also denoted as quantum correlation dynamics -- and presented in detail. The on-shell 2-body collision integral is briefly reviewed and the on-shell 3-body collision integral is derived on the basis of the same two-body interaction in leading order. The resulting equations obey particle number as well as energy-momentum conservation. For a quantification of the relative impact of 3-body interactions we employ a model study for a homogeneous system in space in a finite box with periodic boundary conditions. We address spin-isospin symmetric nuclear matter systems with momentum distributions that are given by shifted Fermi spheres (without overlap) as encountered in the initial phase of nucleus-nucleus collisions after contact. The results for the relaxation times -- employing an effective 2-body interaction -- are compared to Boltzmann-Uehling-Uhlenbeck (BUU) transport calculations in the continuum limit for the same bombarding energies and are found to agree on the level of a few percent. We find that the additional 3-body interactions reduce the relaxation times up to a factor of 3 at 130 A$\cdot$MeV.
Furthermore, it is shown in BUU transport calculations that an enhanced stopping by 3$\leftrightarrow$3 collisions shows up in the angular distribution of energetic nucleons ($>$ 60 MeV) e.g. in central $^{97}_{45}Rh$ collisions  at 40 A$\cdot$MeV that lead to the formation of a compound nucleus. The angular distribution of the energetic nucleons changes from a slightly forward peaked angular distribution to a slightly sidewards peaked angular distribution which might be controlled experimentally.
\end{abstract}


\maketitle

\section{Introduction}
\label{introduction}

The dynamics of quantum many-body systems is of particular interest in the context of the nuclear many-body problem and in heavy-ion collisions, which probe the matter at high nuclear densities in and out-off equili\-brium. While at low bombarding energies below about 10 A$\cdot$MeV the dynamics of the heavy-ion collision can reasonably be described by Time-Dependent-Hartree-Fock (TDHF) \cite{Bonche:1976zz,Negele:1981tw} the impact of 2-body collisions  becomes important with increasing bombarding energies since the Pauli-blocking effect weakens for higher energies.

In this energy range extended TDHF approaches have been introduced such as the Time-Dependent-Density-Matrix (TDDM) approach \cite{tohyama1989small,gong1990application}, which accounts for the additional impact of binary elastic nucleon-nucleon collisions that are not blocked by the Pauli-principle \cite{Nordheim:1928,uehling1933transport}. However, this approach is limited to light nuclear systems and still low bombarding energies due to the limited number of basis states. On the other hand semiclassical transport models have been formulated in the past \cite{Bertsch:1984gb,Aichelin:1985zz,Bertsch:1988ik}, that can be solved within the testparticle approximation and also be applied for heavy systems at higher energy. These models are denoted by Boltzmann-Uehling-Uhlenbeck (BUU) or Vlasov--Uehling-Uhlenbeck (VUU) type approaches \cite{Cassing:2021fkc} and have been employed by many transport groups worldwide for the analysis of heavy-ion collision data in particular for the production of hard photons, pions or kaons \cite{Cassing:1990dr}.

While at bombarding energies below about 130 A$\cdot$MeV the production of pions is subthreshold, the role of inelastic nucleon-nucleon collisions (e.g. $NN \rightarrow N \Delta, \Delta \rightarrow N \pi $) rises with increasing bombarding energy. Such processes can easily be incorporated in semiclassical transport simulations because the excited states (nucleon resonances or produced mesons) have different quantum numbers than the nucleons and also have low or moderate  phase-space densities. The role of inelastic reaction channels becomes important for the stopping and equilibration in heavy-ion collisions since in particular the elastic nucleon-nucleon collisions become forward peaked and decrease in magnitude for higher invariant energies. However, the question remains about the role of higher-order interactions in the energy domain below about 130 A$\cdot$MeV where the degrees of freedom are just identical fermions (nucleons) and the dominant processes are elastic scattering.

It is the aim of this study to formulate/derive a corresponding collision term and to investigate quantitatively the impact of additional 3$\leftrightarrow$3 interactions while keeping the full antisymmetry of the problem.
The derivation of transport theories either conveniently starts from the BBGKY\footnote{Bogolyubov, Born, Green, Kirkwood, Yvon} hierarchy in case of on-shell theories \cite{Cassing:2021fkc} or from the Kadanoff-Baym equations for $n$-body propagators \cite{KadanoffBaym,Danielewicz:1982ca}, which are the counterparts of the reduced density matrices. Such derivations have been performed in the past in the non-relativistic and relativistic case \cite{Cassing:1990dr,Ko:1987gp,Blattel:1988zz,Botermans:1990qi}\footnote{For a comparison of various transport approaches and model results at intermediate energies we refer the reader to the review \cite{TMEP:2022xjg}.}  as well as for $2 \leftrightarrow n$ reactions ($n$=3,4...) in case of high density systems at relativistic energies \cite{Cassing:2001ds}, where Pauli-blocking or Bose-enhancement factors play a minor role since almost the full 'zoo' of hadronic states is involved. As mentioned above a rigid formulation for 3$\leftrightarrow$3 processes in the medium is still lacking and will be presented here. This is of particular relevance since there is practically no information from experimental data on 3-body scattering events especially in the nuclear physics context. Information about 3-body scattering is practically only available from chiral perturbation theory (ChPT) or lattice QCD (lQCD) calculations \cite{Dawid:2025zxc}.

The outline of this work is as follows: In Section 2 we will briefly recall the cluster decomposition of the reduced density matrices $\rho_n$ and recall the derivation of the equations of motion for 2-body and 3-body correlation functions on the basis of Ref. \cite{Wang:1984cg}. These equations will be expanded in a single-particle basis and the physical interpretation of the various terms be discussed for the 2-body and 3-body correlation functions. Section 3 is devoted to the derivation of on-shell collision terms in the 2-body and 3-body case (in leading order) within a natural basis $|\alpha \rangle$ that diagonalizes the 1-body density matrix. Explicit results for the matrix elements of the 2-body and 3-body correlations will be given  while shifting the numerous interaction terms to the Appendix. In Section 4 we present a first model study for spin-isospin symmetric nuclear matter systems with momentum distributions that are given by shifted Fermi spheres (without overlap) as encountered in the initial phase of nucleus-nucleus collisions after contact. The explicit time evolution of the quadrupole moment in momentum space will be presented for different bombarding energies and the corresponding relaxation times be extracted. For comparison we provide BUU transport calculations in a finite box in the infinite nuclear matter limit employing the same (simple) 2-body interaction. This will allow to quantify the role of the 3-body interactions which appear on top of the 2-body scatterings.

The results of the model study are primarily academic and the question arises if a higher stopping power by additional 3$\leftrightarrow$3 collisions could be seen in experiments. To this aim we  perform in Section 5 actual BUU transport calculations (with enhanced cross sections) for finite systems at 40 A$\cdot$MeV - leading to a fusing system with particle emission - and compute the quadrupole moment in momentum space of very energetic nucleons that are emitted early in the system. A summary of the results and its implications will be given in Section 6 while the full list of interaction terms is given in the Appendix.

\section{Equations of motion for correlation functions}

The dynamical description of the nuclear many-body problem in the
nonrelativistic limit is based on the von-Neumann equation for the A-body
density matrix $\rho _A$ \cite{balescu:1975}, $i.e$. $(\hbar = 1)$
\beq \label{2.1}
i\frac{\partial}{\partial t}\ \rho _A = [H_A , \rho _A]
\eeq
where $H_A$ denotes the A-body Hamiltonian in the restriction to two-body interactions $v(ij)$
\beq \label{2.2}
H_A = \sum_{i=1}^A  t(i) + \sum_{i<j}^{A-1}  v(ij) .
\eeq
The quantity $t(i)$ denotes the kinetic energy of particle $i$, which in case of
nucleons is characterized by position ${\bf r}_i$ (or momentum ${\bf p}_i$), spin $\sigma _i$ and isospin
$\tau _i$, while $v(ij)$ denotes the two-body interaction. Equation (\ref{2.1}) is equivalent to a coupled set of equations for the reduced density matrices of rank $n$,
\bea \label{2.3}
&&\rho _n(1\cdots n, 1' \cdots n') =  \\
&&\frac{1}{(A-n)!} \  Tr_{(n+1, \cdots ,A)}  \rho _A(1 \cdots A,1' \cdots n',n+1 \cdots A) ,
\nonumber
\eea
known as the BBGKY hierarchy \cite{KadanoffBaym,balescu:1975}, i.e.
\bea \label{2.4}
&&i\frac{\partial}{\partial t}\ \rho _n = \\
&&\left[\sum_{i=1}^n  t(i) , \rho _n\right]
 + \left[\sum_{i<j}^{n-1}  v(ij) , \rho _n \right] + \sum_{i=1}^n \left[v(i,n+1) ,
\rho _{n+1} \right]
\nonumber
\eea
for $1 \leq  n \leq  A$ with $\rho _{A+1}=0$. In these equations the time evolution of $\rho_n$ couples to $\rho_{n+1}$ and truncation schemes are necessary to obtain adequate solutions to the set of equations (\ref{2.4}). Different expansion and truncation schemes have been proposed and investigated in the past
\cite{Krieg:1991,Gherega:1993exu,Tohyama:2010hj,Tohyama:2013eta,Schuck:2020kpj} and compared to the exact solution in case of simple few-body or few-level systems \cite{Litvinova:2019kwg,Tohyama:2019qro,Tohyama:2017ssf,Schuck:2016ted}.
Depending on the model systems different approximation schemes are found to be suited in each case.

In order to obtain a suitable truncation scheme for the nuclear many-body problem and in particular for the dynamics of low energy heavy-ion reactions the density matrices $\rho _n$ are rewritten in terms of
antisymmetrized products of density matrices of lower order, i.e. \cite{Wang:1984cg}
	\begin{widetext}
\beq \label{2.5a}
\rho _1(11') = \rho (11') ,
\eeq
\beq \label{2.5b}
\rho _2(12,1'2') = {\cal A}_{1'2'}\rho (11')\rho (22') + c_2(12,1'2') =
\rho _{20}(12,1'2') + c_2(12,1'2')
= {\cal A}_{12}\rho (11')\rho (22') + c_2(12,1'2') ,
\eeq
$$
\rho _3(123,1'2'3') =
{\cal S}_{1'} \rho (11') \rho_{20} (23,2'3') +
{\cal S}_{1'}\rho (11')c_2(23,2'3') +
{\cal S}_{2'}\rho (22')c_2(13,1'3') + {\cal S}_{3'}\rho (33')c_2(12,1'2') +
c_3(123,1'2'3')
$$
\beq \label{2.5c}
= {\cal S}_1 \rho (11')\rho_{20} (23,2'3') + {\cal S}_1\rho (11')c_2(23,2'3')
+ {\cal S}_2\rho (22')c_2(13,1'3') + {\cal S}_3\rho (33')c_2(12,1'2') +
c_3(123,1'2'3')
\eeq
with the two- and three-body antisymmetrization operators
\beq \label{2.5d}
{\cal A}_{ij} = 1 - P_{ij} ; \hp   {\cal A}_{i'j'} = 1 - P_{i'j'}
\eeq
$$
{\cal S}_{1'} = 1-P_{1'2'}-P_{1'3'} ; \hp  {\cal S}_1 = 1-P_{12}-P_{13}
$$
$$
{\cal S}_{2'} = 1-P_{1'2'}-P_{2'3'} ; \hp  {\cal S}_2 = 1-P_{12}-P_{23}
$$
\beq \label{2.5e}
{\cal S}_{3'} = 1-P_{1'3'}-P_{2'3'} ; \hp  {\cal S}_3 = 1-P_{13}-P_{23}
\eeq
where $P_{ij}$ denotes the exchange operator between particle $i$ and $j$. The
cluster decomposition for $\rho _4$ -- that is needed for the equation of
motion for $\rho _3$ -- is given by
$$
\rho _4(1234,1'2'3'4') =
{\cal A}_{1'2'}{\cal S}_{3'}\Lambda _{4'}\rho (11')\rho (22')\rho (33')\rho (44')
+ \Gamma_{1'} \rho_{20} (12,1'2') c_2(34,3'4')$$
$$
+ \Gamma_{2'} \rho_{20} (13,1'3') c_2(24,2'4')
+ \Gamma_{3'} \rho_{20} (14,1'4') c_2(23,2'3')+ \Gamma_{3'} \rho_{20} (23,2'3') c_2(14,1'4')
$$
$$
+ \Gamma_{2'} \rho_{20} (24,2'4') c_2(13,1'3')+ \Gamma_{1'} \rho_{20} (34,3'4') c_2(12,1'2')
+ \Gamma _{1'}c_2(12,1'2')c_2(34,3'4')$$
$$
 + \Gamma _{2'}c_2(13,1'3')c_2(24,2'4')
+ \Gamma _{3'}c_2(14,1'4')c_2(23,2'3')
+ \Lambda _{1'}\rho (11')c_3(234,2'3'4')$$
$$
 +
\Lambda _{2'}\rho (22')c_3(134,1'3'4')
+ \Lambda _{3'}\rho (33')c_3(124,1'2'4') +
\Lambda _{4'}\rho (44')c_3(123,1'2'3')
$$
\beq \label{A.1}
+ c_4(1234,1'2'3'4')
\eeq
with the four-body antisymmetrization operators
$$
\Gamma _{1'} = (1-P_{1'3'}-P_{1'4'}-P_{2'3'}-P_{2'4'}+P_{1'3'}P_{2'4'}); \hp
\Gamma _1= (1-P_{13}-P_{14}-P_{23}-P_{24}+P_{13}P_{24})
$$
$$
\Gamma _{2'} = (1-P_{1'2'}-P_{1'4'}-P_{2'3'}-P_{3'4'}+P_{1'2'}P_{3'4'}); \hp
\Gamma _2= (1-P_{12}-P_{14}-P_{23}-P_{34}+P_{12}P_{34})
$$
$$
\Gamma _{3'} = (1-P_{1'2'}-P_{1'3'}-P_{2'4'}-P_{3'4'}+P_{1'3'}P_{2'4'}); \hp
\Gamma _3= (1-P_{12}-P_{13}-P_{24}-P_{34}+P_{13}P_{24})
$$
$$
\Lambda _{1'} = (1-P_{1'2'}-P_{1'3'}-P_{1'4'});  \hp  \Lambda _1 =
(1-P_{12}-P_{13}-P_{14});
$$
$$
\Lambda _{2'} =
(1-P_{2'1'}-P_{2'3'}-P_{2'4'}); \hp   \Lambda _2 = (1-P_{21}-P_{23}-P_{24});
$$
$$
\Lambda _{3'} = (1-P_{3'1'}-P_{3'2'}-P_{3'4'}); \hp   \Lambda _3 =
(1-P_{31}-P_{32}-P_{34});
$$
\beq \label{A.2}
\Lambda _{4'} = (1-P_{4'1'}-P_{4'2'}-P_{4'3'});  \hp  \Lambda _4 =
(1-P_{41}-P_{42}-P_{43})
\eeq
We obtain (after some lengthy algebra)
$$
i\frac{\partial}{\partial t}\ \rho (11';t) = [t(1)-t(1')]\rho (11';t)
 + Tr_{(2=2')} [v(12){\cal A}_{12}-v(1'2'){\cal A}_{1'2'}]
 \rho (11';t)\rho (22';t)
$$
\beq \label{2.6}
+ Tr_{(2=2')} [v(12)-v(1'2')] c_2(12,1'2';t)
\eeq
for the time-evolution of the 1-body density $\rho (11';t)$ and
$$
i\frac{\partial}{\partial t}\ c_2(12,1'2';t) = [t(1)+t(2)-t(1')-t(2')]c_2(12,1'2';t)
$$
$$
+ Tr_{(3=3')}
[v(13){\cal A}_{13}+v(23){\cal A}_{23}-v(1'3'){\cal A}_{1'3'}- v(2'3')%
{\cal A}_{2'3'}] \rho (33';t) c_2(12,1'2';t)
$$
$$
+ [v(12)-v(1'2')]\rho _{20}(12,1'2')
$$
$$
- Tr_{(3=3')} [v(13)\rho (23';t)\rho _{20}(13,1'2';t) -
v(1'3')\rho (32';t)\rho _{20}(12,1'3';t)
$$
\beq \label{2.7a}
+ v(23)\rho (13';t)\rho _{20}(32,1'2';t) -
v(2'3')\rho (31';t)\rho _{20}(12,3'2';t)]
\eeq
$$
+ [v(12)-v(1'2')]c_2(12,1'2';t)
$$
$$
- Tr_{(3=3')}
[v(13)\rho (23';t)c_2(13,1'2';t) - v(1'3')\rho (32';t)c_2(12,1'3';t)
$$
\beq \label{2.7b}
+ v(23)\rho (13';t)c_2(32,1'2';t) - v(2'3')\rho (31';t)c_2(12,3'2';t)]
\eeq
$$
+ Tr_{(3=3')}
([v(13){\cal A}_{13}{\cal A}_{1'2'}-v(1'3'){\cal A}_{1'3'}{\cal A}_{12}]%
\rho (11';t) c_2(22,2'3';t)
$$
\beq \label{2.7c}
+ [v(23){\cal A}_{23}{\cal A}_{1'2'}-
v(2'3'){\cal A}_{2'3'}{\cal A}_{12}] \rho (22';t) c_2(13,1'3';t))
\eeq
\beq \label{2.7d}
+ Tr_{(3=3')} [v(13)+v(23)-v(1'3')-v(2'3')] c_3(123,1'2'3';t)
\eeq
for the time-evolution of the 2-body correlation $c_2(12,1'2';t)$.

The equation of motion for the 3-body correlation $c_3$ amounts to
$$
i\frac{\partial}{\partial t}\ c_3(123,1'2'3';t) =
[t(1)+t(2)+t(3)-t(1')-t(2')-t(3')] c_3(123,1'2'3';t)
$$
$$
+ [v(12)+v(13)+v(23)-v(1'2')-v(1'3')-v(2'3')] c_3(123,1'2'3';t)
$$
$$
+ [(v(12)+v(13)){\cal S}_{1'}-(v(1'2')+v(1'3')){\cal S}_1]
\rho (11';t) c_2(23,2'3';t)
$$
$$
+ [(v(12)+v(23)){\cal S}_{2'}-(v(1'2')+v(2'3')){\cal S}_2]
\rho (22';t) c_2(13,1'3';t)
$$
\beq \label{2.8a}
+ [(v(13)+v(23)){\cal S}_{3'}-(v(1'3')+v(2'3')){\cal S}_3]
\rho (33';t) c_2(12,1'2';t)
\eeq
$$
- Tr_{(4=4')}
\{[v(14){\cal S}_{1'}-v(1'4'){\cal S}_1]
(P_{24}+P_{34}+P_{2'4'}+P_{3'4'}) \rho (11';t)\rho (44';t)c_2(23,2'3';t)
$$
$$
+ [v(24){\cal S}_{2'}-v(2'4'){\cal S}_2] (P_{1'4'}+P_{3'4'}+P_{14}+P_{34})
\rho (22';t)\rho (44';t)c_2(13,1'3';t)
$$
$$
+ [v(34){\cal S}_{3'}-v(3'4'){\cal S}_3] (P_{14}+P_{24}+P_{1'4'}+P_{2'4'})
\rho (33';t)\rho (44';t)c_2(12,1'2';t)
$$
$$
+ [(v(24)+v(34)){\cal S}_{1'}-(v(2'4')+v(3'4')){\cal S}_1]
\rho (41';t)\rho (14';t)c_2(23,2'3';t)
$$
$$
+ [(v(14)+v(34)){\cal S}_{2'}-(v(1'4')+v(3'4')){\cal S}_2]
\rho (42';t)\rho (24';t)c_2(13,1'3';t)
$$
\beq \label{2.8b}
+ [(v(14)+v(24)){\cal S}_{3'}-(v(1'4')+v(2'4')){\cal S}_3]
\rho (43';t)\rho (34';t)c_2(12,1'2';t)\}
\eeq
$$
+ Tr_{(4=4')}
\{[v(14)+v(24)+v(34)-v(1'4')-v(2'4')-v(3'4')]
\cdot \{{\cal S}_{3'}c_2(12,3'4';t)c_2(34,1'2';t)
$$
$$
+{\cal S}_{2'}c_2(13,2'4';t)c_2(24,1'3';t) +
{\cal S}_{1'}c_2(14,2'3';t)c_2(23,1'4';t)\}
$$
$$
+ [(v(24)+v(34)){\cal S}_{1'}-(v(2'4')+v(3'4')){\cal S}_1]
c_2(14,1'4';t)c_2(23,2'3';t)
$$
$$
+ [(v(14)+v(34)){\cal S}_{2'}-(v(1'4')+v(3'4')){\cal S}_2]
c_2(13,1'3';t)c_2(24,2'4';t)
$$
\beq \label{2.8c}
+ [(v(14)+v(24)){\cal S}_{3'}-(v(1'4')+v(2'4')){\cal S}_3]
c_2(12,1'2';t)c_2(34,3'4';t)\}
\eeq
$$
+ Tr_{(4=4')}
\{[v(14)+v(24)+v(34)-v(1'4')-v(2'4')-v(3'4')]
$$
$$
\cdot (1-P_{14}-P_{24}-P_{34}-P_{1'4'}-P_{2'4'}-P_{3'4'})
\rho (44';t)c_3(123,1'2'3';t)
$$
$$
+ [v(14){\cal S}_{1'}-v(1'4'){\cal S}_1] \rho (11';t)c_3(234,2'3'4';t)
$$
$$
+ [v(24){\cal S}_{2'}-v(2'4'){\cal S}_2] \rho (22';t)c_3(134,1'3'4';t)
$$
\beq \label{2.8d}
+ [v(34){\cal S}_{3'}-v(3'4'){\cal S}_3] \rho (33';t)c_3(124,1'2'4';t)\}
\eeq
\beq \label{2.8e}
+ Tr_{(4=4')}
[v(14)+v(24)+v(34)-v(1'4')-v(2'4')-v(3'4')] c_4(1234,1'2'3'4';t)
\eeq
and no longer depends on interaction terms of order $v\rho \rho \rho $ and
$Tr_4 \ (v\rho \rho \rho \rho)$ which drop out.

Following Refs. \cite{cassing1987numerical,cassing1988dynamical,cassing1990correlation,cassing1992self} we introduce the bare mean-field
hamiltonian by
$$
h(i) = t(i) + U(i) = t(i) + Tr_{(n=n')} v(in){\cal A}_{in}\rho (nn';t);
$$
\beq \label{2.9}
h(i') = t(i') + U(i') = t(i') + Tr_{(n=n')}
v(i'n'){\cal A}_{i'n'}\rho (nn';t),
\eeq
the particle-particle and hole-hole (pp-hh) blocking operator by
\beq \label{2.10}
Q^=_{ij} = 1 - Tr_{(n=n')} (P_{in}+P_{jn})\rho (nn';t); \hp
Q^=_{i'j'} = 1 - Tr_{(n=n')} (P_{i'n'}+P_{j'n'})\rho (nn';t) ,
\eeq
and the effective (pp-hh) in-medium interaction $V^=(ij)$ via
\beq \label{2.11}
V^=(ij) = Q^=_{ij} v(ij); \hp
V^=(i'j') = Q^=(i'j') v(i'j')
\eeq
which act on all terms to the right.

\vp
The equations of motion for $\rho $, $c_2$ and $c_3$ then reduce to
\beq \label{2.12}
i\frac{\partial}{\partial t}\ \rho (11';t) = [h(1)-h(1')]\rho (11';t)
 + Tr_{(2=2')} [v(12)-v(1'2')]c_2(12,1'2';t) ,
\eeq
$$
i\frac{\partial}{\partial t}\ c_2(12,1'2';t) = [\sum_{i=1}^2  h(i)
- \sum^{2'}_{i'=1'} h(i')] c_2(12,1'2';t)
$$
\beq \label{2.13a}
+ [V^=(12) - V^=(1'2')] \rho _{20}(12,1'2';t)
\eeq
\beq \label{2.13b}
+ [V^=(12) - V^=(1'2')] c_2(12,1'2';t)
\eeq
$$
+ Tr_{(3=3')}
\{[v(13){\cal A}_{13}{\cal A}_{1'2'} -
v(1'3'){\cal A}_{1'3'}{\cal A}_{12}] \rho (11';t) c_2(23,2'3';t)
$$
\beq \label{2.13c}
+ [v(23){\cal A}_{23}{\cal A}_{1'2'} - v(2'3'){\cal A}_{2'3'}{\cal A}_{12}]
\rho (22';t) c_2(13,1'3';t)\}
\eeq
\beq \label{2.13d}
+ Tr_{(3=3')} [v(13)+v(23) - v(1'3')-v(2'3')] c_3(123,1'2'3';t) ,
\eeq

and
$$
i\frac{\partial}{\partial t}\ c_3(123,1'2'3';t) = [\sum_{i=1}^3 h(i)
- \sum_{i'=1'}^{3'} h(i')] c_3(123,1'2'3';t)
$$
$$
+  [V^=(12){\cal S}_{1'}+V^=(13){\cal S}_{1'} -
V^=(1'2'){\cal S}_1-V^=(1'3'){\cal S}_1] \rho (11';t)c_2(23,2'3';t)
$$
$$
+  [V^=(12){\cal S}_{2'}+V^=(23){\cal S}_{2'} -
V^=(1'2'){\cal S}_2-V^=(2'3'){\cal S}_2] \rho (22';t)c_2(13,1'3';t)
$$
\beq \label{2.14a}
+  [V^=(13){\cal S}_{3'}+V^=(23){\cal S}_{3'} -
V^=(1'3'){\cal S}_3-V^=(2'3'){\cal S}_3] \rho (33';t)c_2(12,1'2';t)
\eeq
$$
- Tr_{(4=4')}
\{[v(14){\cal S}_{1'}(P_{2'4'}+P_{3'4'})-v(1'4'){\cal S}_1(P_{24}+P_{34})%
] \rho (44';t)\rho (11';t)c_2(23,2'3';t)
$$
$$
+ [v(24){\cal S}_{2'}(P_{1'4'}+P_{3'4'})-v(2'4'){\cal S}_2(P_{14}+P_{34})]
\rho (44';t)\rho (22';t)c_2(13,1'3';t)
$$
\beq \label{2.14d}
+ [v(34){\cal S}_{3'}(P_{1'4'}+P_{2'4'})-v(3'4'){\cal S}_3(P_{14}+P_{24})]
\rho (44';t)\rho (33';t)c_2(12,1'2';t)\}
\eeq
$$
+ Tr_{(4=4')}
\{[v(14)+v(24)+v(34)-v(1'4')-v(2'4')-v(3'4')]\cdot \{{\cal S}_{1'}c_2(14,%
2'3';t)c_2(23,1'4';t)
$$
$$
+{\cal S}_{2'}c_2(13,2'4';t)c_2(24,1'3';t)+{\cal S}_{3'}c_2(12,3'4';t)c_2(34,%
1'2';t)\}
$$
$$
+ [(v(24)+v(34)){\cal S}_{1'}-(v(2'4')+v(3'4')){\cal S}_1]
c_2(14,1'4';t)c_2(23,2'3';t)
$$
$$
+ [(v(14)+v(34)){\cal S}_{2'}-(v(1'4')+v(3'4')){\cal S}_2]
c_2(13,1'3';t)c_2(24,2'4';t)
$$
\beq \label{2.14e}
+ [(v(14)+v(24)){\cal S}_{3'}-(v(1'4')+v(2'4')){\cal S}_3]
c_2(12,1'2';t)c_2(34,3'4';t)\}
\eeq
\beq \label{2.14b}
+  [V^=(12) + V^=(13) + V^=(23) - V^=(1'2') - V^=(1'3') - V^=(2'3')]
 c_3(123,1'2'3';t)
\eeq
$$
+ Tr_{(n=n')}
\{[v(1n){\cal A}_{1n}{\cal S}_{1'} - v(1'n'){\cal A}_{1'n'}{\cal S}_1]
\rho (11';t) c_3(n23,n'2'3';t)
$$
$$
+[v(2n){\cal A}_{2n}{\cal S}_{2'} - v(2'n'){\cal A}_{2'n'}{\cal S}_2]
\rho (22';t) c_3(1n3,1'n'3';t)
$$
\beq \label{2.14c}
+[v(3n){\cal A}_{3n}{\cal S}_{3'} - v(3'n'){\cal A}_{3'n'}{\cal S}_3]
\rho (33';t) c_3(12n,1'2'n';t)\}
\eeq
\beq \label{2.14f}
+ Tr_{(4=4')}
[v(14)+v(24)+v(34)-v(1'4')-v(2'4')-v(3'4')] c_4(1234,1'2'3'4';t)
\eeq
where a couple of interaction terms have been rearranged.

\vp
These equations suggest to define the additional operators
\beq \label{2.15}
V^\perp _2(i,i') = Tr_{(n=n')}
[v(in){\cal A}_{in}{\cal A}_{1'2'} - v(i'n'){\cal A}_{i'n'}{\cal A}_{12}]
\rho (ii';t)P_{in}P_{i'n'}
\eeq
for the two-body correlation case and
$$
V^\perp _3(i,i') = Tr_{(n=n')}
[v(in){\cal A}_{in}{\cal S}_{i'} - v(i'n'){\cal A}_{i'n'}{\cal S}_i]
\rho (ii';t)P_{in}P_{i'n'} ,
$$
\beq \label{2.16}
V^\perp _4(i,i') = Tr_{(n=n')}
[v(in){\cal A}_{in}\Lambda _{i'} - v(i'n'){\cal A}_{i'n'}\Lambda _i]
\rho (ii';t)P_{in}P_{i'n'}
\eeq
for the three- and four-body correlation case, where all permutation and
antisymmetrization operators act to the right. The operators (\ref{2.15}) and (\ref{2.16})
describe interactions in the various particle-hole $(p-h)$ channels \cite{cassing1990correlation,cassing1992self} (see
below).

\vp
Introducing, furthermore, adjoint operators via the relations
\beq \label{2.17}
(O(ij)c_n)^{\dagger}  = c_n O(ij)^{\dagger}  = O(i'j')c_n
\eeq
and a commutator via
\beq \label{2.18}
[O , c_n] = Oc_n - c_n O^{\dagger}
\eeq
the equations of motion can be written in more compact form, i.e.
\beq \label{2.19}
i\frac{\partial}{\partial t}\ \rho  - [h(1) , \rho ] =  Tr_{(2=2')}
[v(12) , c_2] ;
\eeq
\beq \label{2.20}
i\frac{\partial}{\partial t}\ c_2 - [\sum_{i=1}^2 h(i) , c_2]
 - [V^=(12) , c_2] - \sum_{i=1}^2  V^\perp _2(i,i') c_2 =
[V^=(12) , \rho _{20}] + Tr_{(3=3')} [ \sum_{i=1}^3 v(i3) , c_3]
\eeq
and
$$
i\frac{\partial}{\partial t}\ c_3 - [\sum_{i=1}^3 h(i) , c_3]
 - [\sum_{i<j}^2 V^=(ij) , c_3] - \sum_{i=1}^3  V^\perp _3(i,i') c_3 =
$$
\beq \label{2.21a}
+ [{\cal S}_3(V^=(13)+V^=(23)){\cal S}_{3'} , \rho (33';t)c_2(12,1'2';t)]
\eeq
\beq \label{2.21b}
- Tr_{(4=4')} [{\cal S}_1v(14){\cal S}_{1'}{\cal A}_{2'3'}P_{24} ,
\rho (11';t)\rho (22';t)c_2(34,3'4';t)]
\eeq
$$
+ Tr_{(4=4')}   \{[{\cal S}_3(v(14)+v(24)+v(34)){\cal S}_{3'} ,
c_2(12,3'4';t) c_2(34,1'2';t)]
$$
\beq \label{2.21c}
+ [{\cal S}_3(v(14)+v(24)){\cal S}_{3'} , c_2(12,1'2';t)c_2(34,3'4';t)]\}
\eeq
\beq \label{2.21d}
+ Tr_{(4=4')} [ \sum_{i=1}^3 v(i4) , c_4(1234,1'2'3'4';t)] .
\eeq

	\end{widetext}

The set of equations (\ref{2.19}) to (\ref{2.21d}) describe the nonperturbative
time-evolution of arbitrary Fermion systems interacting via two-body forces and
avoid any double counting of interactions.

\vp
Various limiting cases of these equations are known \cite{cassing1987numerical, cassing1988dynamical,cassing1990correlation,cassing1992self}.
 When neglecting two-body correlations $c_2$ in (\ref{2.19}) we recover the time-dependent Hartree-Fock theory ({TDHF}) for weakly interacting systems, while $-i  Tr_{(2=2')}
[v(12) , c_2]$ gives a collision term. In Eq. (\ref{2.20}) the terms $[V^=(12) , \rho _{20}]$  describe
residual nucleon-nucleon scattering in the Born approximation ({BA}),
the additional terms   $[V^=(12) , c_2]$ lead to
time-dependent $G$-matrix theory ({TDGMT}),  while the contribution with $V^\perp _2(i,i') c_2$ lead to additional $2p-2h$ interactions
that are important e.g. for the damping of giant resonances \cite{de1992nonperturbative}.
The physical contents of these equations becomes more transparent within an arbitrary single-particle basis although the compact form of (40) to (47) gets partly lost.

\subsection{Expansion in a single-particle basis}

Following  \cite{cassing1987numerical,cassing1988dynamical,cassing1990correlation,cassing1992self} we expand $\rho $, $c_2$ and $c_3$  in an arbitrary complete single-particle basis $\psi _\alpha $, i.e.
	\begin{widetext}
\beq \label{3.1a}
\rho (11';t) = \sum_{\alpha \alpha '}
 \rho _{\alpha \alpha '}(t) \psi _\alpha (1) \psi ^*_{\alpha '}(1') ,
\eeq
\beq \label{3.1b}
c_2(12,1'2';t) = \sum_{\alpha \beta \alpha ' \beta '}
 C_{\alpha \beta \alpha '\beta '}(t)
\psi _\alpha (1) \psi _\beta (2) \psi ^*_{\alpha '}(1')
\psi ^*_{\beta '}(2') ,
\eeq
\beq \label{3.1c}
c_3(123,1'2'3';t) = \sum_{\alpha \beta \gamma \alpha '\beta '\gamma '}
C^3_{\alpha \beta \gamma ,\alpha '\beta '\gamma '}(t)
\psi _\alpha (1) \psi _\beta (2) \psi _\gamma (3) \psi ^*_{\alpha '}(1')
\psi ^*_{\beta '}(2') \psi ^*_{\gamma '}(3') .
\eeq
A straight forward (but lengthy) calculation then gives
\beq \label{3.2} 
i\frac{\partial}{\partial t}\ \rho _{\alpha \alpha '} =
\sum_{\lambda } [h_{\alpha \lambda } \rho _{\lambda \alpha '} -
\rho _{\alpha \lambda } h_{\lambda \alpha '}]
+ \sum_{\beta } \sum_{\lambda \gamma}
\{\langle \alpha \beta |v|\lambda \gamma \rangle  C_{\lambda \gamma \alpha '\beta } -
C_{\alpha \beta \lambda \gamma } \langle \lambda \gamma |v|\alpha '\beta \rangle \}
\eeq
for the one-body density matrix while the time evolution of the two-body
correlation matrix is given by
$$
i\frac{\partial}{\partial t}\ C_{\alpha \beta \alpha '\beta '} =
\sum_{\lambda }
\{h_{\alpha \lambda } C_{\lambda \beta \alpha '\beta '} + h_{\beta \lambda }
C_{\alpha \lambda \alpha '\beta '} -
C_{\alpha \beta \lambda \beta '} h_{\lambda \alpha '} -
C_{\alpha \beta \alpha '\lambda } h_{\lambda \beta '}\}
$$
\beq \label{3.3a}
+ \sum_{\lambda \lambda '\gamma \gamma '}
\{Q^=_{\alpha \beta \lambda '\gamma '}
\langle \lambda '\gamma '|v|\lambda \gamma \rangle
(\rho _{20})_{\lambda \gamma \alpha '\beta '} -
(\rho _{20})_{\alpha \beta \lambda '\gamma '}
\langle \lambda '\gamma '|v|\lambda \gamma \rangle  Q^=_{\lambda \gamma \alpha '\beta '}\}
\eeq
\beq \label{3.3b}
+ \sum_{\lambda \lambda ' \gamma \gamma '}
Q^=_{\alpha \beta \lambda '\gamma '} \langle \lambda '\gamma '|v|\lambda \gamma \rangle
C_{\lambda \gamma \alpha '\beta '} - C_{\alpha \beta \lambda '\gamma '}
\langle \lambda '\gamma '|v|\lambda \gamma \rangle  Q^=_{\lambda \gamma \alpha '\beta '}\}
\eeq

\medskip
\beq \label{3.3c}
+ {\cal A}_{\alpha \beta} \sum_{\lambda \lambda '\gamma \gamma '}
\langle \lambda '\gamma '|v|\lambda \gamma \rangle _{\cal A}
[Q^\perp _{\alpha \lambda \alpha '\lambda '} C_{\gamma \beta \gamma '\beta '} +
Q^\perp _{\beta \lambda \beta '\lambda '} C_{\alpha \gamma \alpha '\gamma '}]
\eeq

\medskip
\beq \label{3.3d}
+ \sum_{\lambda \lambda ' \gamma \gamma '}
\langle \lambda '\gamma '|v|\lambda \gamma \rangle
\{\delta _{\alpha \lambda '}C^3_{\lambda \beta \gamma ,\alpha '\beta '\gamma '}
+ \delta _{\beta \lambda '}C^3_{\alpha \lambda \gamma ,\alpha '\beta '\gamma '}
-\delta _{\lambda \alpha '}C^3_{\alpha \beta \gamma ,\lambda '\beta '\gamma '}
-\delta _{\lambda \beta '}C^3_{\alpha \beta \gamma ,\alpha '\lambda '\gamma '}\} ,
\eeq
where we have suppressed the explicit time dependence of the expansion
coefficients. We have, furthermore, introduced the particle-hole $(p-h)$
blocking operator
\beq \label{3.4}
Q^\perp _{\alpha \lambda \alpha '\lambda '} =  \delta _{\alpha \lambda '}
\rho _{\lambda \alpha '} -  \rho _{\alpha \lambda '} \delta _{\alpha '\lambda }
\equiv  (1\cdot \rho  - \rho \cdot 1)_{\alpha \lambda \alpha '\lambda '} = -
Q^{\perp \dag } ,
\eeq
the (pp-hh) blocking operator (cf. (\ref{2.10}))
\beq \label{3.5}
Q^=_{\alpha \beta \lambda '\gamma '} =
\delta _{\alpha \lambda '}\delta _{\beta \gamma '} -
\delta _{\alpha \lambda '}\rho _{\beta \gamma '} -
\rho _{\alpha \lambda '}\delta _{\beta \gamma '} \equiv  1\cdot 1 -
1\cdot \rho  - \rho \cdot 1 = Q^{=\dag },
\eeq
and the mean-field hamiltonian (cf. (\ref{2.9}))
\beq \label{3.6}
h_{\alpha \lambda } = \langle \alpha |t|\lambda \rangle  + \sum_{\gamma \gamma '}
\langle \alpha \gamma '|v|\lambda \gamma \rangle _{\cal A} \ \rho _{\gamma \gamma '} ,
\eeq
while the uncorrelated two-body density matrix is given by (cf.(\ref{2.5b}))
\beq \label{3.7}
(\rho _{20})_{\alpha \beta \alpha '\beta '} = \rho _{\alpha \alpha '}
\rho _{\beta \beta '} - \rho _{\alpha \beta '} \rho _{\beta \alpha '} .
\eeq
The index ${\cal A}$ denotes antisymmetrization of the two-body interaction
$v$, i.e.
 $$
\langle \alpha ' \beta '|v|\alpha \beta \rangle _{\cal A} =
\langle \alpha ' \beta ' |v|\alpha \beta \rangle  - \langle \alpha ' \beta ' |v|\beta \alpha \rangle ,
$$
$$
\langle \alpha ' \beta ' |v|\alpha \beta \rangle = \int d1 \int d2 \ \psi^*_{\alpha '}(1) \psi^*_{\beta '}(2) \ v(12) \ \psi_\alpha(1) \psi_\beta(2) .
$$
\vp
The time evolution for the expansion coefficient
$C^3_{\alpha \beta \gamma ,\alpha '\beta '\gamma '}$ , furthermore, reads
$$
i\frac{\partial}{\partial t}\ C^3_{\alpha \beta \gamma ,\alpha '\beta '\gamma '} 
- \sum_{\lambda }
\{h_{\alpha \lambda } C^3_{\lambda \beta \gamma ,\alpha '\beta '\gamma '} +
h_{\beta \lambda } C^3_{\alpha \lambda \gamma ,\alpha '\beta '\gamma '} +
h_{\gamma \lambda } C^3_{\alpha \beta \lambda ,\alpha '\beta '\gamma '}$$ $$
- C^3_{\alpha \beta \gamma ,\lambda \beta '\gamma '}h_{\lambda \alpha '} -
C^3_{\alpha \beta \gamma ,\alpha '\lambda \gamma '}h_{\lambda \beta '} -
C^3_{\alpha \beta \gamma ,\alpha '\beta '\lambda }h_{\lambda \gamma '}\} =
$$
$$
 \sum_{\lambda \eta}
 \{\langle \alpha \beta |Q^=v|\lambda \eta \rangle _{\cal A} {\cal S}_{\alpha '}
\rho _{\lambda \alpha '} C_{\eta \gamma \beta '\gamma '} - {\cal S}_\alpha
C_{\beta \gamma \eta \gamma '} \rho _{\alpha \lambda }
\langle \lambda \eta |vQ^{=\dagger }|\alpha '\beta '\rangle _{\cal A}
$$
$$
+ \langle \beta \gamma |Q^=v|\lambda \eta \rangle _{\cal A} {\cal S}_{\beta '}
\rho _{\lambda \beta '} C_{\alpha \eta \alpha '\gamma '} - {\cal S}_\beta
C_{\alpha \gamma \alpha '\eta } \rho _{\beta \lambda }
\langle \lambda \eta |vQ^{=\dagger }|\beta '\gamma '\rangle _{\cal A}
$$
\beq \label{3.8a}
+ \langle \gamma \alpha |Q^=v|\lambda \eta \rangle _{\cal A} {\cal S}_{\gamma '}
\rho _{\lambda \gamma '} C_{\eta \beta \alpha '\beta '} - {\cal S}_\gamma
C_{\alpha \beta \eta \beta '} \rho _{\gamma \lambda }
\langle \lambda \eta |vQ^{=\dagger }|\gamma '\alpha '\rangle _{\cal A}\}
\eeq
$$
+ \sum_{\lambda \lambda ' \eta \eta'} \{{\cal S}_\gamma
\rho _{\alpha \lambda '}\rho _{\beta \eta '}
\langle \lambda '\eta '|v|\alpha '\eta \rangle _{\cal A}
\delta _{\lambda \alpha '}C_{\eta \gamma \beta '\gamma '}
- {\cal S}_{\gamma '}C_{\beta \gamma \eta '\gamma '}\delta _{\alpha \lambda '}
\langle \alpha \eta '|v|\lambda \eta \rangle _{\cal A}
\rho _{\lambda \alpha '}\rho _{\eta \beta '}
$$
$$
+ {\cal S}_\alpha
\rho _{\beta \lambda '}\rho _{\gamma \eta '}
\langle \lambda '\eta '|v|\beta '\eta \rangle _{\cal A}
\delta _{\lambda \beta '}C_{\alpha \eta \alpha '\gamma '}- {\cal S}_{\alpha '}C_{\alpha \gamma \alpha '\eta '}\delta _{\beta \lambda '}
\langle \beta \eta '|v|\lambda \eta \rangle _{\cal A}
\rho _{\lambda \beta '}\rho _{\eta \gamma '}
$$
\beq \label{3.8d}
+ {\cal S}_\beta
\rho _{\gamma \lambda '}\rho _{\alpha \eta '}
\langle \lambda '\eta '|v|\gamma '\eta \rangle _{\cal A}
\delta _{\lambda \gamma '}C_{\eta \beta \alpha '\beta '}
- {\cal S}_{\beta '}C_{\alpha \beta \eta '\beta '}\delta _{\gamma \lambda '}
\langle \gamma \eta '|v|\lambda \eta \rangle _{\cal A}
\rho _{\lambda \gamma '}\rho _{\eta \alpha '}  \}
\eeq
$$
+ \sum_{\lambda \eta} \{\langle \alpha \beta |Q^=v|\lambda \eta \rangle
C^3_{\lambda \eta \gamma ,\alpha '\beta '\gamma '} -
C^3_{\alpha \beta \gamma ,\lambda \eta \gamma '}
\langle \lambda \eta |vQ^{=\dagger }|\alpha '\beta '\rangle
+ \langle \alpha \gamma |Q^=v|\lambda \eta \rangle
C^3_{\lambda \beta \eta ,\alpha '\beta '\gamma '} $$
\beq \label{3.8b}
- C^3_{\alpha \beta \gamma ,\lambda \beta '\eta }
\langle \lambda \eta |vQ^{=\dagger }|\alpha '\gamma '\rangle
+ \langle \beta \gamma |Q^=v|\lambda \eta \rangle
C^3_{\alpha \lambda \eta ,\alpha '\beta '\gamma '} -
C^3_{\alpha \beta \gamma ,\alpha '\lambda \eta }
\langle \lambda \eta |vQ^{=\dagger }|\beta '\gamma '\rangle \}
\eeq
\beq \label{3.8c}
+ \sum_{\lambda \lambda ' \eta \eta '}
\langle \lambda '\eta '|v|\lambda \eta \rangle _{\cal A}
\{{\cal S}_\alpha Q^\perp _{\alpha \lambda \alpha '\lambda '}C^3_{\eta \beta \gamma ,\eta '\beta '\gamma '}
+ {\cal S}_\beta  Q^\perp _{\beta \lambda \beta '\lambda '}
C^3_{\alpha \eta \gamma ,\alpha '\eta '\gamma '} + {\cal S}_\gamma
Q^\perp _{\gamma \lambda \gamma '\lambda '}
C^3_{\alpha \beta \eta ,\alpha '\beta '\eta '}\}
\eeq
$$
+ \sum_{\eta \delta \lambda}
\{{\cal S}_{\alpha '}C_{\beta \gamma \alpha '\eta }
\langle \alpha \eta |v|\delta \lambda\rangle  C_{\delta \lambda \beta '\gamma '} - {\cal S}_\alpha
C_{\beta \gamma \delta \lambda} \langle \delta \lambda|v|\alpha '\eta \rangle  C_{\alpha \eta \beta '\gamma '}
+ {\cal S}_{\beta '}C_{\alpha \gamma \beta '\eta }
\langle \beta \eta |v|\delta \lambda\rangle  C_{\delta \lambda \alpha '\gamma '} $$
\beq \label{3.8e}
- {\cal S}_\beta
C_{\alpha \gamma \delta \lambda} \langle \delta \lambda|v|\beta '\eta \rangle  C_{\beta \eta \alpha '\gamma '}
+ {\cal S}_{\gamma '}C_{\alpha \beta \gamma '\eta }
\langle \gamma \eta |v|\delta \lambda\rangle  C_{\delta \lambda \alpha '\beta '} - {\cal S}_\gamma
C_{\alpha \beta \delta \lambda}\langle \delta \lambda|v|\gamma '\eta \rangle  C_{\gamma \eta \alpha '\beta '}\}
\eeq
$$
+ \sum_{\lambda \lambda ' \eta \eta '}
 \{{\cal S}_{\alpha '} (1 + P_{\gamma \beta }P_{\gamma '\beta '})
C_{\lambda \gamma \beta '\eta '} \langle \alpha \eta '|v|\lambda \eta \rangle
C_{\beta \eta \alpha '\gamma '} \delta _{\alpha \lambda '}
- {\cal S}_\alpha  (1 + P_{\gamma \beta }P_{\gamma '\beta '})
C_{\alpha \gamma \beta '\eta '} \langle \lambda '\eta '|v|\alpha '\eta \rangle
C_{\beta \eta \lambda '\gamma '} \delta _{\lambda \alpha '}
$$
$$
+ {\cal S}_{\beta '} (1 + P_{\alpha \gamma }P_{\alpha '\gamma '})
C_{\lambda \gamma \alpha '\eta '} \langle \beta \eta '|v|\lambda \eta \rangle
C_{\alpha \eta \beta '\gamma '} \delta _{\beta \lambda '}
- {\cal S}_\beta  (1 + P_{\alpha \gamma }P_{\alpha '\gamma '})
C_{\beta \gamma \alpha '\eta '} \langle \lambda '\eta '|v|\beta '\eta \rangle
C_{\alpha \eta \lambda '\gamma '} \delta _{\lambda \beta '}
$$
\beq \label{3.8f}
+ {\cal S}_{\gamma '} (1 + P_{\alpha \beta }P_{\alpha '\beta '})
C_{\lambda \beta \alpha '\eta '} \langle \gamma \eta '|v|\lambda \eta \rangle
C_{\alpha \eta \gamma '\beta '} \delta _{\gamma \lambda '}
- {\cal S}_\gamma  (1 + P_{\alpha \beta }P_{\alpha '\beta '})
C_{\gamma \beta \alpha '\eta '} \langle \lambda '\eta '|v|\gamma '\eta \rangle
C_{\alpha \eta \lambda '\beta '} \delta _{\lambda \gamma '}\}
\eeq
$$
+ \sum_{\lambda \lambda ' \eta \eta '}
\langle \lambda '\eta '|v|\lambda \eta \rangle  \{[{\cal S}_{\alpha '}
{\cal A}_{\beta \gamma } C_{\beta \lambda \beta '\gamma '}
\delta _{\gamma \lambda '} - {\cal S}_\alpha  {\cal A}_{\beta '\gamma '}
C_{\beta \gamma \beta '\lambda '} \delta _{\lambda \gamma '}]
C_{\alpha \eta \alpha '\eta '}
$$
\beq \label{3.8g}
+ [{\cal S}_{\beta '} {\cal A}_{\alpha \gamma }
C_{\lambda \gamma \alpha '\gamma '} \delta _{\alpha \lambda '} -
{\cal S}_\beta  {\cal A}_{\alpha '\gamma '} C_{\alpha \gamma \lambda '\gamma '}
\delta _{\lambda \alpha '}] C_{\beta \eta \beta '\eta '}
+ [{\cal S}_{\gamma '} {\cal A}_{\alpha \beta }
C_{\alpha \lambda \alpha '\beta '} \delta _{\beta \lambda '} - {\cal S}_\gamma
{\cal A}_{\alpha '\beta '} C_{\alpha \beta \alpha '\lambda '}
\delta _{\lambda \beta '}] C_{\gamma \eta \gamma '\eta '}\} ,
\eeq

	\end{widetext}
where we have neglected the coupling to explicit four-body correlations $c_4$ such that the set of equations is closed.
The three-body antisymmetrization operators ${\cal S}_\alpha $ are defined as
in (\ref{2.5e}) by replacing $(1 \rightarrow  \alpha )$, $(2 \rightarrow  \beta )$,
$(3 \rightarrow  \gamma )$, $(1' \rightarrow  \alpha ')$, $(2' \rightarrow
\beta ')$ and $(3' \rightarrow  \gamma ').$

\vp
The physical contents of (\ref{3.2}) to (\ref{3.8g}) now becomes transparent when
adopting a 'natural' basis $\psi _\alpha $ that diagonalizes the one-body
density matrix, i.e. $\rho _{\alpha \alpha '} = n_\alpha
\delta _{\alpha \alpha '}$, where $n_\alpha $ denotes the occupation number of
the state $\psi _\alpha $. The term $h_{\alpha \alpha '}n_{\alpha '}$ then
describes the conventional one-loop Hartree-Fock propagation whereas
$\sum _\beta  \langle \alpha \beta |vc_2|\alpha '\beta \rangle  $ and
$\sum _\beta  \langle \alpha \beta |c_2v|\alpha '\beta \rangle  $ describe the
coupling to more complicated configurations and will be attributed to 2-body scatterings.

\subsection{Two-body correlation dynamics}

The equation of motion for the two-body correlation function
$C_{\alpha \beta \alpha '\beta '}(t)$ provides more precise information on the
higher order configurations considered. Whereas the terms with
$h_{\alpha \lambda }$  describe the evolution of the state in the
time-dependent mean field, the limit (\ref{3.3a}) accounts for interactions in the
Born approximation, i.e.
\beq \label{3.9}
\sum_{\lambda \gamma}  \{\langle \alpha \beta |Q^=v|\lambda \gamma \rangle
(\rho _{20})_{\lambda \gamma \alpha '\beta '} -
(\rho _{20})_{\alpha \beta \lambda \gamma }
\langle \lambda \gamma |vQ^{=\dag }|\alpha '\beta '\rangle \}
= \eeq $$
\langle \alpha \beta |v|\alpha '\beta '\rangle _{\cal A} [n_{\alpha '} n_{\beta '}
(1-n_\alpha )(1-n_\beta ) - n_\alpha  n_\beta
(1-n_{\alpha '})(1-n_{\beta '})] ,
$$
where we have used that $Q^=$ becomes diagonal in the 'natural' basis,
\beq \label{3.10}
Q^=_{\alpha \beta \lambda \gamma } = \delta _{\alpha \lambda }
\delta _{\beta \gamma } [1 - n_\alpha  - n_\beta ] .
\eeq
In the Hartree-Fock (HF) limit $(n_\alpha = 0$ or 1) the operator $Q^=$
vanishes for $ph$ or $hp$ states and thus acts as a projector on $pp$ and $hh$ states.

The interaction terms in (\ref{3.3b}) correspond to a resummation of ladder diagrams
\cite{cassing1987numerical} - \cite{cassing1990correlation} which becomes obvious when considering the {stationary  limit of the particular approach including the interaction terms in (\ref{3.3a}) and (\ref{3.3b}))}, i.e.
 	\begin{widetext}
$$
(\omega  - (\epsilon _\alpha +\epsilon _\beta ))
C_{\alpha \beta \alpha '\beta '} - (\omega  -
(\epsilon _{\alpha '}+\epsilon _{\beta '})) C_{\alpha \beta \alpha '\beta '} =
$$
\beq \label{3.11}
\sum_{\lambda \gamma }
(1-n_\alpha -n_\beta ) \langle \alpha \beta |v|\lambda \gamma \rangle  (\rho_2)_{\lambda
\gamma \alpha '\beta '} - \sum_{\lambda \gamma }
(\rho_2)_{\alpha \beta \lambda \gamma } \langle \lambda \gamma |v|\alpha '\beta '\rangle
(1-n_{\alpha '}-n_{\beta '})
\eeq
assuming $h_{\alpha \lambda } = \epsilon _\alpha  \delta _{\alpha \lambda }$,
while $(\rho_2)_{\alpha \beta \alpha '\beta '}$ is given by
\beq \label{3.12}
(\rho_2)_{\alpha \beta \alpha '\beta '} =
 (\rho _{20})_{\alpha \beta \alpha '\beta '} +
C_{\alpha \beta \alpha '\beta '}.
\eeq
Equation (\ref{3.11}) is fulfilled for $(\epsilon  \rightarrow  0^+)$
\beq \label{3.13}
C_{\alpha \beta \alpha '\beta '} = [\omega  -
(\epsilon _\alpha +\epsilon _\beta ) +i\epsilon ]^{-1}(1-n_\alpha -n_\beta )
\sum_{\lambda \gamma}  \langle \alpha \beta |v|\lambda \gamma \rangle  (\rho_2)_{\lambda \gamma \alpha '\beta '}
\eeq
and its hermitian conjugate equation. Defining a $G$-matrix via
\beq \label{3.14}
\langle \alpha \beta |v \rho_2|\alpha '\beta '\rangle  =
\langle \alpha \beta |G \rho _{20}|\alpha '\beta '\rangle
\eeq
and using (\ref{3.12}) we obtain
\beq \label{3.15}
\sum_{\lambda ' \gamma '} \langle \alpha \beta |G|\lambda '\gamma '\rangle
(\rho _{20})_{\lambda '\gamma '\alpha '\beta '} = \eeq 
$$
\sum _{\lambda ' \gamma '} \langle \alpha \beta |v|\lambda '\gamma '\rangle
(\rho _{20})_{\lambda '\gamma '\alpha '\beta '}
+ \sum_{\lambda \gamma}
\langle \alpha \beta |v|\lambda \gamma \rangle
{{(1-n_\lambda -n_\gamma )} \over
 {[\omega  - (\epsilon _\lambda +\epsilon _\gamma ) + i\epsilon ]}}
 \sum_{\lambda ' \gamma '}
 \langle \lambda \gamma |G|\lambda '\gamma '\rangle
(\rho _{20})_{\lambda '\gamma '\alpha '\beta '} ,
$$
which for arbitrary matrix elements of $\rho _{20}$ gives the result
\beq \label{3.16}
\langle \alpha \beta |G|\alpha '\beta '\rangle  = \langle \alpha \beta |v|\alpha '\beta '\rangle  +
\sum_{\lambda \gamma } \langle \alpha \beta |v|\lambda \gamma \rangle
{{(1-n_\lambda -n_\gamma )} \over
 {[\omega  - (\epsilon _\lambda +\epsilon _\gamma ) + i\epsilon ]}}
 \langle \gamma \lambda |G|\alpha '\beta '\rangle ,
\eeq
which is the $G$-matrix equation \cite{Day:1978zz,Brueckner:1955zze}.  Thus the terms (\ref{3.3b}) account for the
'short-range' correlations in the traditional sense. Note that in conventional
Brueckner theory the matrix element of $Q^=$ is approximated by $(1 - n_a - n_b
+ n_\alpha n_\beta ) = (1 - n_\alpha )(1 - n_\beta )$, i.e. $hh$ scattering is
neglected.

The terms in (\ref{3.3c}) account for $ph$ interaction diagrams (fully antisymmetrized)
since also $Q^\perp $ becomes diagonal in the 'natural' basis,
\beq \label{3.18}
Q^\perp _{\alpha \lambda ,\alpha '\lambda '} = \delta _{\alpha \lambda '}
\delta _{\lambda \alpha '} [n_{\alpha '} - n_\alpha ]
\eeq
and excludes $pp$ or $hh$ interactions in the HF limit.

When neglecting the antisymmetrization ${\cal A}_{\alpha \beta }$ in (\ref{3.3c}) and
assuming $Q^=v \equiv  0$, i.e. neglecting $pp$ and $hh$ interactions, the
two-body correlation matrix becomes separable, i.e.
\beq \label{3.19}
C_{\alpha \beta ,\alpha '\beta '}(t) = f_{\alpha \alpha '}(t)
f_{\beta \beta '}(t)
\eeq
and we have
$$
f_{\beta \beta '} \{i\frac{\partial}{\partial t}\ f_{\alpha \alpha '} -
(\epsilon _\alpha  - \epsilon _{\alpha '}) f_{\alpha \alpha '}
 - \sum_{\gamma \gamma '}
(n_{\alpha '} - n_\alpha )\langle \alpha \gamma '|v|\alpha '\gamma \rangle _{\cal A}
f_{\gamma \gamma '}\}
$$
\beq \label{3.20}
+ f_{\alpha \alpha '} \{i\frac{\partial}{\partial t}\ f_{\beta \beta '} -
(\epsilon _\beta  - \epsilon _{\beta '}) f_{\beta \beta '}
 - \sum_{\gamma \gamma '}
(n_{\beta '} - n_\beta )\langle \beta \gamma '|v|\beta '\gamma \rangle _{\cal A}
f_{\gamma \gamma '}\} = 0
\eeq
when assuming again $h_{\alpha \lambda } = \epsilon _\alpha
\delta _{\alpha \lambda }$. Equation (\ref{3.20}) is fulfilled if
$f_{\alpha \alpha '}$ and $f_{\beta \beta '}$ each are the solution of the RPA\footnote{Random Phase Approximation}
equation \cite{cassing1990correlation} 
\beq \label{3.21}
[\omega  - (\epsilon _\alpha  - \epsilon _{\alpha '})]
f_{\alpha \alpha '}(\omega ) = (n_{\alpha '} - n_\alpha ) \sum_{\gamma
\gamma'} \langle \alpha \gamma '|v|\alpha '\gamma \rangle _{\cal A}
 f_{\gamma \gamma '}(\omega )
\eeq
when performing a Fourier transformation in time $(t \rightarrow  \omega )$.
Thus (\ref{3.19}) describes RPA correlations in the specific limit considered and
accounts for 'long-range' correlations or 'density fluctuations' {whereas when including all interaction terms in 
(\ref{3.3a}), (\ref{3.3b}) and (\ref{3.3c}) simultaneously} we describe 'long' and 'short' range correlation phenomena on the same footing without double counting of interactions. We note
that the theory dynamically accounts for the lowest order parquet diagrams \cite{Krieg:1991}.

\subsection{Three-body correlation dynamics}
Within the natural basis ($\rho_{\alpha \beta} = \delta_{\alpha \beta} \ n_\alpha$) the equations for the expansion coefficients read:
$$
i\frac{\partial}{\partial t}\ C^3_{\alpha \beta \gamma ,\alpha '\beta '\gamma '} - (\epsilon_\alpha + \epsilon_\beta + \epsilon_\gamma
- \epsilon_{\alpha '} - \epsilon_{\beta '} - \epsilon_{\gamma '})\ C^3_{\alpha \beta \gamma ,\alpha '\beta '\gamma '} =
$$
$$
 \sum_{\eta} \{
((1-n_\alpha -n_\beta) [\langle \alpha \beta |v|\alpha ' \eta \rangle _{\cal A} \ n_{\alpha '}\  C_{\eta \gamma \beta '\gamma '} - \langle \alpha \beta |v|\beta ' \eta \rangle _{\cal A} \ n_{\beta '}\   C_{\eta \gamma \alpha '\gamma '} -  \langle \alpha \beta |v|\gamma ' \eta \rangle _{\cal A} \ n_{\gamma '}\  C_{\eta \gamma \beta '\alpha '})$$ $$
- (C_{\beta \gamma \eta \gamma '} \ n_{\alpha}\ \langle \alpha \eta |v|\alpha '\beta '\rangle _{\cal A}- C_{\alpha \gamma \eta \gamma '} \ n_{\beta}\ \langle \beta \eta |v|\alpha '\beta '\rangle _{\cal A}
- C_{\beta \alpha \eta \gamma '} \ n_{\gamma}\ \langle \gamma \eta |v|\alpha '\beta '\rangle _{\cal A} ](1-n_{\alpha '} -n_{\beta '}))
$$
$$
+ ( (1-n_{\beta} -n_{\gamma}) [\langle \beta \gamma |v|\beta ' \eta \rangle _{\cal A}\  n_{\beta '}\  C_{\alpha \eta \alpha '\gamma '}
-  \langle \beta \gamma |v|\alpha ' \eta \rangle _{\cal A} \ n_{\alpha '}\  C_{\alpha \eta \beta '\gamma '} -  \langle \beta \gamma |v|\gamma ' \eta \rangle _{\cal A} \ n_{\gamma '}\  C_{\alpha \eta \alpha '\beta '} ])
$$ $$ - ([C_{\alpha \gamma \alpha '\eta } \ n_{\beta}\ \langle \beta \eta |v|\beta '\gamma '\rangle _{\cal A}- C_{\beta \gamma \alpha '\eta } \ n_{\alpha}\ \langle \alpha \eta |v|\beta '\gamma '\rangle _{\cal A}
- C_{\alpha \beta \alpha '\eta } \ n_{\gamma}\ \langle \gamma \eta |v|\beta '\gamma '\rangle _{\cal A}) ] (1-n_{\beta '} -n_{\gamma '})
$$
$$
+ ((1-n_{\alpha} -n_{\gamma}) [\langle \gamma \alpha |v|\gamma ' \eta \rangle _{\cal A} \ n_{\gamma '}\  C_{\eta \beta \alpha '\beta '}
- \langle \gamma \alpha |v|\beta ' \eta \rangle _{\cal A} \ n_{\beta '}\  C_{\eta \beta \alpha '\gamma '} - \langle \gamma \alpha |v|\alpha ' \eta \rangle _{\cal A} \ n_{\alpha '}\  C_{\eta \beta \gamma '\beta '}])
$$ $$ - ([C_{\alpha \beta \eta \beta '} \ n_{\gamma }\ \langle \gamma \eta |v|\gamma '\alpha '\rangle _{\cal A} - C_{\gamma \beta \eta \beta '}\  n_{\alpha}\ \langle \alpha \eta |v|\gamma '\alpha '\rangle _{\cal A}
- C_{\alpha \gamma \eta \beta '} \ n_{\beta}\ \langle \beta \eta |v|\gamma '\alpha '\rangle _{\cal A}) ] (1-n_{\alpha '} -n_{\gamma '})) \}
$$
$$
- \sum_{\eta} \{
(C_{\beta \gamma \eta \gamma '}\langle \alpha \eta |v|\alpha ' \beta '\rangle _{\cal A} n_{\alpha '} n_{\beta '}
- C_{\beta \gamma \eta \beta '}\langle \alpha \eta |v|\alpha ' \gamma '\rangle _{\cal A} n_{\alpha '} n_{\gamma '}
- C_{\beta \gamma \eta \alpha '}\langle \alpha \eta |v|\gamma ' \beta '\rangle _{\cal A} n_{\gamma '} n_{\beta '})$$
$$  -
(n_{\alpha} n_{\beta}\langle \alpha \beta|v|\alpha '\eta \rangle _{\cal A} C_{\eta \gamma \beta '\gamma '}
- n_{\alpha} n_{\gamma}\langle \alpha \gamma|v|\alpha '\eta \rangle _{\cal A} C_{\eta \beta \beta '\gamma '}
- n_{\gamma} n_{\beta} \langle \gamma \beta |v|\alpha '\eta \rangle _{\cal A} C_{\eta \alpha \beta '\gamma '})$$
$$+
(C_{\alpha \gamma \alpha '\eta }\langle \beta \eta |v|\beta ' \gamma '\rangle _{\cal A} n_{\beta '} n_{\gamma '}
- C_{\alpha \gamma \beta '\eta }\langle \beta \eta |v|\alpha ' \gamma '\rangle _{\cal A} n_{\alpha '} n_{\gamma '}
- C_{\alpha \gamma \gamma '\eta }\langle \beta \eta |v|\beta ' \alpha ' \rangle _{\cal A} n_{\beta '} n_{\alpha '}) $$
$$- (
n_{\beta} n_{\gamma}\langle \beta \gamma|v|\beta '\eta \rangle _{\cal A} C_{\alpha \eta \alpha '\gamma '}
- n_{\alpha} n_{\gamma}\langle \alpha \gamma|v|\beta '\eta \rangle _{\cal A} C_{\beta \eta \alpha '\gamma '}
- n_{\beta} n_{\alpha}\langle \beta \alpha|v|\beta '\eta \rangle _{\cal A} C_{\gamma \eta \alpha '\gamma '}) $$
$$+
(C_{\alpha \beta \eta \beta '} \langle \gamma \eta |v|\gamma ' \alpha ' \rangle _{\cal A} n_{\gamma '} n_{\alpha '}
- C_{\alpha \beta \eta \alpha '} \langle \gamma \eta |v|\gamma ' \beta ' \rangle _{\cal A} n_{\gamma '} n_{\beta '}
- C_{\alpha \beta \eta \gamma '} \langle \gamma \eta |v|\beta ' \alpha ' \rangle _{\cal A} n_{\beta '} n_{\eta \alpha '}) $$
\beq \label{3.21a}
- (n_{\gamma} n_{\alpha} \langle \gamma \alpha|v|\gamma '\eta \rangle _{\cal A} C_{\eta \beta \alpha '\beta '}
- n_{\gamma} n_{\beta}\langle \gamma \beta\eta '|v|\gamma '\eta \rangle _{\cal A} C_{\eta \alpha \alpha '\beta '}
- n_{\beta} n_{\alpha}\langle \beta \alpha|v|\gamma '\eta \rangle _{\cal A}  C_{\eta \gamma \alpha '\beta '})
\}
\eeq
$$
+ \sum_{\lambda \eta} \{(1-n_\alpha - n_\beta)\langle \alpha \beta |v|\lambda \eta \rangle
C^3_{\lambda \eta \gamma ,\alpha '\beta '\gamma '} -
C^3_{\alpha \beta \gamma ,\lambda \eta \gamma '}
\langle \lambda \eta |v|\alpha '\beta '\rangle (1-n_{\alpha '}-n_{\beta '})
$$
$$
+ (1-n_\gamma - n_\alpha) \langle \alpha \gamma |v|\lambda \eta \rangle
C^3_{\lambda \beta \eta ,\alpha '\beta '\gamma '} -
C^3_{\alpha \beta \gamma ,\lambda \beta '\eta }
\langle \lambda \eta |v|\alpha '\gamma '\rangle (1-n_{\alpha '}-n_{\gamma '})
$$
\beq \label{3.21b}
+ (1-n_\gamma - n_\beta)\langle \beta \gamma |v|\lambda \eta \rangle
C^3_{\alpha \lambda \eta ,\alpha '\beta '\gamma '} -
C^3_{\alpha \beta \gamma ,\alpha '\lambda \eta }
\langle \lambda \eta |v|\beta '\gamma '\rangle (1-n_{\beta '}-n_{\gamma '})\}
\eeq
$$
+ \sum_{\eta \eta '}
\{\langle \alpha '\eta '|v|\alpha \eta \rangle _{\cal A} (n_\alpha - n_{\alpha '}) C^3_{\eta \beta \gamma ,\eta '\beta '\gamma '}
-\langle \alpha '\eta '|v|\beta \eta \rangle _{\cal A} (n_{\beta} - n_{\alpha '}) C^3_{\eta \alpha \gamma ,\eta '\beta '\gamma '} 
-\langle \alpha '\eta '|v|\gamma \eta \rangle _{\cal A} (n_{\gamma} - n_{\alpha '}) C^3_{\eta \beta \alpha ,\eta '\beta '\gamma '} $$ 
$$
+ \langle \beta '\eta '|v|\alpha \eta \rangle _{\cal A} (n_{\alpha} - n_{\beta '}) C^3_{\beta \eta \gamma ,\alpha '\eta '\gamma '}
-\langle \beta '\eta '|v|\beta \eta \rangle _{\cal A} (n_{\beta} - n_{\beta '}) C^3_{\alpha \eta \gamma ,\alpha '\eta '\gamma '} 
-\langle \beta '\eta '|v|\gamma \eta \rangle _{\cal A} (n_{\gamma} - n_{\beta '}) C^3_{\alpha \eta \beta ,\alpha '\eta '\gamma '} $$
\beq \label{3.21c} +\langle \gamma '\eta '|v|\gamma \eta \rangle _{\cal A} (n_{\gamma} - n_{\gamma '}) C^3_{\alpha \beta \eta ,\alpha '\beta '\eta '}
-\langle \gamma '\eta '|v|\beta \eta \rangle _{\cal A} (n_{\beta} - n_{\gamma '}) C^3_{\alpha \gamma \eta ,\alpha '\beta '\eta '}
-\langle \gamma '\eta '|v|\alpha \eta \rangle _{\cal A} (n_{\alpha} - n_{\gamma '}) C^3_{\gamma \beta \eta ,\alpha '\beta '\eta '} \}
\eeq
where we have discarded terms of order $v c_2 c_2$. The terms of order $v \rho c_3$ describe corrections to the propagator for $c_3$ in the $pp-hh$ channel as well as in the $p-h$ channel while the terms of order $v \rho c_2$ and $v \rho \rho c_2$ are the leading order contributions for the 3-body collision term. Thus -- neglecting the terms with $v \rho c_3$ and $v c_2 c_2$ -- we obtain in leading order
\beq \label{3.22}
\left(i\frac{\partial}{\partial t} - (\epsilon_\alpha + \epsilon_\beta + \epsilon_\gamma
- \epsilon_{\alpha '} - \epsilon_{\beta '} - \epsilon_{\gamma '})\right) \ C^3_{\alpha \beta \gamma ,\alpha '\beta '\gamma '} =
\eeq
$$
 \sum_{\eta} \{
((n_\alpha n_\beta {\bar n}_{\alpha '}+ {\bar n}_\alpha {\bar n}_\beta {n}_{\alpha '})  \langle \alpha \beta |v|\alpha ' \eta \rangle _{\cal A} \  C_{\eta \gamma \beta '\gamma '} - (n_\alpha n_\beta {\bar n}_{\beta '}+ {\bar n}_\alpha {\bar n}_\beta {n}_{\beta '})\langle \alpha \beta |v|\beta ' \eta \rangle _{\cal A} \   C_{\eta \gamma \alpha '\gamma '} $$ $$-  (n_\alpha n_\beta {\bar n}_{\gamma '}+ {\bar n}_\alpha {\bar n}_\beta {n}_{\gamma '})\langle \alpha \beta |v|\gamma ' \eta \rangle _{\cal A} \  C_{\eta \gamma \beta '\alpha '})
- (C_{\beta \gamma \eta \gamma '} \ \langle \alpha \eta |v|\alpha '\beta '\rangle _{\cal A}(n_{\alpha '}n_{\beta '} {\bar n}_\alpha +
{\bar n}_{\alpha '} {\bar n}_{\beta '} {n}_\alpha)$$ $$- C_{\alpha \gamma \eta \gamma '} \ \langle \beta \eta |v|\alpha '\beta '\rangle _{\cal A} (n_{\alpha '}n_{\beta '} {\bar n}_\beta +{\bar n}_{\alpha '} {\bar n}_{\beta '} {n}_\beta)
- C_{\beta \alpha \eta \gamma '} \ \langle \gamma \eta |v|\alpha '\beta '\rangle _{\cal A} ](n_{\alpha '} n_{\beta '} {\bar n}_\gamma +
{\bar n}_{\alpha '} {\bar n}_{\beta '} {n}_\gamma))
$$
$$
+ ( (n_{\beta} n_{\gamma} {\bar n}_{\beta '} + {\bar n}_{\beta} {\bar n}_{\gamma} {n}_{\beta '}) \langle \beta \gamma |v|\beta ' \eta \rangle _{\cal A}\  \  C_{\alpha \eta \alpha '\gamma '}
-  (n_{\beta} n_{\gamma} {\bar n}_{\alpha '} + {\bar n}_{\beta} {\bar n}_{\gamma} {n}_{\alpha '}) \langle \beta \gamma |v|\alpha ' \eta \rangle _{\cal A} \  C_{\alpha \eta \beta '\gamma '} $$ $$ - (n_{\beta} n_{\gamma} {\bar n}_{\gamma '} + {\bar n}_{\beta} {\bar n}_{\gamma} {n}_{\gamma '}) \langle \beta \gamma |v|\gamma ' \eta \rangle _{\cal A} \  C_{\alpha \eta \alpha '\beta '} )
 - (C_{\alpha \gamma \alpha '\eta }  \ \langle \beta \eta |v|\beta '\gamma '\rangle _{\cal A} ({n}_{\beta '} {n}_{\gamma '}  {\bar n}_{\beta}+{\bar n}_{\beta '} {\bar n}_{\gamma '} n_{\beta})$$ $$ - C_{\beta \gamma \alpha '\eta } \ \langle \alpha \eta |v|\beta '\gamma '\rangle _{\cal A} (n_{\beta'} {n}_{\gamma '} {\bar n}_{\alpha}+{\bar n}_{\beta '} {\bar n}_{\gamma '} n_{\alpha})
- C_{\alpha \beta \alpha '\eta } \ \langle \gamma \eta |v|\beta '\gamma '\rangle _{\cal A})  ({n}_{\beta '} {n}_{\gamma '} {\bar n}_{\gamma}+{\bar n}_{\beta '} {\bar n}_{\gamma '} n_{\gamma}))
$$
$$
+ (n_{\alpha} n_{\gamma} {\bar n}_{\gamma '}+ {\bar n}_{\alpha} {\bar n}_{\gamma} {n}_{\gamma '}) \langle \gamma \alpha |v|\gamma ' \eta \rangle _{\cal A} \   C_{\eta \beta \alpha '\beta '}
- (n_{\alpha} n_{\gamma} {\bar n}_{\beta '}+ {\bar n}_{\alpha} {\bar n}_{\gamma} {n}_{\beta '})\langle \gamma \alpha |v|\beta ' \eta \rangle _{\cal A} \  C_{\eta \beta \alpha '\gamma '}$$ $$ - (n_{\alpha} n_{\gamma} {\bar n}_{\alpha '}+ {\bar n}_{\alpha} {\bar n}_{\gamma} {n}_{\alpha '})\langle \gamma \alpha |v|\alpha ' \eta \rangle _{\cal A} \  C_{\eta \beta \gamma '\beta '})
 - (C_{\alpha \beta \eta \beta '} \ \langle \gamma \eta |v|\gamma '\alpha '\rangle _{\cal A}(n_{\alpha '} n_{\gamma '} {\bar n}_{\gamma}+ {\bar n}_{\alpha '} {\bar n}_{\gamma '} {n}_{\gamma}  )$$ $$ - C_{\gamma \beta \eta \beta '}\ \langle \alpha \eta |v|\gamma '\alpha '\rangle _{\cal A}(n_{\alpha '} n_{\gamma '} {\bar n}_{\alpha}+ {\bar n}_{\alpha '} {\bar n}_{\gamma '} {n}_{\alpha}  )
- C_{\alpha \gamma \eta \beta '} \  \langle \beta \eta |v|\gamma '\alpha '\rangle _{\cal A})  (n_{\alpha '} n_{\gamma '} {\bar n}_{\beta}+ {\bar n}_{\alpha '} {\bar n}_{\gamma '} {n}_{\beta}  )) \}
$$
$$
=: {\hat V}^3_{\alpha \beta \gamma, \alpha' \beta' \gamma'}(t)
$$
	\end{widetext}
using the identity
 \beq \label{3.23}
 n_{\beta} n_{\gamma}  + (1-{n}_{\beta} - {n}_{\gamma}) {n}_{\beta '} =n_{\beta} n_{\gamma} {\bar n}_{\beta '} + {\bar n}_{\beta} {\bar n}_{\gamma} {n}_{\beta '}  \eeq
with ${\bar n}_\beta = 1- n_\beta$, which clearly demonstrates the Pauli-blocking. Note that the r.h.s. of (\ref{3.22}) is entirely a function of the occupation numbers and the two-body correlations, which will be determined in leading order below.

\section{Collision terms}

In the context of transport theories the 2-body and 3-body collision terms are of primary interest. In order to obtain transparent results we will consider the leading order terms for $C_{\alpha \beta \alpha' \beta'}(t)$ and $C^3_{\alpha \beta \gamma, \alpha' \beta' \gamma '}(t)$.

\subsection{Two-body case}
In leading order we consider
\beq \label{11.19}
i\frac{\partial}{\partial t}\ \rho  - [h(1) , \rho ] =  Tr_{(2=2')} [v(12) , c_2] ;
\eeq
and
\beq \label{11.20}
i\frac{\partial}{\partial t}\ c_2 - [\sum_{i=1}^2 h(i) , c_2] = [V^=(12) , \rho _{20}] .
\eeq
The collision term in (\ref{11.19}) requires to compute
\beq
\label{14.46} I^2(11';t) := -i Tr_{(2=2')} [v(12), c_2(12,1'2';t)]
\eeq
or in a single-particle basis
\bea
\label{14.46b}
I^2_{\alpha \alpha '}(t) = -i \sum_{\beta } \sum_{\lambda \gamma}
\{\langle \alpha \beta |v|\lambda \gamma \rangle  C_{\lambda \gamma \alpha '\beta }(t) \\
- C_{\alpha \beta \lambda \gamma }(t) \langle \lambda \gamma |v|\alpha '\beta \rangle \},
\nonumber
\eea
i.e. the explicit knowledge of the 2-particle correlation function in an arbitrary
basis\ $|\alpha \rangle$.

To calculate the 2-particle correlation function in leading order we  use a
discrete basis in which the single-particle Hamiltonian
$h_{\alpha \lambda }(t)$ and in particular $\rho_{\alpha \alpha '}(t)$
is diagonal, i.e.
\beq
\label{14.xx} h_{\alpha \lambda }(t) \approx
\epsilon _\alpha (t)\delta _{\alpha \lambda }; \hp \rho _{\alpha \alpha '}(t)
= n_\alpha (t)\delta _{\alpha \alpha '} .
\eeq
In this basis the equation of motion for the expansion coefficients
$C_{\alpha \beta \alpha '\beta '}(t)$  then reduces  to
(omitting the explicit time dependence of all quantities):
\bea
\label{14.47} \{i \frac{\partial}{\partial t} - [\epsilon _\alpha +
\epsilon _\beta - \epsilon _{\alpha '} -\epsilon _{\beta '}]\}
C_{\alpha \beta \alpha '\beta '}(t) =
\eea
$$ \sum_{\lambda \gamma} \{\langle \alpha \beta |Q^=v|\lambda \gamma \rangle
(\rho _{20})_{\lambda \gamma \alpha '\beta '} -
(\rho _{20})_{\alpha \beta \lambda \gamma }
\langle \lambda \gamma |vQ^=|\alpha '\beta '\rangle \}
$$
$$
= \langle \alpha \beta |v|\alpha '\beta '\rangle _{\cal A} [n_{\alpha '} n_{\beta '}
{\bar n}_\alpha  {\bar n}_\beta ) - n_\alpha n_\beta {\bar n}_{\alpha '} {\bar n}_{\beta '})]$$
$$=: \langle \alpha \beta |V_B(t)|\alpha '\beta '\rangle  ,$$
where we have taken advantage of the fact that in this basis $Q^=$ is diagonal
\beq
\label{14.48} Q^=_{\alpha \beta \lambda \gamma } = \delta _{\alpha \lambda }
\delta _{\beta \gamma } [1 - n_\alpha - n_\beta ] .
\eeq
Furthermore, we have used the identity
\bea \label{14.48b}
n_{\alpha '} n_{\beta '} (1-n_\alpha -n_\beta ) - n_\alpha n_\beta (1-n_{\alpha '}-n_{\beta '})
\\ 
= n_{\alpha '} n_{\beta '}
{\bar n}_\alpha {\bar n}_\beta - n_\alpha n_\beta {\bar n}_{\alpha '} {\bar n}_{\beta '}
\nonumber
\eea
with ${\bar n}_\gamma = 1-n_\gamma$.

\begin{figure}[t!]
\begin{center}
  {\includegraphics[width=7cm]{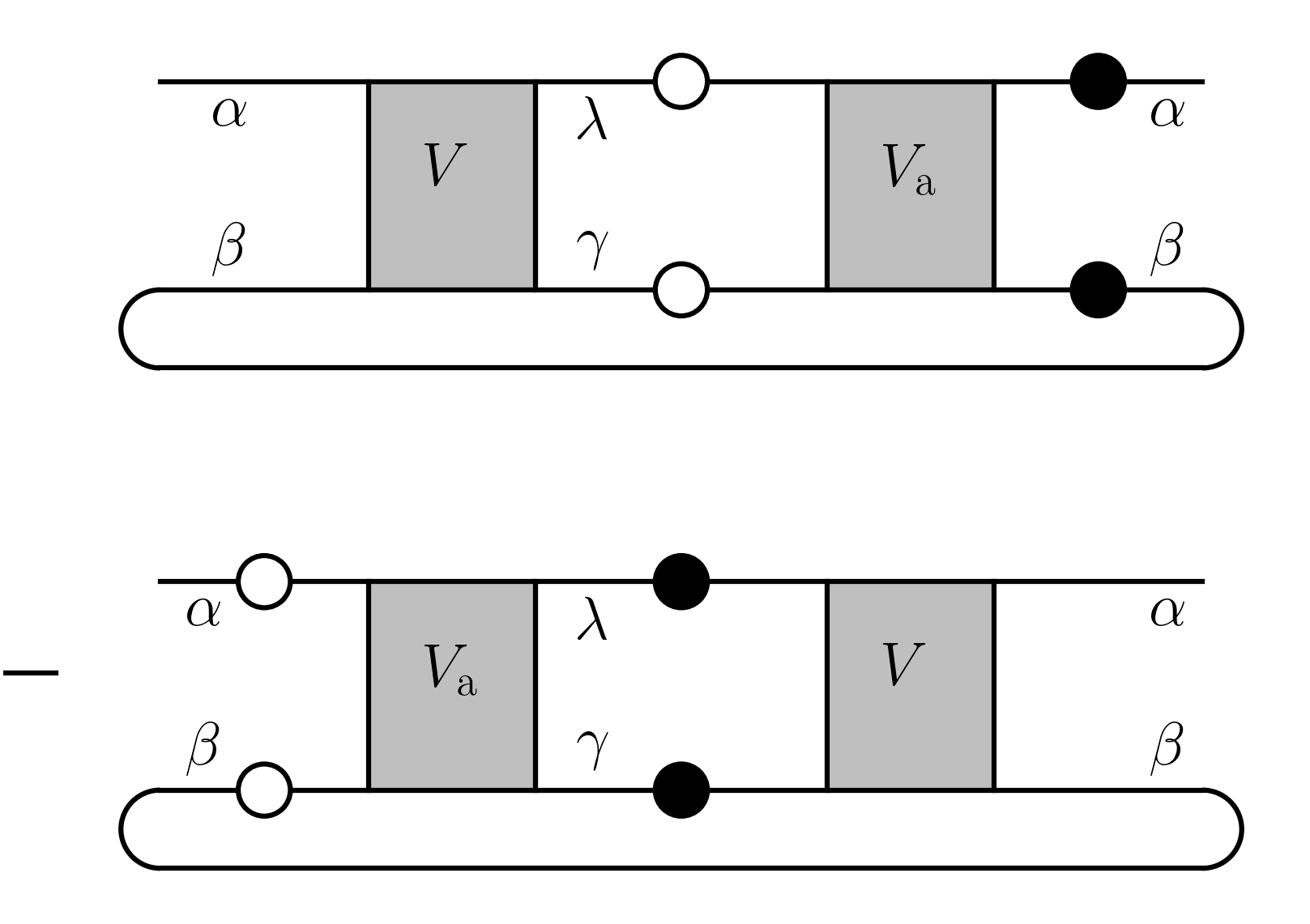} }
 \end{center}
         \caption{Illustration of the collision integral (\ref{14.50}), where the open circles denote a factor $n$ for the occupation number while the full circles denote a factor ${\bar n}$. The closed line corresponds to a summation over particle 2 (or index $\beta$). Here $V_a$ denotes the antisymmetrized matrix element of the interaction $V$. } \label{Fig1}
       \end{figure}

Equation (\ref{14.47}) is a differential equation of first order in time that
can be integrated directly. With regards to approximations to be carried out later (within the framework of the energy conservation in 2-particle collisions), we assume in the following that in particular the single-particle energies $\epsilon _\alpha (t) \approx \epsilon _\alpha $ are weakly varying functions of time.  This approximation is particularly well or exactly fulfilled for the electron states in solids as well as for the single-particle states in a sufficiently large normalization volume.  For a vanishing homogeneous solution of
(\ref{14.47}) $C_{\alpha \beta \alpha '\beta '}(t)$ then is  given by
	\begin{widetext}
\bea
\label{14.49} C_{\alpha \beta \alpha '\beta '}(t) = 
-i \int_{t_0}^t dt'\ \exp \{ -i[\epsilon _\alpha +\epsilon _\beta -\epsilon_{\alpha '}
-\epsilon _{\beta '}](t-t')\}
\cdot \langle \alpha \beta |V_B(t')|\alpha '\beta'\rangle ,
\eea
as one can easily verify by insertion in (\ref{14.47}). Due to the function in the occupation numbers the matrix elements $C_{\alpha \beta \alpha \beta}$ vanish identically and do not contribute to the normalization of the 2-body density matrix $\rho_2$, i.e. 
\beq \label{norm2}
\sum_{\alpha \beta} \ (\rho_{\alpha \alpha} \rho_{\beta  \beta}- \rho_{\alpha \beta} \rho_{\beta  \alpha} + C_{\alpha \beta \alpha \beta}) =  \sum_{\alpha \beta} (n_\alpha n_\beta- n_\alpha \delta_{\alpha \beta}) = A (A-1) 
\eeq
with $A$ denoting the total particle number.

For the diagonal element of the 2-body collision term (\ref{14.46b}) we obtain with
(\ref{14.49})
\beq
\label{14.50} I^2_{\alpha \alpha }(t) = -i \sum_\beta \sum_{\lambda \gamma}
 \{\langle \alpha \beta |v|\lambda \gamma \rangle  C_{\lambda \gamma \alpha \beta }(t) -
C_{\alpha \beta \lambda \gamma }(t) \langle \lambda \gamma |v|\alpha \beta \rangle \}
\eeq
$$
= -\sum_\beta \sum_{\lambda\gamma}
\int^t_{t_0} dt' \{\exp\{-i[\epsilon _\lambda +\epsilon _\gamma -\epsilon _\alpha -\epsilon_\beta ](t-t')\}
\cdot
\langle \alpha \beta |v|\lambda \gamma \rangle \langle \lambda \gamma |V_B(t')|\alpha \beta \rangle
$$
$$
- \exp \{-i[\epsilon _\alpha +\epsilon _\beta -\epsilon _\lambda -\epsilon _\gamma ](t-t')%
\} \langle \alpha \beta |V_B(t')|\lambda \gamma \rangle  \langle \lambda \gamma |v|\alpha \beta \rangle \}
$$
$$
= \sum_\beta \sum_{\lambda \gamma} \int^t_{t_0} dt'\
2\  \cos \{[\epsilon _\alpha +\epsilon _\beta -\epsilon _\lambda -\epsilon _\gamma ](t-t')\}
$$
$$
\cdot \langle \alpha \beta |v|\lambda \gamma \rangle \langle \lambda \gamma |v|\alpha \beta \rangle _%
{\cal A}[n_\lambda(t')n_\gamma(t')\bar n_\alpha(t')\bar n_\beta(t')-n_\alpha(t')n_\beta(t')\bar n_\lambda(t')\bar n_\gamma(t')]
$$
with  $V_B(t')$ from (\ref{14.47}). In (\ref{14.50}) we have used
\beq \label{symm1}
 \langle \alpha \beta |V_B(t')|\lambda \gamma \rangle  \langle \lambda \gamma |v|\alpha \beta \rangle \} = -
 \langle \alpha \beta |v|\lambda \gamma \rangle \langle \lambda \gamma |V_B(t')|\alpha \beta \rangle
 \eeq
	\end{widetext} 
since the function with the occupation numbers changes sign. The collision integral (\ref{14.50}) then can be illustrated as shown in Fig. \ref{Fig1}, where the open circles denote a factor $n$ for the occupation number while the full circles denote a factor ${\bar n}$.

\begin{figure}[t!]
\begin{center}
 {\includegraphics[width=7cm]{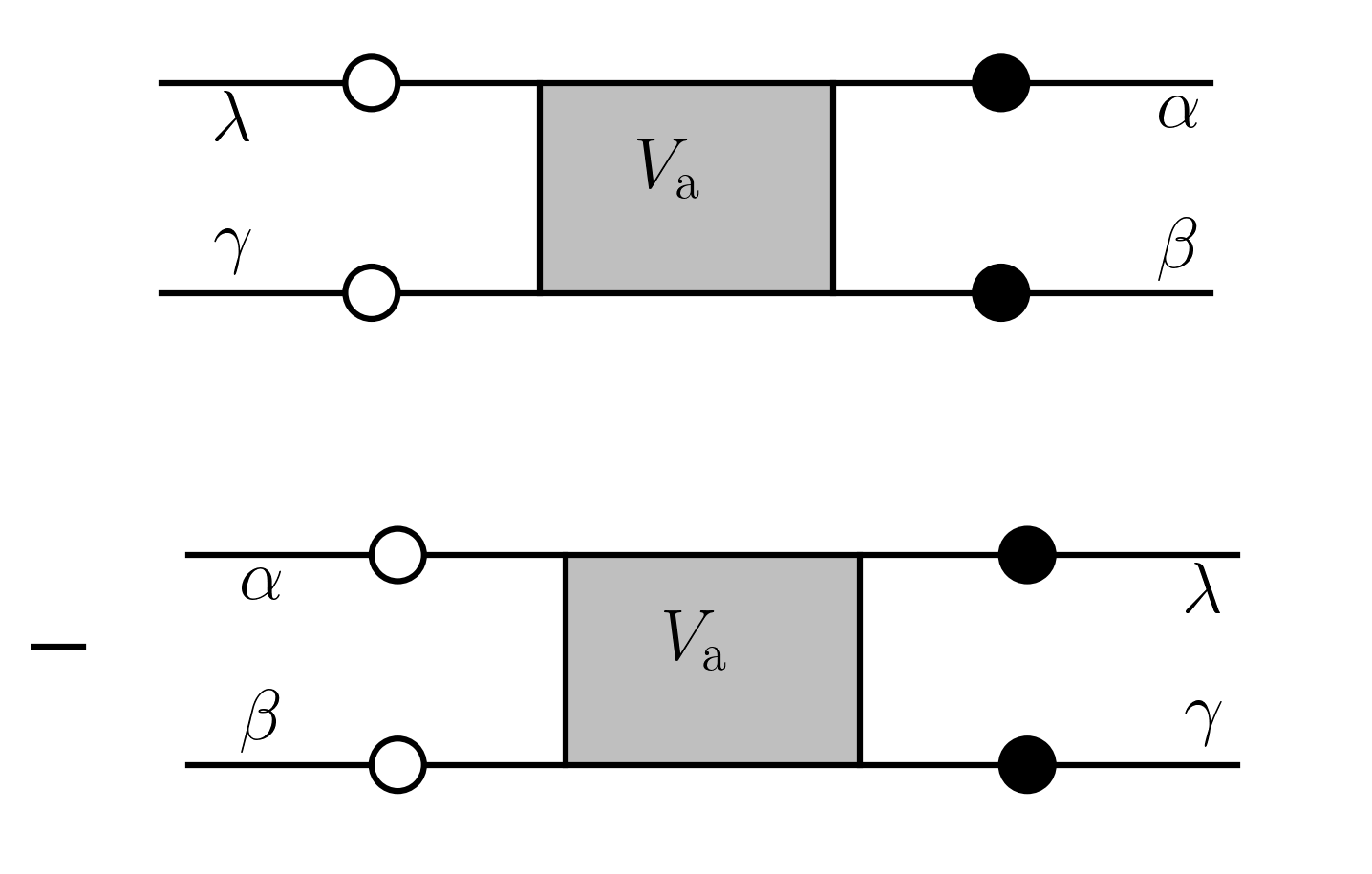} }  
 \end{center}
         \caption{Illustration of the matrix element $C_{\lambda \gamma \alpha \beta}$ in the on-shell limit, where the open circles denote a factor $n$ for the occupation number while the full circles denote a factor ${\bar n}$. Here the box with $V_a$ denotes the antisymmetrized matrix element of the interaction $V$.  } \label{Fig2}
       \end{figure}

The further assumption is that the occupation numbers  $n_\alpha(t') \approx n_\alpha(t)$ are approximately constant in time. In this case we can carry out the time integration in (\ref{14.50}) for systems of low density or weak residual interaction and get:
\bea
\label{14.51}\int^t_{t_0} dt'\ \cos([\epsilon _\alpha +\epsilon _\beta -\epsilon _\lambda -\epsilon _\gamma ](t-t')%
)  \\
\approx \pi\ \delta(\epsilon _\alpha +\epsilon _\beta -\epsilon _\lambda -\epsilon_\gamma) ,
\noindent
\eea
i.e. the energy conservation in the 2-body collisions. Formally this result is obtained by taken the limit $t \rightarrow \infty$. -- Equation (\ref{14.51}) implies that  the time between two subsequent collisions $\tau _s$ is large compared to the actual collision time $\tau _c$, such
that the energy uncertainty associated to $\tau_s$, i.e. $\Delta \epsilon \approx
\hbar/\tau_s$ will be small. -- For the diagonal elements of the collision term we then obtain
	\begin{widetext}
\beq
\label{14.52} I^2_{\alpha \alpha }(t) \approx 2\pi \sum_\beta \sum_{\lambda\gamma}
\delta(\epsilon_\alpha +\epsilon_\beta -\epsilon _\lambda -\epsilon _\gamma %
) \langle \alpha \beta |v|\lambda \gamma \rangle  \langle \lambda \gamma |v|\alpha \beta \rangle _{\cal A}
\cdot [n_\lambda n_\gamma \bar n_\alpha \bar n_\beta - n_\alpha n_\beta \bar n_\lambda \bar n_\gamma ](t) \eeq
in the basis $|\alpha \rangle$ in which the matrix $\rho _{\alpha \alpha '}$ is diagonal. Note that the result (\ref{14.52}) is obtained directly when approximating $C_{\lambda \gamma \alpha \beta}$ by
\beq \label{onshell}
C_{\lambda \gamma \alpha \beta}(t) = i \pi\  \delta(\epsilon _\alpha +\epsilon _\beta -\epsilon _\lambda -\epsilon _\gamma) \langle \lambda \gamma |v|\alpha \beta \rangle _{\cal A} [n_\lambda n_\gamma \bar n_\alpha \bar n_\beta - n_\alpha n_\beta \bar n_\lambda \bar n_\gamma ](t) 
\eeq
$$ = : -i \pi \ \langle \lambda \gamma |{\tilde V}_B(t)|\alpha \beta \rangle _{\cal A}, $$
	\end{widetext}    
i.e. by the on-shell matrix element of the (antisymmetrized) interaction $v$ with its characteristic function in the associated occupation numbers (except for a factor $-i \pi$). Accordingly, $C_{\lambda \gamma \alpha \beta}$ in the on-shell limit can be illustrated by the diagrams shown in Fig. \ref{Fig2} where the open circles denote occupation numbers and the full circles blocking factors.

We now evaluate (\ref{14.52}) in the basis of plane waves  $|\alpha \rangle \sim
\exp \{i{\bf p}_\alpha \cdot{\bf r}\}$, such that the  matrix $\rho$ becomes diagonal in momentum space:
\beq
\label{14.53} \rho({\bf p},{\bf p}') =  (2\pi )^3 \delta^3({\bf p-p}') n({\bf p}) .
\eeq
In (\ref{14.53}) then $n({\bf p})$ has the physical
interpretation of the occupation number of a plane wave with wave number (or momentum)  ${\bf p}$.

Next, we assume for simplicity that the matrix elements of the
interaction $v$ in (\ref{14.52}) are independent of spin $\sigma_i $ (and isospin
$\tau_i )$ and in spatial representation given by:
	\begin{widetext}
\bea
\label{14.54}
\langle {\bf r}_1 {\bf r}_2 \sigma_1\sigma_2\tau_1\tau_2|v|\sigma_{1'}\sigma_{2'}\tau_{1'}%
\tau_{2'}{\bf r}_{1'}{\bf r}_{2'}\rangle  = 
\delta_{\sigma_1\sigma_{1'}}\delta_{\sigma_2\sigma_{2'}}\delta_{\tau_1%
\tau_{1'}}\delta_{\tau_2\tau_{2'}}
\cdot \delta^3({\bf r}_1-{\bf r}_{1'})\delta^3({\bf r}_2-{\bf r}_{2'})
v({\bf r}_1-{\bf r}_2) ,
\eea
or in momentum representation by
\beq
\label{14.54'}
\langle {\bf p}_1 {\bf p}_2\sigma_1\sigma_2\tau_1\tau_2|v|\sigma_{1'}\sigma_{2'}\tau_{1'}%
\tau_{2'}{\bf p}_{1'}{\bf p}_{2'}\rangle  =
\delta_{\sigma_1\sigma_{1'}}\delta_{\sigma_2\sigma_{2'}}\delta_{\tau_1%
\tau_{1'}}\delta_{\tau_2\tau_{2'}}
\eeq
$$\cdot(2\pi)^3\delta^3({\bf p}_1+{\bf p}_2-{\bf p}_{1'}-{\bf p}_{2'})
v({\bf p}_2-{\bf p}_{2'}), $$
which expresses the conservation of momentum in the 2-body collision.

With these approximations we can evaluate (\ref{14.52})  and obtain with $\epsilon({\bf p}) = {\bf p}^2/(2m)$
(non-relativistically)
\beq
\label{14.55} I^2({\bf p}_1,{\bf p}_1;t) = (2s+1)(2\tau +1)
 \int \frac{d^3p_2}{(2\pi )^{3}} \frac{d^3p_3}{(2\pi )^{3}} \frac{d^3p_4}{(2\pi )^{3}}
\eeq
$$
 2\pi \delta (\frac{{1}}{2m} [p^2_1+p^2_2-p^2_3-p^2_4])\ (2\pi )^3
\delta^3 ({\bf p}_1+{\bf p}_2-{\bf p}_3-{\bf p}_4) \ 
v({\bf p}_2-{\bf p}_4) v_{\cal A}({\bf p}_4-{\bf p}_2)
$$
$$
\times \{n({\bf p}_3;t)n({\bf p}_4;t)\bar{n}({\bf p}_1;t) \bar{n}({\bf p}_2;t) -
n({\bf p}_1;t)n({\bf p}_2;t)\bar{n}({\bf p}_3;t)\bar{n}({\bf p}_4;t)\} .
$$
Equation (\ref{14.55}) describes scattering processes ${\bf p}_1+{\bf p}_2
\rightarrow {\bf p}_3+{\bf p}_4$
('loss' term) and ${\bf p}_3+{\bf p}_4 \rightarrow {\bf p}_1+{\bf p}_2$
('gain' term) with the conservation of energy and momentum.
Furthermore, the momentum states ${\bf p}_3$, ${\bf p}_4$ (for the 'loss' terms) or ${\bf p}_1$, ${\bf p}_2$
(for the 'gain' terms) cannot be completely occupied due to the factors $\bar{n}({\bf p}_i;t)$, which contain the Pauli blocking for fermions.

We note in passing that using the Born approximation for the elastic differential cross section
\beq \label{Born}
\frac{d \sigma_{el}}{d \Omega} = \frac{m^2}{16 \pi^2 \hbar^4} v v_{\cal A}^\dagger
\eeq
for 2-body scattering of particles (of mass $m$) the collision integral -- after integration over the $\delta$-functions -- becomes (cf. \cite{Cassing:2021fkc} p. 62)
\beq \label{vuu}
I^2({\bf p}_1,{\bf p}_1;t) =  (2s+1)(2\tau +1)
 \int \frac{d^3 p_2}{(2\pi )^{3}} \int d \Omega \ v_{rel} \frac{d \sigma_{el}}{d \Omega}({\bf p}_1+{\bf p}_2,{\bf p}_2-{\bf p}_4)
\eeq
$$
\times \{n({\bf p}_3;t)n({\bf p}_4;t)\bar{n}({\bf p}_1;t) \bar{n}({\bf p}_2;t) -
n({\bf p}_1;t)n({\bf p}_2;t)\bar{n}({\bf p}_3;t)\bar{n}({\bf p}_4;t)\}
$$
with the relative velocity
\beq
v_{rel} = \frac{|{\bf p}_1-{\bf p}_2|}{m} ,
\eeq
which is of the familiar Uehling-Uhlenbeck form \cite{uehling1933transport}. Note that in (\ref{vuu}) the 4 momenta are constrained by energy and momentum conservation.

\subsection{The 3-body case}
We now consider the following set of equations for $\rho(t)$, $c_2(t)$ and $c_3(t)$:
\beq \label{12.19}
i\frac{\partial}{\partial t}\ \rho  - [h(1) , \rho ] =  Tr_{(2=2')} [v(12) , c_2] ;
\eeq
\beq \label{12.20}
i\frac{\partial}{\partial t}\ c_2 - [\sum_{i=1}^2 h(i) , c_2] =
[V^=(12) , \rho _{20}] +  Tr_{(3=3')} [ v(13)+v(23), c_3]
\eeq
and
$$
i\frac{\partial}{\partial t}\ c_3 - [\sum_{i=1}^3 h(i) , c_3] =
 [{\cal S}_3(V^=(13)+V^=(23)){\cal S}_{3'} , \rho (33';t)c_2(12,1'2';t)]
$$
\beq \label{12.21}
- Tr_{(4=4')} [{\cal S}_1v(14){\cal S}_{1'}{\cal A}_{2'3'}P_{24} ,
\rho (11';t)\rho (22';t)c_2(34,3'4';t)] .
\eeq
Here the l.h.s. describes the propagation of all particles (initial and final) in the time-dependent mean field, whereas the r.h.s.
presents the individual inhomogeneous terms of lower order, which represent the source terms. In analogy to the previous Section we obtain for
$c_3(t)$ in the natural basis in the limit (\ref{12.21}):
\beq \label{12.22}
C^3_{\alpha \beta \gamma, \alpha ' \beta ' \gamma '}(t)= -i \int_{t_0}^t dt' \ \exp(-i [ \epsilon_{\alpha}+ \epsilon_{\beta}+ \epsilon_{\gamma} - \epsilon_{\alpha '} -\epsilon_{\beta '} - \epsilon_{\gamma '}](t-t'))
\eeq
$$
 \times \sum_{\eta} \{
((n_\alpha n_\beta {\bar n}_{\alpha '}+ {\bar n}_\alpha {\bar n}_\beta {n}_{\alpha '})  \langle \alpha \beta |v|\alpha ' \eta \rangle _{\cal A} \  C_{\eta \gamma \beta '\gamma '} - (n_\alpha n_\beta {\bar n}_{\beta '}+ {\bar n}_\alpha {\bar n}_\beta {n}_{\beta '})\langle \alpha \beta |v|\beta ' \eta \rangle _{\cal A} \   C_{\eta \gamma \alpha '\gamma '} $$ $$-  (n_\alpha n_\beta {\bar n}_{\gamma '}+ {\bar n}_\alpha {\bar n}_\beta {n}_{\gamma '})\langle \alpha \beta |v|\gamma ' \eta \rangle _{\cal A} \  C_{\eta \gamma \beta '\alpha '})
- (C_{\beta \gamma \eta \gamma '} \ \langle \alpha \eta |v|\alpha '\beta '\rangle _{\cal A}(n_{\alpha '}n_{\beta '} {\bar n}_\alpha +
{\bar n}_{\alpha '} {\bar n}_{\beta '} {n}_\alpha)$$ $$- C_{\alpha \gamma \eta \gamma '} \ \langle \beta \eta |v|\alpha '\beta '\rangle _{\cal A} (n_{\alpha '}n_{\beta '} {\bar n}_\beta +{\bar n}_{\alpha '} {\bar n}_{\beta '} {n}_\beta)
- C_{\beta \alpha \eta \gamma '} \ \langle \gamma \eta |v|\alpha '\beta '\rangle _{\cal A} (n_{\alpha '} n_{\beta '} {\bar n}_\gamma +
{\bar n}_{\alpha '} {\bar n}_{\beta '} {n}_\gamma))
$$
$$
+ ( (n_{\beta} n_{\gamma} {\bar n}_{\beta '} + {\bar n}_{\beta} {\bar n}_{\gamma} {n}_{\beta '}) \langle \beta \gamma |v|\beta ' \eta \rangle _{\cal A}\  \  C_{\alpha \eta \alpha '\gamma '}
-  (n_{\beta} n_{\gamma} {\bar n}_{\alpha '} + {\bar n}_{\beta} {\bar n}_{\gamma} {n}_{\alpha '}) \langle \beta \gamma |v|\alpha ' \eta \rangle _{\cal A} \  C_{\alpha \eta \beta '\gamma '} $$ $$ - (n_{\beta} n_{\gamma} {\bar n}_{\gamma '} + {\bar n}_{\beta} {\bar n}_{\gamma} {n}_{\gamma '}) \langle \beta \gamma |v|\gamma ' \eta \rangle _{\cal A} \  C_{\alpha \eta \alpha '\beta '} )
 - (C_{\alpha \gamma \alpha '\eta }  \ \langle \beta \eta |v|\beta '\gamma '\rangle _{\cal A} ({n}_{\beta '} {n}_{\gamma '}  {\bar n}_{\beta}+{\bar n}_{\beta '} {\bar n}_{\gamma '} n_{\beta})$$ $$ - C_{\beta \gamma \alpha '\eta } \ \langle \alpha \eta |v|\beta '\gamma '\rangle _{\cal A} (n_{\beta'} {n}_{\gamma '} {\bar n}_{\alpha}+{\bar n}_{\beta '} {\bar n}_{\gamma '} n_{\alpha})
- C_{\alpha \beta \alpha '\eta } \ \langle \gamma \eta |v|\beta '\gamma '\rangle _{\cal A})  ({n}_{\beta '} {n}_{\gamma '} {\bar n}_{\gamma}+{\bar n}_{\beta '} {\bar n}_{\gamma '} n_{\gamma}))
$$
$$
+ (n_{\alpha} n_{\gamma} {\bar n}_{\gamma '}+ {\bar n}_{\alpha} {\bar n}_{\gamma} {n}_{\gamma '}) \langle \gamma \alpha |v|\gamma ' \eta \rangle _{\cal A} \   C_{\eta \beta \alpha '\beta '}
- (n_{\alpha} n_{\gamma} {\bar n}_{\beta '}+ {\bar n}_{\alpha} {\bar n}_{\gamma} {n}_{\beta '})\langle \gamma \alpha |v|\beta ' \eta \rangle _{\cal A} \  C_{\eta \beta \alpha '\gamma '}$$ $$ - (n_{\alpha} n_{\gamma} {\bar n}_{\alpha '}+ {\bar n}_{\alpha} {\bar n}_{\gamma} {n}_{\alpha '})\langle \gamma \alpha |v|\alpha ' \eta \rangle _{\cal A} \  C_{\eta \beta \gamma '\beta '})
 - (C_{\alpha \beta \eta \beta '} \ \langle \gamma \eta |v|\gamma '\alpha '\rangle _{\cal A}(n_{\alpha '} n_{\gamma '} {\bar n}_{\gamma}+ {\bar n}_{\alpha '} {\bar n}_{\gamma '} {n}_{\gamma}  )$$ $$ - C_{\gamma \beta \eta \beta '}\ \langle \alpha \eta |v|\gamma '\alpha '\rangle _{\cal A}(n_{\alpha '} n_{\gamma '} {\bar n}_{\alpha}+ {\bar n}_{\alpha '} {\bar n}_{\gamma '} {n}_{\alpha}  )
- C_{\alpha \gamma \eta \beta '} \  \langle \beta \eta |v|\gamma '\alpha '\rangle _{\cal A})  (n_{\alpha '} n_{\gamma '} {\bar n}_{\beta}+ {\bar n}_{\alpha '} {\bar n}_{\gamma '} {n}_{\beta}  )) \}
$$
where the explicit dependence of all terms on $t'$ has been suppressed. It is of advantage to rewrite (\ref{12.22}) in more compact form using e.g.
\beq \label{V3}
\langle \alpha \beta |{\hat V}|\alpha ' \eta \rangle _{\cal A}: = (n_\alpha n_\beta {\bar n}_{\alpha '}+ {\bar n}_\alpha {\bar n}_\beta {n}_{\alpha '})  \langle \alpha \beta |v|\alpha ' \eta \rangle _{\cal A}, \eeq
$$
\langle \beta \eta |{\hat V}^\dagger|\gamma '\alpha '\rangle _{\cal A} : =  \langle \beta \eta |v|\gamma '\alpha '\rangle _{\cal A} (n_{\alpha '} n_{\gamma '} {\bar n}_{\beta}+ {\bar n}_{\alpha '} {\bar n}_{\gamma '} {n}_{\beta}  ) = \langle \eta \beta |{\hat V}^\dagger|\alpha ' \gamma '\rangle _{\cal A},
$$
which explicitly depend on time $t'$ via the occupation numbers involved. This leads to
\beq \label{12.22b}
C^3_{\alpha \beta \gamma, \alpha ' \beta ' \gamma '}(t)= -i \int_{t_0}^t dt' \ \exp(-i [ \epsilon_{\alpha}+ \epsilon_{\beta}+ \epsilon_{\gamma} - \epsilon_{\alpha '} -\epsilon_{\beta '} - \epsilon_{\gamma '}](t-t'))
\eeq
$$
 \times \sum_{\eta} \{
 (1-P_{\alpha' \beta'}-P_{\alpha' \gamma'}) \langle \alpha \beta |{\hat V}|\alpha ' \eta \rangle _{\cal A} \  C_{\eta \gamma \beta '\gamma '}
- (1-P_{\alpha \beta}-P_{\alpha \gamma}) C_{\beta \gamma \eta \gamma '} \ \langle \alpha \eta |{\hat V}^\dagger|\alpha '\beta '\rangle _{\cal A}
$$
$$
+  (1-P_{\alpha' \beta'}-P_{\beta' \gamma'}) \langle \beta \gamma |{\hat V}|\beta ' \eta \rangle _{\cal A}\  \  C_{\alpha \eta \alpha '\gamma '}
 - (1-P_{\alpha \beta}-P_{\beta \gamma}) C_{\alpha \gamma \alpha '\eta }  \ \langle \beta \eta |{\hat V}^\dagger|\beta '\gamma '\rangle _{\cal A}
$$
$$
+ (1-P_{\alpha' \gamma'}-P_{\beta' \gamma'}) \langle \gamma \alpha |{\hat V}|\gamma ' \eta \rangle _{\cal A} \   C_{\eta \beta \alpha '\beta '}
 - (1-P_{\alpha \gamma}-P_{\beta \gamma}) C_{\alpha \beta \eta \beta '} \ \langle \gamma \eta |{\hat V}^\dagger|\gamma '\alpha '\rangle_{\cal A} \}
$$
in more compact form with ${\hat V}(t')$ and $C(t')$. Note that $\sum_\eta$ does not involve an intermediate propagator! An illustration of a single term in $C_{\alpha \beta \gamma, \alpha' \beta' \gamma'}$ is presented in Fig. \ref{Fig3}.

       \begin{figure*}[h!]
\begin{center}
 {\includegraphics[width=15cm]{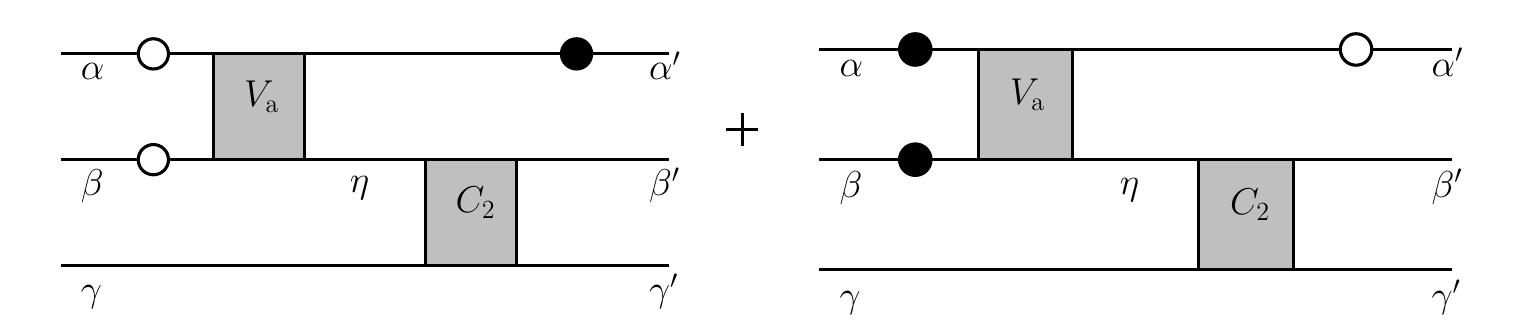} }
 \end{center}
         \caption{Illustration of a term in the 3-body correlation ${\tilde C}^3_{\alpha \beta \gamma ,\alpha '\beta '\gamma '}(t)$ where the initial and final lines are constrained by energy conservation.} \label{Fig3}
       \end{figure*}

Now we use the result (\ref{14.49}) for the two-body correlations, i.e.
\beq
\label{12.49} C_{\alpha \beta \alpha '\beta '}(t') = -i \int_{t_0}^{t'} dt"\ \exp \{
-i[\epsilon _\alpha +\epsilon _\beta -\epsilon_{\alpha '}-\epsilon _{\beta '}](t'-t")\}
\cdot \langle \alpha \beta |V_B(t")|\alpha '\beta '\rangle \eeq
and insert in (\ref{12.22b}). Since this is straight forward (but lengthy) we discard an explicit representation.

The next step is to insert  $C_{\alpha \beta \gamma, \alpha '\beta ' \gamma '}(t')$ (\ref{12.22b}) in
$$
\langle \alpha \beta|V_C^3| \alpha ' \beta ' \rangle (t'): =  \sum_{\gamma=\gamma' } \sum_{\lambda  \mu} \{
\langle \alpha \gamma |v|\lambda \mu \rangle
\ C^3_{\lambda \beta \mu ,\alpha '\beta '\gamma' }(t')
+ \langle \beta \gamma |v|\lambda \mu \rangle \ C^3_{\alpha \lambda \mu ,\alpha '\beta '\gamma' }(t') \}
$$
\beq \label{12.23} - \sum_{\gamma=\gamma'} \sum_{\lambda  \mu} \ \{
C^{3 \dagger}_{\alpha \beta \gamma ,\lambda \beta '\mu}(t') \ \langle \lambda  \mu|v|\alpha ' \gamma' \rangle
+ C^{3 \dagger }_{\alpha \beta \gamma ,\alpha '\lambda \mu}(t')\ \langle \lambda  \mu |v|\beta ' \gamma' \rangle \} ,
\eeq
which is the additional inhomogeneous term in (\ref{12.20}) ($Tr_{(3=3')} [(v(13)+v(23)), c_3]$).

\medskip
We obtain the extended two-body correlation ${\hat C}_{\alpha \beta \alpha ' \beta '}(t)$
with the additional inhomogeneous term (\ref{12.23}) by a further time integration  as
\beq \label{12.24}
{\hat C}_{\alpha \beta \alpha '\beta '}(t) = -i
\int_{t_0}^t dt'\ \exp \{-i[\epsilon _\alpha +\epsilon _\beta -\epsilon_{\alpha '}
-\epsilon _{\beta '}](t-t')\} \cdot \langle \alpha \beta |V_B(t')+V_C^3(t')|\alpha '\beta'\rangle ,
\eeq
which now includes the effects from 3-body correlations via (\ref{12.23}).

\vp

For the diagonal element of the collision term we proceed as in Section 3.1 and obtain
\beq
\label{11.50} {\hat I}_{\alpha \alpha }(t) = -i \sum_\beta \sum_{\delta \nu}
 \{\langle \alpha \beta |v|\delta \nu \rangle  {\hat C}_{\delta \nu \alpha \beta }(t) -
{\hat C}_{\alpha \beta \delta \nu }(t) \langle \delta \nu |v|\alpha \beta \rangle \}
\eeq
$$
= I^2_{\alpha \alpha}(t) -\sum_\beta \sum_{\delta \nu}
\int^t_{t_0} dt' \{\exp\{-i[\epsilon _\delta +\epsilon _\nu -\epsilon _\alpha
-\epsilon_\beta ](t-t')\}
\ \langle \alpha \beta |v|\delta \nu \rangle \langle \delta \nu |V_C^3(t')|\alpha \beta \rangle
$$
$$
- \exp \{-i[\epsilon _\alpha +\epsilon _\beta -\epsilon _\delta -\epsilon _\nu ](t-t')%
\} \langle \alpha \beta |V_C^{3 \dagger}(t')|\delta \nu \rangle  \langle \delta \nu |v|\alpha \beta \rangle \}
$$
which includes the 2-body contribution $I^2_{\alpha \alpha}(t)$ from Section 3.1 and the additional 3-body contributions by $V_C^3(t)$.
The other two time integrals -- appearing in the matrix elements of the 2-body correlations and 3-body correlations -- are treated in the same fashion
and approximately give additional factors $- i \pi \delta$(s.p. energies). 
 Although one might question this approximation, which is invalid for short time intervals,
this approximation leads to energy conservation in each individual interaction matrix element.

\medskip

We then  get for the first 2 contributions in the 3-body collision part:
\beq \label{12.38}
{\tilde I}^3_{\alpha \alpha}(t) \approx - \pi^3 \sum_\beta
\sum_{\delta \nu} \ \delta(\epsilon _\alpha +\epsilon _\beta -\epsilon _\delta -\epsilon _\nu) \langle \alpha \beta |v|\delta \nu \rangle
\cdot \sum_{\gamma = \gamma'} \sum_{\lambda  \mu} \sum_\eta  \eeq
$$
 \times  \{\delta(\epsilon _\lambda +\epsilon _\nu +\epsilon _\mu -\epsilon_{\alpha}-\epsilon _{\beta}-\epsilon _{\gamma}) \delta(\epsilon _\eta +\epsilon _\mu -\epsilon_{\beta}-\epsilon _{\gamma}) \  \langle \delta \gamma |v|\lambda \mu \rangle \   \langle \lambda \nu |{\hat V}(t)|\alpha \eta \rangle_{\cal A}
\  \langle \eta \mu |V_B|\beta \gamma'  \rangle
$$
$$
 +   \delta(\epsilon _\delta +\epsilon _\lambda +\epsilon _\mu -\epsilon_{\alpha}-\epsilon _{\beta}-\epsilon _{\gamma}) \delta(\epsilon _\eta +\epsilon _\mu -\epsilon_{\beta}-\epsilon _{\gamma}) \  \times  \langle \nu \gamma |v|\lambda \mu \rangle \  \langle \delta \lambda |{\hat V}|\alpha \eta \rangle_{\cal A}
\  \langle \eta \mu |V_B|\beta \gamma'  \rangle \},
$$
\beq \label{12.38b}
= -  \pi^3 \sum_\beta
\sum_{\delta \nu} \ \delta(\epsilon _\alpha +\epsilon _\beta -\epsilon _\delta -\epsilon _\nu) \langle \alpha \beta |v|\delta \nu \rangle
\cdot \sum_{\gamma=\gamma'} \sum_{\lambda  \mu} \sum_\eta  \eeq
$$
 \times  \{\delta(\epsilon _\lambda +\epsilon _\nu  -\epsilon_{\alpha}-\epsilon _{\eta}) \delta(\epsilon _\eta +\epsilon _\mu -\epsilon_{\beta}-\epsilon _{\gamma}) \langle \delta \gamma |v|\lambda \mu \rangle \   \langle \lambda \nu |{\hat V}(t)|\alpha \eta \rangle_{\cal A}
\  \langle \eta \mu |V_B|\beta \gamma'  \rangle
$$
$$
 +   \delta(\epsilon _\delta +\epsilon _\lambda  -\epsilon_{\alpha}-\epsilon _{\eta}) \delta(\epsilon _\eta +\epsilon _\mu -\epsilon_{\beta}-\epsilon _{\gamma})  \langle \nu \gamma |v|\lambda \mu \rangle \  \langle \delta \lambda |{\hat V}(t)|\alpha \eta \rangle_{\cal A}
\  \langle \eta \mu |V_B|\beta \gamma'  \rangle \}.
$$
The 3 $\delta$-functions in energy ensure that all 4 matrix elements of the interaction are taken on-shell. Defining
\beq \label{12.38d}
\langle \alpha \beta |{\tilde v}| \mu \nu \rangle : = \delta(\epsilon_\alpha + \epsilon_\beta - \epsilon_\mu - \epsilon_\nu) \langle \alpha \beta |{v}| \mu \nu \rangle , \eeq
$$ \langle \alpha \beta |{\tilde V}(t)| \mu \nu \rangle_{\cal A} : = \delta(\epsilon_\alpha + \epsilon_\beta - \epsilon_\mu - \epsilon_\nu) \langle \alpha \beta |{\hat V}(t)| \mu \nu \rangle_{\cal A} ,
$$
$$
 \langle \alpha \beta |{\tilde V}_B(t)| \mu \nu \rangle_{\cal A} : = \delta(\epsilon_\alpha + \epsilon_\beta - \epsilon_\mu - \epsilon_\nu) \langle \alpha \beta |{V_B}(t)| \mu \nu \rangle ,
$$
we can rewrite (\ref{12.38b}) in more compact form as
\beq \label{12.38c}
{\tilde I}^3_{\alpha \alpha}(t) \approx -  \pi^3 \sum_\beta
\sum_{\delta \nu} \  \langle \alpha \beta |{\tilde v}|\delta \nu \rangle
\cdot \sum_{\gamma} \sum_{\lambda  \mu} \sum_\eta  \eeq
$$
 \times  \{ \langle \delta \gamma |v|\lambda \mu \rangle \   \langle \lambda \nu |{\tilde V}(t)|\alpha \eta \rangle_{\cal A}
\  \langle \eta \mu |{\tilde V}_B|\beta \gamma  \rangle
 +   \langle \nu \gamma |v|\lambda \mu \rangle \  \langle \delta \lambda |{\tilde V}(t)|\alpha \eta \rangle_{\cal A}
\  \langle \eta \mu |{\tilde V}_B|\beta \gamma  \rangle \}.
$$
	\end{widetext}

\begin{figure}[h!]
 \begin{center}
 {\includegraphics[width=8cm]{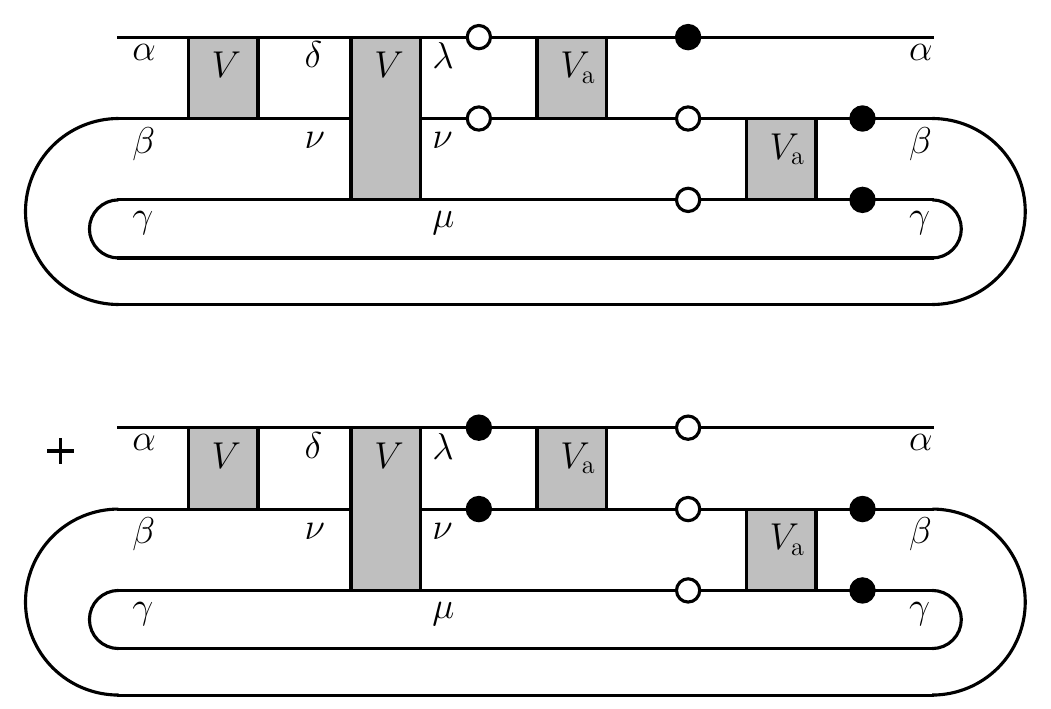} }
 \end{center}
         \caption{Illustration of the interaction terms in (\ref{12.38c}) for the 'loss terms', i.e. including only the occupation factors $-n_\beta n_\gamma {\bar n}_\eta {\bar n}_\mu$ in the matrix element $\langle \eta \mu |{\tilde V}_B|\beta \gamma'  \rangle$. The open circles denote occupation numbers while the full circles stand for blocking factors.  The closed lines corresponds to a summation over particle 2 (or index $\beta$) and particle 3 (or index $\gamma$).} \label{Fig4}
       \end{figure}

Here $\sum_\beta \equiv Tr_2$ denotes the summation over particle 2 while $\sum_\gamma \equiv Tr_3$ denotes the summation over an additional particle 3.
Fig. \ref{Fig4} shows an illustration of the interaction terms in (\ref{12.38c}) for the 'loss terms', i.e. including only the occupation factors $-n_\beta n_\gamma {\bar n}_\eta {\bar n}_\mu$ in the matrix element $\langle \eta \mu |{\tilde V}_B|\beta \gamma'  \rangle$. The open circles denote occupation numbers while the full circles stand for blocking factors.
The full list of interaction terms is given in the Appendix.

\section{Equilibration in a box with periodic boundary conditions}

In this Section we aim for a numerical analysis of the relative impact of 3-body contributions in model systems that are related to shifted Fermi spheres in a finite or infinite box with periodic boundary conditions. Such model systems reflect the initial overlap phase of heavy-ion collisions at low or intermediate energies where elastic collisions and Pauli-blocking dominate the relaxation processes.

\medskip
In case of periodic boundary conditions for a homogeneous system in a finite volume of size $V=a^3$ the single-particle eigenstates in coordinate space (for nucleons) are
	\begin{widetext}
\beq
<{\bf r}|n_x n_y n_z, n_s, n_{\tau}> = < {\bf r}|\alpha> = \frac{1}{a^{3/2}} \exp(i \frac{2 \pi}{a}[n_x x+n_y y+n_z z]) \chi(n_s) \tau(n_\tau) , 
\eeq
where $\chi(n_s)$ denote the two orthogonal eigenstates for the spin and $\tau(n_\tau)$ for isospin, respectively. By construction these states are orthonormal in all quantum numbers.  The momentum components of particles are given by ($i=\pm 1,\pm 2, \pm 3, \cdots $)
\beq p_i(n_i) = \frac{2 \pi \hbar}{a} n_i  \eeq
and the single-particle energies by
\beq \label{p1} \epsilon(n_x,n_y,n_z) = \frac{2 \hbar^2 \pi^2}{M_N a^2} (n_x^2+n_y^2+n_z^2) = : {\bar \omega}  (n_x^2+n_y^2+n_z^2) \eeq
when assuming no contributions from spin and isospin projections. Here $M_N$ stands for the nucleon mass.
Within this basis the conservation of energy and momentum in 2-body transitions can be strictly fulfilled and easily controlled via integer numbers $n_i$ for momentum or $n_i^2$ for energy. Furthermore, the matrix elements of a local $\delta^3({\bf r}_1- {\bf r}_2)$-type interaction, i.e.

\beq \label{interaction}
v({\bf r}_1- {\bf r}_2) = V_0\  \delta^3({\bf r}_1- {\bf r}_2) ,
\eeq 
with interaction strength $V_0$ are - assuming the interactions again to be diagonal in spin and isospin  -
\beq \label{p2}
<n^3_x,n^3_y,n^3_z;n^4_x,n^4_y,n^4_z|v|n^1_x,n^1_y,n^1_z;n^2_x,n^2_y,n^2_z> \eeq $$= \frac{V_0}{a^6} \int_{-a/2}^{a/2} dx
\int_{-a/2}^{a/2} dy \int_{-a/2}^{a/2} dz \ \exp(-i\frac{2 \pi}{a}(n^1_x+n^2_x-n^3_x-n^4_x)x) $$ $$ \times \exp(-i\frac{2 \pi}{a}(n^1 _y+n^2_y-n^3_y-n^4_y)y) \ \exp(-i\frac{2 \pi}{a}(n^1 _z+n^2_z-n^3_z-n^4_z)z) $$ $$= \frac{V_0}{a^3} \ \delta(n^1_x+n^2_x-n^3_x-n^4_x) \ \delta(n^1 _y+n^2_y-n^3_y-n^4_y) \ \delta(n^1 _z+n^2_z-n^3_z-n^4_z) ,$$ which implies momentum conservation in the transitions $1+2 \leftrightarrow 3+4$.

\subsection{The two-body case}

In case of a diagonal density matrix $\rho_{\alpha \alpha'}(t) = \delta_{\alpha \alpha'} n_\alpha(t)$ the change in the occupation numbers $n_\alpha(t)$  is given by the 2-body collision term $I^2_{\alpha \alpha}(t)$,
\beq \label{collterm}
\frac{d}{dt} n_\alpha (t)= I^2_{\alpha \alpha}(t) =  -\frac{i}{\hbar} \sum_\beta \sum_{\lambda \gamma} \ [<\alpha \beta|v|\lambda \gamma> C_{\lambda \gamma \alpha \beta}(t) -  C_{ \alpha \beta \lambda \gamma}(t) <\lambda \gamma|v|\alpha \beta>] . \eeq
When inserting the on-shell solution for $C_2(t)$ (\ref{onshell}) we end up with
\beq \label{collterm2}
 I^2_{\alpha \alpha}(t) =  \frac{2 \pi}{\hbar} \sum_\beta \sum_{\lambda \gamma} \ \delta(\epsilon_\alpha + \epsilon_\beta - \epsilon_\lambda - \epsilon_\gamma)  <\alpha \beta |v| \lambda \gamma> <\lambda \gamma|v| \alpha \beta>_{\cal A}\eeq $$ \times [n_\lambda(t) n_\gamma(t) {\bar n}_\alpha(t) {\bar n}_\beta(t) -  n_\alpha(t) n_\beta(t) {\bar n}_\lambda(t) {\bar n}_\gamma(t)] . $$
 These expressions can be worked out further in case of periodic boundary conditions by using (\ref{p1}) and (\ref{p2}) giving (in the on-shell case)
\beq \label{collterm2b}
 I^2_{\alpha \alpha}(t) =  \frac{2 \pi}{\hbar {\bar \omega}} \frac{3}{4} \frac{V_0^2}{a^6} \sum_\beta \sum_{\lambda \gamma} \  \delta(n^1_x+n^2_x-n^3_x-n^4_x) \ \delta(n^1 _y+n^2_y-n^3_y-n^4_y) \ \delta(n^1 _z+n^2_z-n^3_z-n^4_z) \eeq $$\times \delta(m^2_\alpha + m^2_\beta - m^2_\lambda - m^2_\gamma) \   [n_\lambda(t) n_\gamma(t) {\bar n}_\alpha(t) {\bar n}_\beta(t) -  n_\alpha(t) n_\beta(t) {\bar n}_\lambda(t) {\bar n}_\gamma(t)] $$
 using the notation
 \beq m^2_\alpha = (n^1_x)^2+(n^1_y)^2+(n^1_z)^2 \eeq  which (except for a factor ${\bar \omega}$) gives the discrete single-particle energies.

\medskip
Since the loss-term in (\ref{collterm2b}) is proportional to $n_\alpha$ we can determine the on-shell collision width for the state $\alpha$ by
\beq \label{collterm3}
 \Gamma_{\alpha} =  \frac{2 \pi \hbar}{\hbar {\bar \omega}} \frac{3}{4} \frac{V_0^2}{a^6} \sum_\beta \sum_{\lambda \gamma}  \delta(n^1_x+n^2_x-n^3_x-n^4_x) \ \delta(n^1 _y+n^2_y-n^3_y-n^4_y)\ \delta(n^1 _z+n^2_z-n^3_z-n^4_z) \eeq $$ \times  \ \delta(m^2_\alpha + m^2_\beta - m^2_\lambda - m^2_\gamma) \  n_\beta {\bar n}_\lambda {\bar n}_\gamma .
 $$

In the continuum limit $a \rightarrow \infty$ the collision width of a particle with momentum ${\bf p}_\alpha$ reads
for isotropic scattering with a constant and isotropic cross section \index{cross section}
\beq \label{cross}
\sigma_{el} = \frac{M_N^2}{4 \pi \hbar^4} \frac{3}{4} V_0^2 \eeq
 ($\approx$ 32.4 mb for $V_0 = \pm$ 300 MeV fm$^3$),
\beq \label{collcont}
\Gamma({\bf p}_\alpha) = \frac{\hbar c}{(2 \pi \hbar)^3} d \int d^3 p_\beta \int d\Omega \ v_{rel}({\bf p}_\alpha - {\bf p}_\beta) \frac{\sigma_{el}}{4 \pi} \  n({\bf p}_\beta) (1-n({\bf p}_3)) (1-n({\bf p}_4)) ,\eeq
 	\end{widetext}
where the final states ${\bf p}_3$ and ${\bf p}_4$ are fixed by energy and momentum conservation except for an angle $\Omega = (\cos (\vartheta), \phi)$ in the individual center-of-mass which has to be integrated over. The relative velocity - in the non-relativistic limit - is given by
\beq
v_{rel}({\bf p}_\alpha - {\bf p}_\beta) = \frac{|{\bf p}_\alpha - {\bf p}_\beta|}{M_N} ,
\eeq
while the factor $d$=4 in (\ref{collcont}) stems from summation over spin and isospin.
The factor $3/4$ accounts for the fact that the anti-symmetrized matrix element of the interaction vanishes in case of a $\delta$-force for nucleons with identical spin and isospin. Thus, in practice we have an effective degeneracy factor $d$=3 for the scattering of non-identical particles.

\begin{figure}[b!]
\begin{center}
 {\includegraphics[width=8cm]{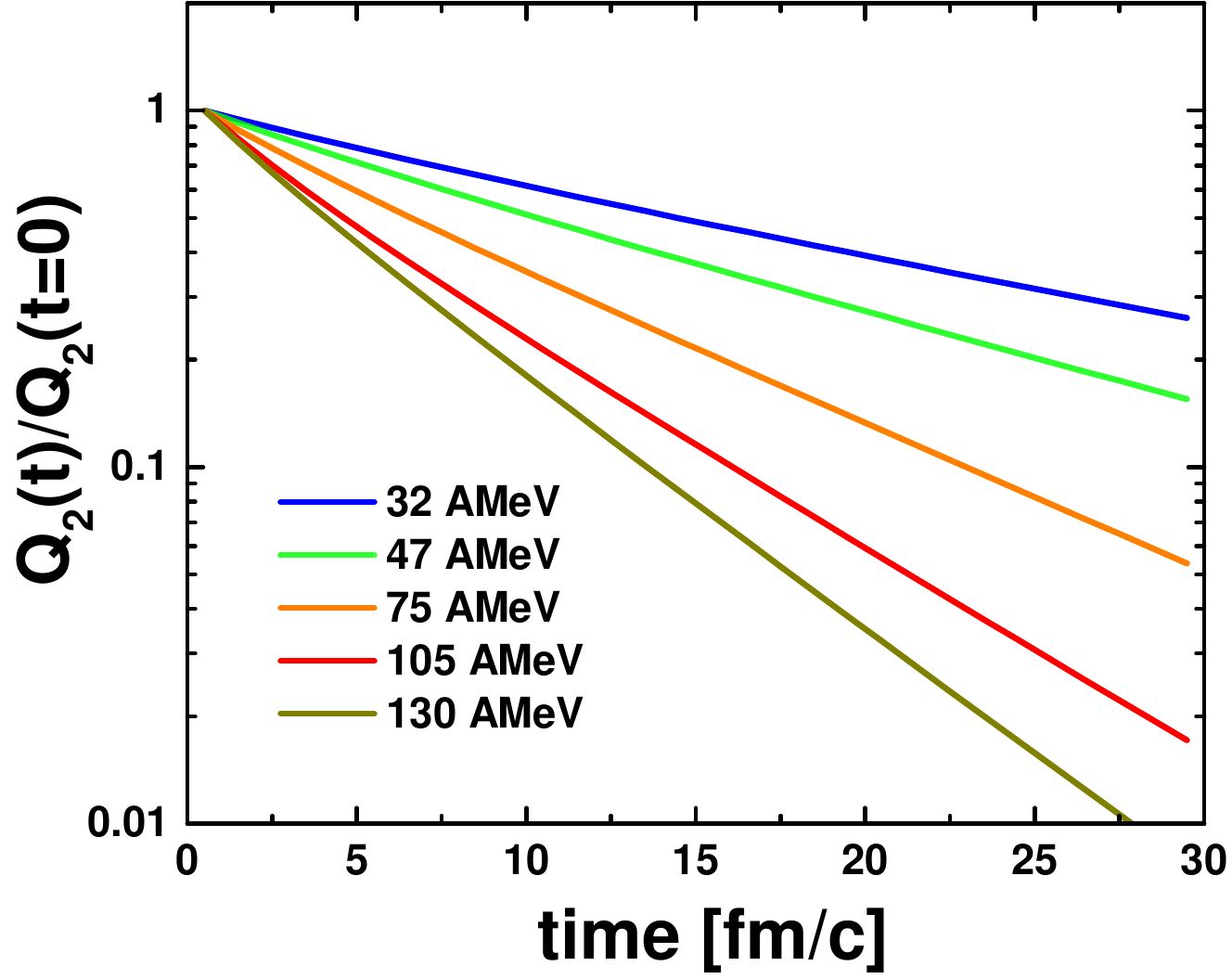} }
 \end{center}
\caption{Numerical results for the time evolution of the quadrupole moment (\ref{quad2}) {-- normalized by $Q_2(t=0)$ --} for collisions at 32, 47, 75, 105 and 130 A$\cdot$MeV.}
\label{Fig10}
\end{figure}
\begin{figure}[h!]
\begin{center}
 {\includegraphics[width=8cm]{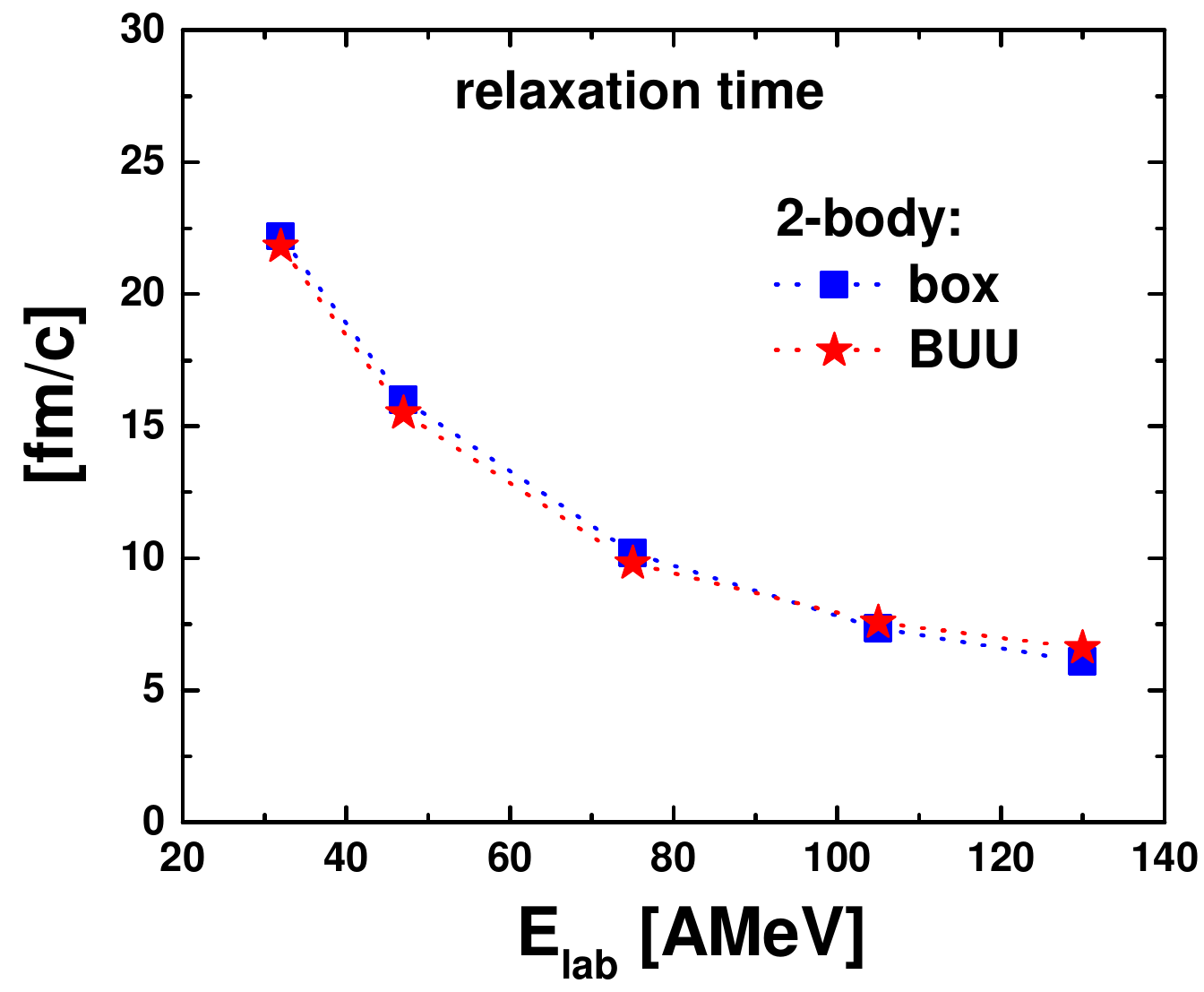} }
 \end{center}
\caption{The relaxation times $\tau$ extracted from the approximation (\ref{relax}) are shown by blue boxes for different bombarding energies. The corresponding results from BUU calculations are displayed by red stars.} \label{Fig11}
\end{figure}

\begin{figure}[h!]
\begin{center}
 {\includegraphics[width=8cm]{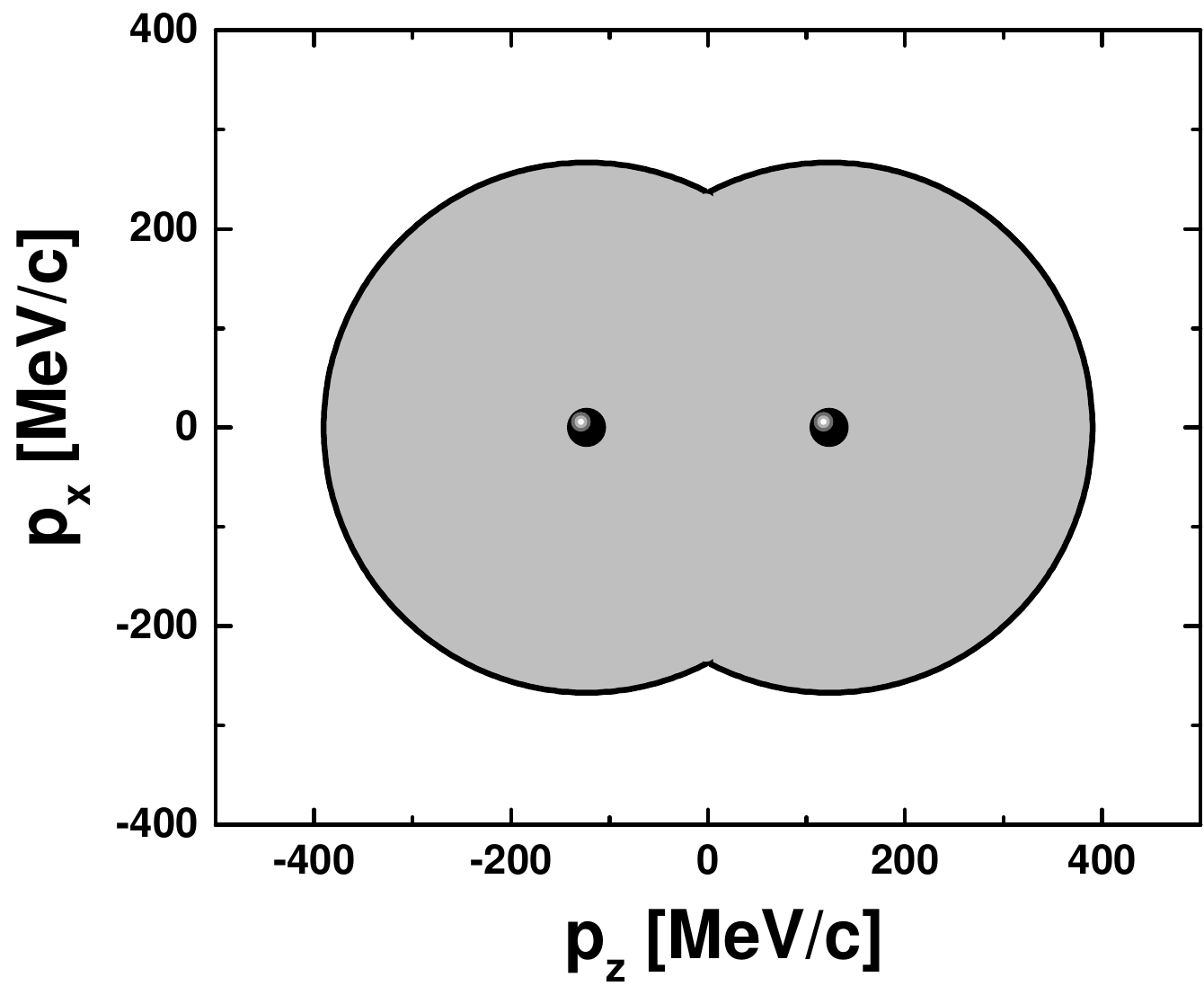} }
 \end{center}
\caption{The initial occupation number as a function of $p_x$ and $p_z$ in case of a bombarding energy of 130 A$\cdot$MeV in the continuum limit. The shaded area here indicates an occupation number of 1 while the outside region in momentum is unoccupied. The Fermi momentum is taken as $p_F$=268 MeV/c which gives a nucleon density of $\sim$ 0.335 fm$^{-3}$.}
 \label{Fig12}
\end{figure}

\noindent
The numerical calculations are carried out in boxes of side-length $a$= 15-18 fm (depending on energy) employing the lowest 1016 basis states. This basis was found to be sufficient since by increasing the number of basis states by a factor of 2 the results obtained for the quadrupole moment in momentum space
\beq \label{quad2}
Q_2(t)= \sum_\alpha (2 p_{\alpha,z}^2 -  p_{\alpha,x}^2 - p_{\alpha,y}^2) n_\alpha (t)
\eeq
do not change within the linewidth. We note in passing that the conservation of energy, momentum and total particle number is fulfilled within the numerical accuracy ($\sim 10^{-6}$).

The actual numerical results for $Q_2(t)$ are displayed in Fig. \ref{Fig10} for a couple of bombarding energies from 32 A$\cdot$MeV to 130 A$\cdot$MeV showing a pronounced exponential decay which can be well approximated by
\beq \label{relax}
Q_2(t)=Q_2(t=0) \exp(-t/\tau)
\eeq
in all cases. The extracted results for the relaxation time $\tau$ are displayed in Fig. \ref{Fig11} for different bombarding energies by the blue boxes and found to drop rapidly with energy due to a weakening of the Pauli-blocking in collisions.

How does this compare to the continuum limit within standard BUU transport calculations? To this end we use the BUU transport code from Ref. \cite{Cassing:1990dr} in a finite box (of similar size) with an isotropic cross section $\sigma_{el} = 32.4$ mb, which corresponds to the interaction strength $V_0 = \pm $300 MeV fm$^3$ according to (\ref{cross}). The initial distribution in occupation number is illustrated in Fig. \ref{Fig12} as a function of $p_x$ and $p_z$ in case of a bombarding energy of 130 A$\cdot$MeV. The shaded area here indicates an occupation number of 1 while the outside region in momentum is unoccupied. {The initial conditions for the BUU calculations are chosen by Monte Carlo (for 1000 parallel ensembles) according to Fig. 7 in momentum space and distributed homogeneously in a box of volume (17 fm)$^3$. The actual densities in the calculations are 0.280 fm$^{-3}$, 0.298  fm$^{-3}$, 0.318 fm$^{-3}$, 0.335 fm$^{-3}$, and 0.34 fm$^{-3}$ for 32, 47, 75, 105, and 130 A MeV, respectively. The densities for the discrete basis amount to 0.274 fm$^{-3}$, 0.294  fm$^{-3}$, 0.32 fm$^{-3}$, 0.33 fm$^{-3}$, and 0.335 fm$^{-3}$ for 32, 47, 75, 105, and 130 A MeV, respectively, while the side-length of the boxes varies from 15 - 18 fm.}

The results of the BUU calculations for the relaxation time $\tau$ are shown in Fig. \ref{Fig11} by the red stars, which are found to agree within a few percent with the discrete finite box calculations (blue boxes). We cannot expect identical results since the continuum limit implies slightly different initial conditions and  slightly different initial densities as quantified above. Nevertheless, the good agreement shows that the on-shell correlation dynamics and standard 2-body collision transport simulations well compare with each other.

\subsection{The 3-body case}
We now turn to the 3-body collision integral.
The 3 $\delta$-functions in energy ensure that all 4 matrix elements of the interaction are taken on-shell. Defining
\beq \label{12.38d}
\langle \alpha \beta |{\hat v}| \mu \nu \rangle : = \delta(\epsilon_\alpha + \epsilon_\beta - \epsilon_\mu - \epsilon_\nu) \langle \alpha \beta |{v}| \mu \nu \rangle \eeq
we can rewrite the first terms in ${\tilde I}^3_{\alpha \alpha}(t)$ as
 	\begin{widetext}
\beq \label{12.38h}
{\tilde I}^3_{\alpha \alpha}(t) \approx - \pi^3 \sum_\beta
\sum_{\delta \nu} \ \langle \delta \nu|v| \alpha \beta \rangle
\cdot \sum_{\gamma} \sum_{\lambda  \mu} \sum_\eta  \eeq
$$
 \{  \langle \alpha \gamma |{\hat v}|\lambda \mu \rangle  [n_\lambda   n_\beta  {\bar n}_{\delta}  + {\bar n}_\lambda  {\bar n}_\beta  {n}_{\delta} ] \langle \lambda \beta |{\hat v}|\delta \eta \rangle_{\cal A}
\  \langle \eta \mu |{\hat v}|\nu \gamma  \rangle_{\cal A} \ [n_{\nu} n_{\gamma } {\bar n}_\eta {\bar n}_\mu -  n_\eta  n_\mu
{\bar n}_{\nu} {\bar n}_{\gamma}]
$$
$$
 +   \langle \beta \gamma |{\hat v}|\lambda \mu \rangle   [n_\alpha   n_\lambda  {\bar n}_{\delta}  + {\bar n}_\alpha  {\bar n}_\lambda  {n}_{\delta} ] \langle \alpha \lambda |{\hat v}|\delta \eta \rangle_{\cal A}
\  \langle \eta \mu |{\hat v}|\nu \gamma  \rangle_{\cal A} \ [n_{\nu} n_{\gamma } {\bar n}_\eta {\bar n}_\mu -  n_\eta  n_\mu
{\bar n}_{\nu} {\bar n}_{\gamma}]
$$
$$
  -    [n_\alpha   n_\beta  {\bar n}_{\lambda}  + {\bar n}_\alpha  {\bar n}_\beta  {n}_{\lambda} ]  \langle \alpha \beta |{\hat v}|\lambda \eta \rangle_{\cal A}
\  \langle \eta \gamma |{\hat v}|\nu \mu  \rangle_{\cal A} \ [n_{\nu} n_{\mu } {\bar n}_\eta {\bar n}_\gamma -  n_\eta  n_\gamma
{\bar n}_{\nu} {\bar n}_{\mu}] \langle \lambda \mu |{\hat v}|\delta \gamma \rangle \
$$
$$
  -    [n_\alpha   n_\beta  {\bar n}_{\delta}  + {\bar n}_\alpha  {\bar n}_\beta  {n}_{\delta} ]  \langle \alpha \beta |{\hat v}|\delta \eta \rangle_{\cal A}
\  \langle \eta \gamma |{\hat v}|\lambda \mu  \rangle_{\cal A} \ [n_{\lambda} n_{\mu } {\bar n}_\eta {\bar n}_\gamma -  n_\eta  n_\gamma
{\bar n}_{\lambda} {\bar n}_{\mu}] \langle \lambda \mu |{\hat v}|\nu \gamma \rangle \ \} \cdots
$$
 	\end{widetext}
Here $\sum_\beta \equiv Tr_2$ denotes the summation over particle 2 while $\sum_\gamma \equiv Tr_3$ denotes the summation over an additional particle 3.

\medskip

\begin{figure}[h!]
\begin{center}
 {\includegraphics[width=8cm]{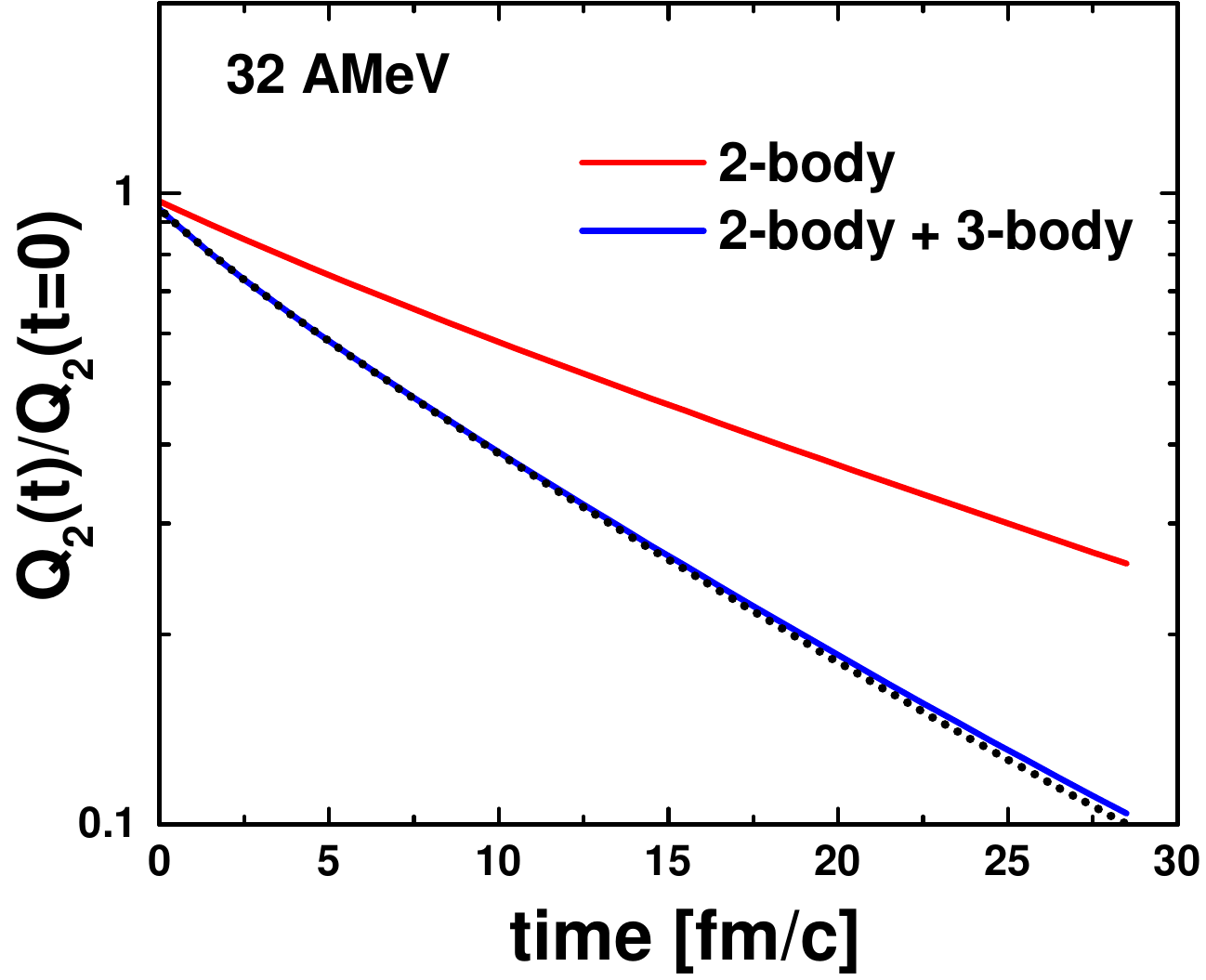} }
 \end{center}
\caption{Comparison of the results for the quadrupole moment $Q_2(t)$ {-- normalized by $Q_2(t=0)$ --} for the 2-body case (upper red line) with the result including the additional 3-body interactions (lower blue line) for 32 A$\cdot$MeV. The result for the 2-body + 3-body case within an enhanced basis of states is shown by the black dotted line and almost identical to the blue line.}
 \label{Fig13}
\end{figure}

The calculations {for the time evolution of the occupation numbers (\ref{finaleq})} including the 3-body collisions -- listed in the Appendix -- are carried in the same basis {(of 1016 or 1568 states)} as for the 2-body case, where the possible final states ($\lambda \mu$) are pre-calculated for all initial states ($\alpha \beta$) and stored in a file. Due to the large number of interactions (see Appendix) and internal loops the required CPU time increases by a factor $\sim 2 \cdot 10^5$ compared to the 2-body case. Again the conservation of energy, momentum and particle number is fulfilled within the numerical accuracy. A typical result for the relaxation of the quadrupole moment is shown in Fig. \ref{Fig13} for 32 A$\cdot$ MeV (lower blue  line), where we compare the results to the 2-body case (upper red line). This result remains practically unchanged when enhancing the number of basis states from 1016 to 1568 (black dotted line), which also increases the CPU time by approximately another order of magnitude.

\begin{figure}[h!]
\begin{center}
 {\includegraphics[width=8cm]{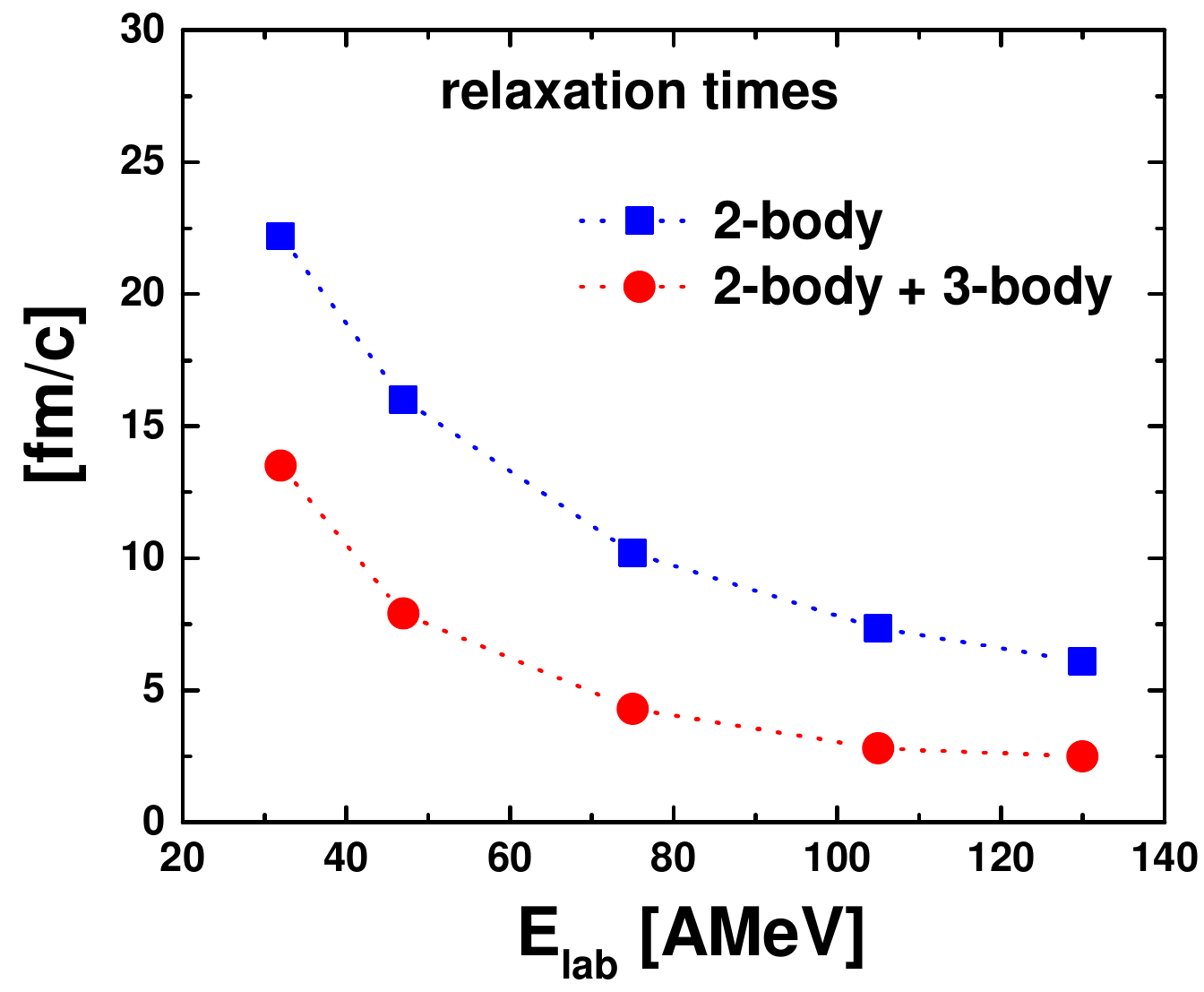} }
 \end{center}
\caption{Comparison of the results for the relaxation time $\tau$ for the 2-body case (blue boxes) with the result including the additional 3-body interactions (red dots) for different bombarding energies.}
 \label{Fig14}
\end{figure}

Extracting again a relaxation time $\tau$ by an exponential fit we obtain the results displayed in Fig. \ref{Fig14} for different bombarding energies which shows that by including the 3-body interactions the relaxation time $\tau$ decreases by about a factor 1.5 at 32 A$\cdot$MeV. This decrease becomes larger (up to a factor of about $\sim$3) with increasing bombarding energy since the systems becomes more dense -- due to the enhanced number of occupied states -- and transverse momentum states are more easily accessed by 3-body scattering than in case of Pauli-blocked 2-body scatterings. These results indicate that standard transport calculations -- only including 2-body scattering -- might underestimate the relaxation in low energy heavy-ion collisions.

Some comments to these results are in order: First of all the effects of 3-body collisions might be overestimated in the model study due to the assumption of isotropic collisions and energy independent cross sections. In reality the individual 2-body collisions get more forward peaked with increasing energy, which leads to a stronger blocking for initial times. {{Moreover, the size of the elastic cross section decreases with collision energy.  Furthermore, the leading order approximation for the 3-body interactions might be questioned. Higher order terms ($\sim Tr (v c_2 c_2$)) -- leading to an additional 'interaction box' (cf. the illustration in Fig. 4 for the collision term) -- might change the present results and also reduce the net interaction cross section. Unfortunately, concrete studies will increase the CPU time by further orders of magnitude if one tries to quantify the influence of these higher order terms. 
Accordingly, this question remains open until increasing CPU/GPU power or new algorithms are available.}

\section{BUU calculations for high energy particle emission in fusion reactions}

The results presented in the previous Section are primarily academic and the question arises if a higher stopping power by elastic 3-body collisions might leave traces in experimental observables. At first sight one might look at the proton rapidity distributions, which should become narrower with increasing cross section, but this will require a systematic study for various target masses as well as a scan in bombarding energy. An alternative is the emission of energetic nucleons which preferentially occurs before kinetic equilibrium is achieved.

\begin{figure}[!]
\begin{center}
 {\includegraphics[width=6.6cm]{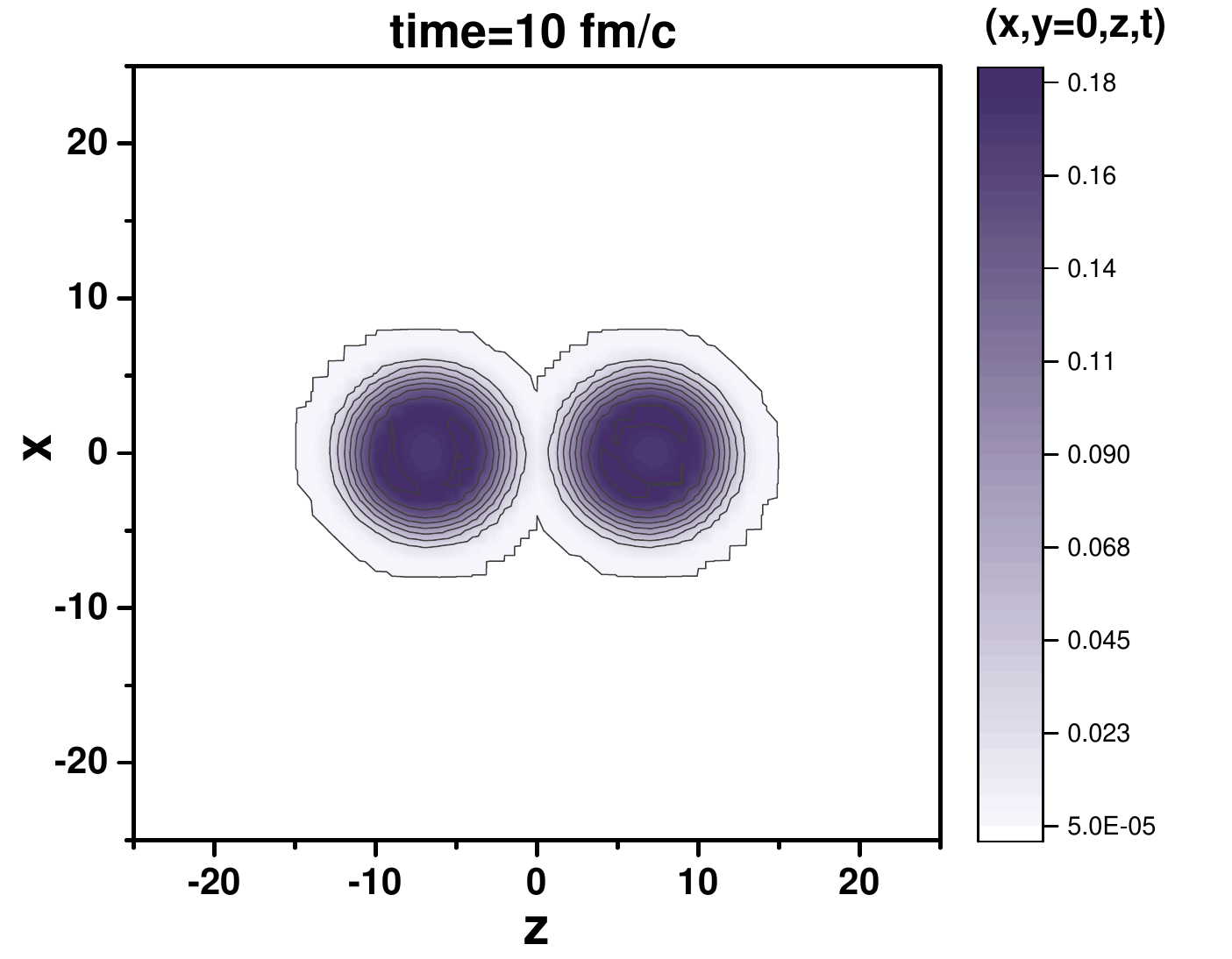} } {\includegraphics[width=6.6cm]{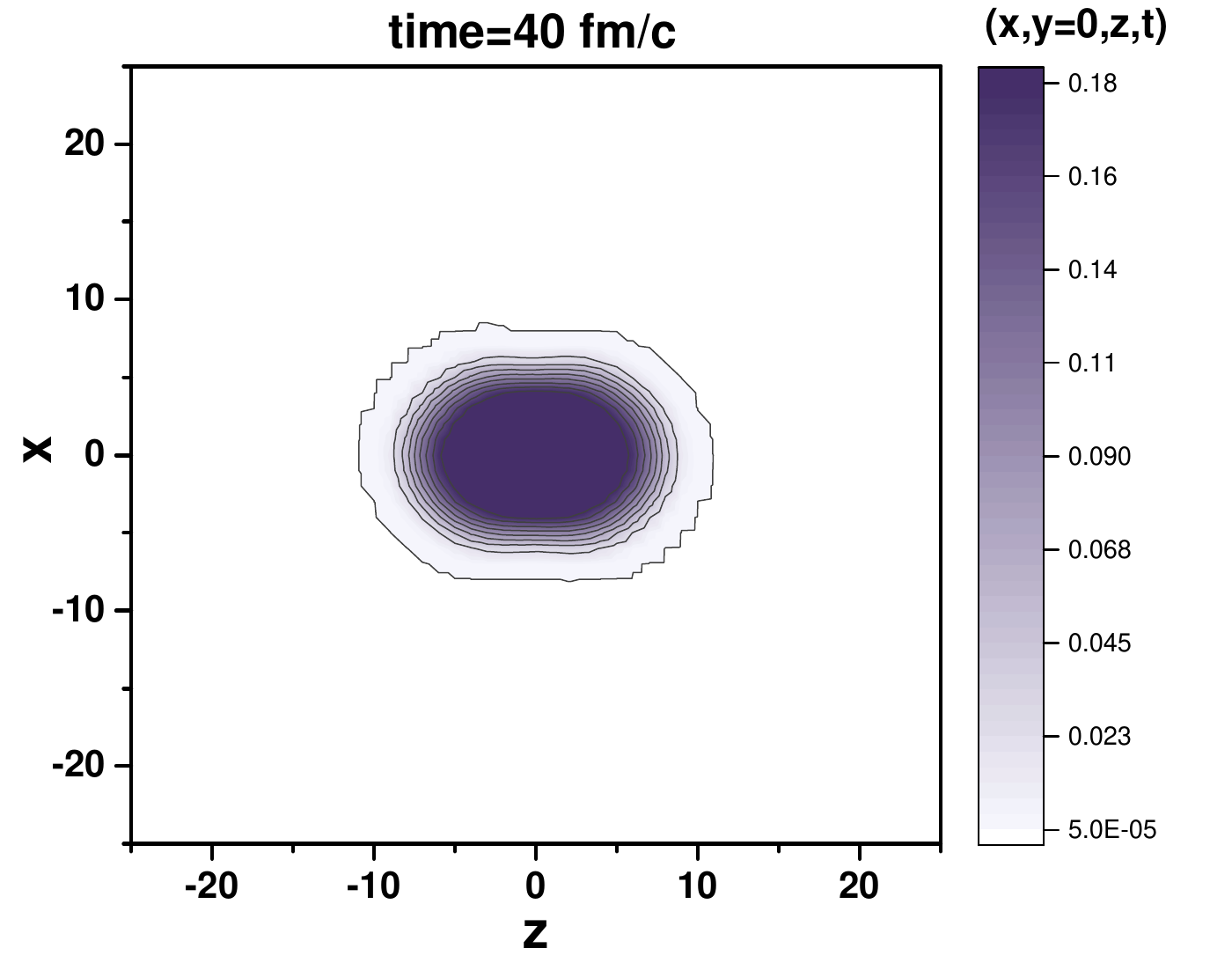} }
 {\includegraphics[width=6.6cm]{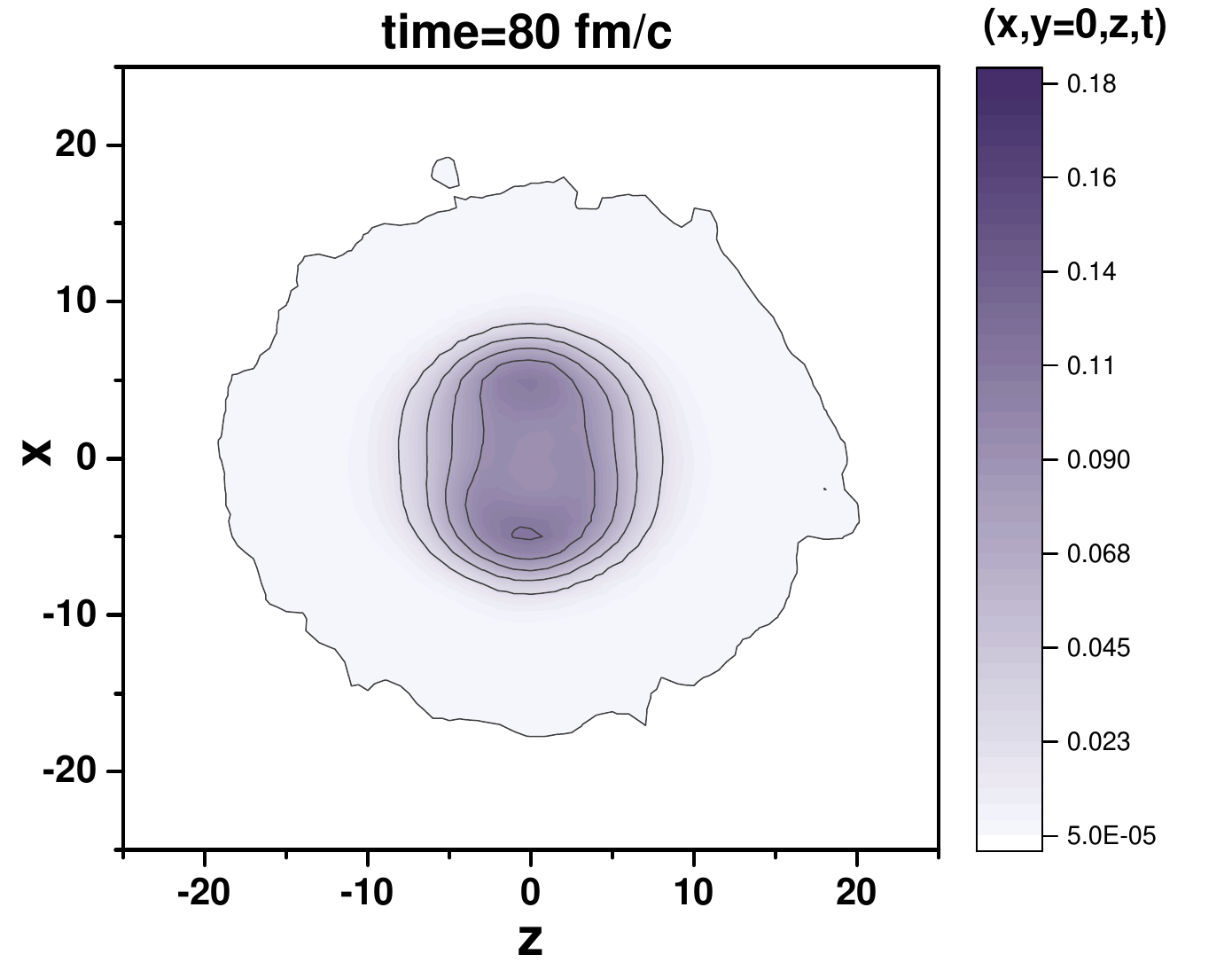} } {\includegraphics[width=6.6cm]{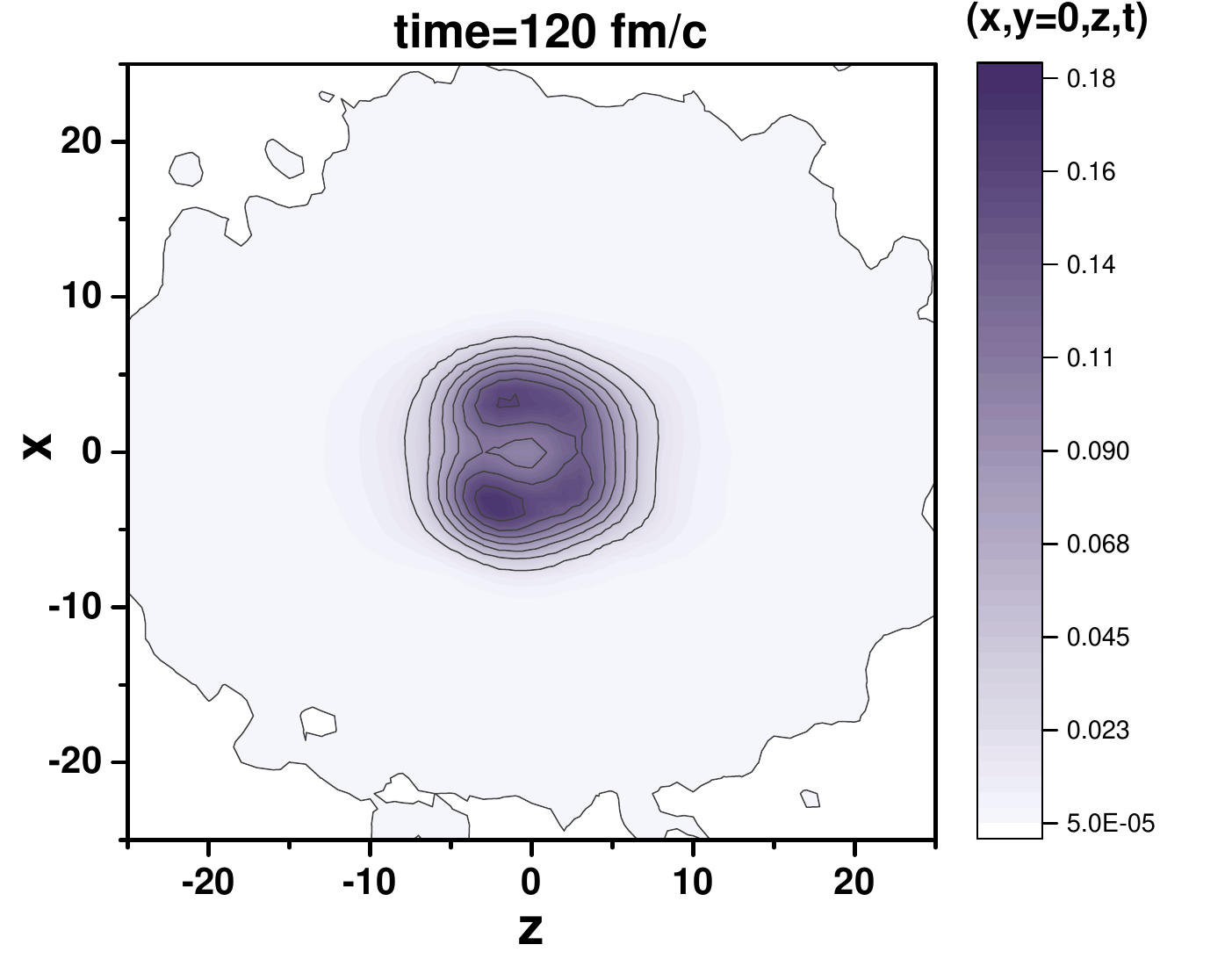} }
  \end{center}
\caption{The nucleon density distribution $\rho(x,y=0,z;t)$ for central collisions of $^{97}_{45}Rh$ + $^{97}_{45}Rh$  at 40 A$\cdot$MeV for times of 10, 40, 80, and 120 fm/c. {The density contours -- specified on the r.h.s. -- are given in units of fm$^{-3}$.}}
 \label{BUUcoll}
\end{figure}

To this end we employ a standard BUU code (described in more detail in \cite{Cassing:1990dr}) with a mean field of Skyrme type and a finite range Yukawa interaction for nucleons. Coulomb forces are included as well as the excitation and decay of $\Delta$-resonances. The elastic scattering cross section $\sigma_{el}$ for binary nucleon-nucleon collisions is provided by the Cugnon parametrization \cite{Cugnon:1980rb} and Pauli-blocking for the final states by calculating the 'local' occupation in phase-space by an average over all ensembles. The simulations are performed within the parallel ensemble method by propagating simultaneously $\sim$ 2000 ensembles. No coalescence of nucleons for the formation of light clusters is included. 

In this BUU model we compute the quadrupole moment in momentum space of the emitted nucleons for different cross sections, i.e. multiplying the actual elastic cross section $\sigma_{el}$ by a factor  5/3 or 2 in fusion reactions of central collisions. Here the bombarding energy should not be too high because the compound system in reality will break into pieces. A suitable system is found to be central collisions of e.g. $^{97}_{45}Rh$ + $^{97}_{45}Rh$  at 40 A$\cdot$MeV, which leads to a fusion process with the emission of particles.
This is demonstrated in Fig. \ref{BUUcoll} where we show the nucleon density distribution $\rho(x,y=0,t:t)$ (in units of $fm^{-3}$) at times of 5, 40, 80, 120 fm/c employing the default elastic cross section $\sigma_{el}$ from \cite{Cugnon:1980rb}. The first contour plot at $t$=10 fm/c displays the two nuclei just before contact while in the second plot for $t$=40 fm/c a maximum overlap of the impinging nuclei is achieved. They form a fusing system as seen from the density distributions at 80 fm/c and 120 fm/c while throughout the reaction nucleons are emitted. The emitted nucleons can clearly be separated from the fusing remnant by a cut in coordinate space for $r >$ 10 fm.

Only very energetic nucleons are emitted early and provide information about the dynamics (or stopping power) at early times. To this end high statistics BUU calculations have been performed and the quadrupole moment in momentum space has been calculated for nucleons outside of the compound nucleus ($>$ 10 fm) for different cutoff kinetic energies $E_c$ for the nucleons in steps of 10 MeV. It is found that only very energetic nucleons above 60 MeV of kinetic energy provide a quadrupole moment that is approximately constant in time after 80 fm/c and can be used as an observable for the stopping of energetic nucleons.


\begin{figure}[h!]
\begin{center}
 {\includegraphics[width=8cm]{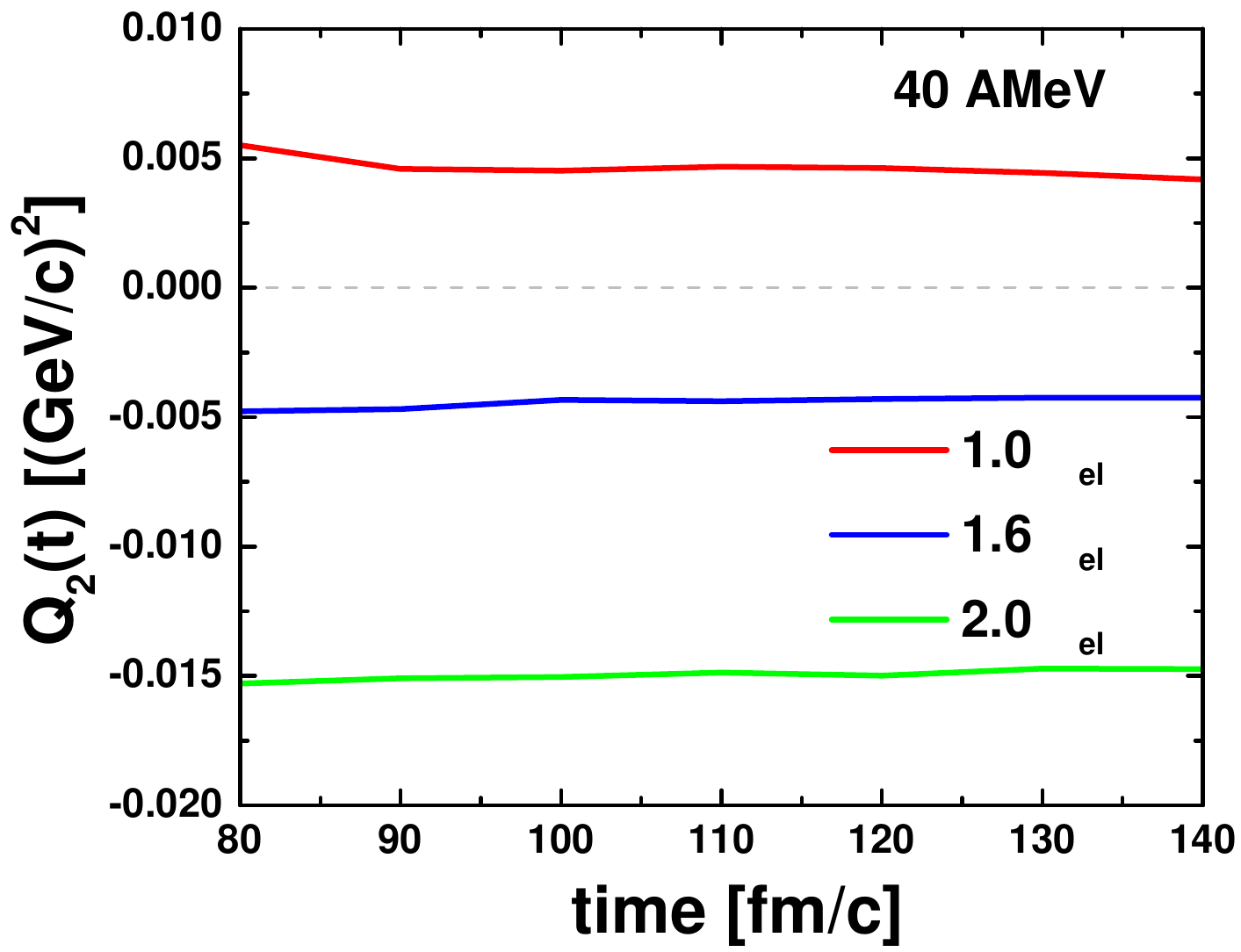} }
 \end{center}
\caption{Comparison of the results for the quadrupole moment in momentum space of energetic nucleons ($>$ 60 MeV) in central collisions of $^{97}_{45}Rh$ + $^{97}_{45}Rh$   at 40 A$\cdot$MeV for different elastic nucleon-nucleon cross sections of 1.0 $\sigma_{el} $, 1.6 $\sigma_{el} $ and 2.0 $\sigma_{el}$, where $\sigma_{el}$ denotes the elastic Cugnon cross section from \cite{Cugnon:1980rb}. }
 \label{Fig15}
\end{figure}

The numerical results for the quadrupole moment in momentum space of the emitted nucleons ($>$ 60 MeV) are presented in Fig. \ref{Fig15} as a function of time for different interaction strength, i.e. multiplication factors of the elastic cross section $\sigma_{el}$  in the BUU code. We note in passing that inelastic reaction channels are found to play no role at the energy of 40 A$\cdot$MeV. The actual numbers in Fig. \ref{Fig15} show that the quadrupole moment of the energetic nucleons is approximately stable in time and positive for the default interaction strength (upper line), which implies a preferential forward emission of these nucleons. With increasing cross section the quadrupole moment turns negative,  which implies a preferential emission in sideward direction. The expected factor at 40 A$\cdot$MeV is about 5/3 according to Fig. \ref{Fig14} and indicates that the emission of these energetic nucleons should be slightly enhanced in sideward direction contrary to the expectation in the default 2-body interaction scenario. Although such experimental studies  might suffer from low statistics these investigations could shed further light on the stopping power of elastic nucleon-nucleon collisions and the role of 3-body collision processes.

{Some comments to these results are in order: One might argue that actual transport simulations for heavy-ion collisions,  that are based
on the two-body level using an effective in-medium cross section for 2$\leftrightarrow$2 scattering, describe actual data on rapidity distributions or collective flow quite well and that there is no need for additional 3$\leftrightarrow$3 interactions. In principle, these effective in-medium cross sections  might be 
derived from $G$-matrix calculations (cf. Eq. (72)), but actually are 'fitted' to experimental data for 'stopping' or 'collective flow' within the two-body approach alone. Even in case of microscopic $G$-matrix calculations the situation is not unique since these calculations start from a bare 2-body interaction, which is iterated for the $T$-matrix that finally describes the scattering phase-shifts. However, different 2-body interactions may give the same on-shell $T$-matrix but differ in the off-shell matrix elements of the interaction that appear in the $G$-matrix equation (72). Thus there is no unique answer from theory for the 'right' in-medium 2-body cross section so far. On the other hand, systematic experimental studies e.g. on pre-equilibrium particle emission as a function of the system size (or mass of projectile and target) -- in comparison to 2-body transport calculations -- might indicate the necessity to incorporate 3$\leftrightarrow$3 interactions.  }

\section{Summary}
In this study we have based the formulation of 3$\leftrightarrow$3 interactions on the equations of motion method for identical fermions which has also been denoted as quantum correlation dynamics. All explicit interactions terms for the 2-body and 3-body correlations have been derived/presented in full detail and discussed in a single-particle basis $|\alpha \rangle$. In this framework the on-shell 2-body collision integral has been reviewed and the on-shell 3-body collision integral been derived in the on-shell limit on the basis of the same two-body interaction in leading order. Here no explicit 3-body interaction has been accounted for. The resulting equations obey particle number as well as energy-momentum conservation and avoid a double-counting of interactions and maintain full antisymmetry with respect to particle exchange.

For a quantification of the relative impact of 3-body interactions we have performed a model study for a homogeneous system in space in a finite box with periodic boundary conditions. The systems investigated are  spin-isospin symmetric nuclear matter configurations with initial occupation distributions that are given by shifted Fermi spheres (without overlap, cf. Fig. 7) as encountered in the initial phase of nucleus-nucleus collisions after contact. The results for the relaxation times -- employing an effective 2-body interaction -- have been compared to Boltzmann-Uehling-Uhlenbeck (BUU) transport calculations in the continuum limit for the same bombarding energies and agree on the level of a few percent. We have found that the additional 3-body interactions reduce the relaxation times up to a factor of $\sim$3 -- depending on bombarding energy -- which indicates that the equilibration in low-energy nucleus-nucleus collisions by elastic nucleon-nucleon scattering might have been underestimated so far. 

{In principle, one expects that the influence of 3-body reactions should become more important with increasing bombarding energy in actual heavy-ion collisions due to
compression effects, but in this case the inelastic channels between hadrons of different quantum numbers (resonances, mesons etc.) become dominant as noted in the Introduction. This particular situation is not in the range of the model calculations addressed here because they only describe elastic interactions of identical fermions including exact Pauli blocking. Thus the model cases considered had to be restricted to lower energies where the exact implementation of Pauli-blocking is important. As shown in Fig. 6 the blocking effect decreases for higher bombarding energies and the proper implementation of Pauli-blocking becomes less important such that classical estimates might be used. The classical collision criterion involves the critical impact parameter $b=\sqrt{\sigma/\pi}$, which in our model case gives $b \approx$ 1 fm. This has to be compared to the average distance between nucleons at density $\rho$ given by $d=\rho^{-1/3}$, i.e. amounts to $d \approx$=1.45 fm at twice nuclear matter density. Note that
the model studies performed here refer to shifted Fermi spheres (cf. Fig. 7) and give a nucleon density which at most gives twice nuclear matter density, i.e. in case the shifted Fermi-spheres just touch each other in momentum space or are fully separated. On the other hand screening effects by e.g. by mesons reduce the probability for 3-body fermion scattering such that the geometrical estimates at higher density have to be taken with care.  A fully quantum mechanically solution to this problem is not available so far.}

\medskip
Furthermore, it has been shown within BUU transport calculations that the enhanced stopping by 3$\leftrightarrow$3 collisions -- modeled by a corresponding increase in the elastic 2-body cross section -- shows up in the angular distribution of energetic nucleons ($>$ 60 MeV of kinetic energy) in e.g. central $^{97}_{45}Rh$ collisions  at 40 A$\cdot$MeV that lead to the formation of a compound nucleus, i.e. in a partial fusion reaction. The angular distribution of the energetic nucleons changes from a slightly forward peaked angular distribution to a slightly sidewards peaked angular distribution which could be controlled experimentally. {However, a single experiment will not provide a final answer on the role of elastic 3-body reactions because the 'correct' in-medium 2-body cross section cannot be determined sufficiently accurate by $G$-matrix calculations. Accordingly, systematic experimental studies e.g. on pre-equilibrium particle emission will be necessary -- in comparison to 2-body transport calculations -- for different sizes of projectile and target to unravel the role of elastic 3-body nucleon interactions.}

\medskip

In principle such 3-body interactions can be implemented in transport codes employing the in-cell method \cite{lang1993new}, which is based on matrix elements and a probabilistic interpretation of the collision integral, and added to the 2-body matrix element squared as in case of the calculations in the discrete basis performed in Section IV. Such concepts have been employed e.g. in Refs. \cite{Cassing:2001ds,Linnyk:2015rco} for 2$\leftrightarrow$3 channels.

\medskip
We note, furthermore, that the energy-independent isotropic cross section employed in the model calculations might overestimate the effect of 3$\leftrightarrow$3 reactions and future studies are required with more realistic interactions. Additionally, the role of higher order interaction terms has to be clarified.

\section*{Acknowledgements}
The author likes to thank Elena Bratkovskaya for a critical reading of the manuscript and valuable comments as well as Marie Cassing for preparing the illustrations presented in the text.

\section*{Appendix}
In order to transparently derive all terms in the 3-body part of the collision integral we recall the result for
$c_3(t)$ in the natural basis in the limit (\ref{12.21}):
 	\begin{widetext}
\beq \label{12.22}
C^3_{\alpha \beta \gamma , \alpha ' \beta ' \gamma '}(t)= -i \int_{t_0}^t dt' \ \exp(-i [ \epsilon_{\alpha}+ \epsilon_{\beta}+ \epsilon_{\gamma} - \epsilon_{\alpha '} -\epsilon_{\beta '} - \epsilon_{\gamma '}](t-t'))
\eeq
$$
 \times \sum_{\eta} \{
((n_\alpha n_\beta {\bar n}_{\alpha '}+ {\bar n}_\alpha {\bar n}_\beta {n}_{\alpha '})  \langle \alpha \beta |v|\alpha ' \eta \rangle _{\cal A} \  C_{\eta \gamma \beta '\gamma '} - (n_\alpha n_\beta {\bar n}_{\beta '}+ {\bar n}_\alpha {\bar n}_\beta {n}_{\beta '})\langle \alpha \beta |v|\beta ' \eta \rangle _{\cal A} \   C_{\eta \gamma \alpha '\gamma '} $$ $$-  (n_\alpha n_\beta {\bar n}_{\gamma '}+ {\bar n}_\alpha {\bar n}_\beta {n}_{\gamma '})\langle \alpha \beta |v|\gamma ' \eta \rangle _{\cal A} \  C_{\eta \gamma \beta '\alpha '})
- (C_{\beta \gamma \eta \gamma '} \ \langle \alpha \eta |v|\alpha '\beta '\rangle _{\cal A}(n_{\alpha '}n_{\beta '} {\bar n}_\alpha +
{\bar n}_{\alpha '} {\bar n}_{\beta '} {n}_\alpha)$$ $$- C_{\alpha \gamma \eta \gamma '} \ \langle \beta \eta |v|\alpha '\beta '\rangle _{\cal A} (n_{\alpha '}n_{\beta '} {\bar n}_\beta +{\bar n}_{\alpha '} {\bar n}_{\beta '} {n}_\beta)
- C_{\beta \alpha \eta \gamma '} \ \langle \gamma \eta |v|\alpha '\beta '\rangle _{\cal A} (n_{\alpha '} n_{\beta '} {\bar n}_\gamma +
{\bar n}_{\alpha '} {\bar n}_{\beta '} {n}_\gamma))
$$
$$
+ ( (n_{\beta} n_{\gamma} {\bar n}_{\beta '} + {\bar n}_{\beta} {\bar n}_{\gamma} {n}_{\beta '}) \langle \beta \gamma |v|\beta ' \eta \rangle _{\cal A}\  \  C_{\alpha \eta \alpha '\gamma '}
-  (n_{\beta} n_{\gamma} {\bar n}_{\alpha '} + {\bar n}_{\beta} {\bar n}_{\gamma} {n}_{\alpha '}) \langle \beta \gamma |v|\alpha ' \eta \rangle _{\cal A} \  C_{\alpha \eta \beta '\gamma '} $$ $$ - (n_{\beta} n_{\gamma} {\bar n}_{\gamma '} + {\bar n}_{\beta} {\bar n}_{\gamma} {n}_{\gamma '}) \langle \beta \gamma |v|\gamma ' \eta \rangle _{\cal A} \  C_{\alpha \eta \alpha '\beta '} )
 - (C_{\alpha \gamma \alpha '\eta }  \ \langle \beta \eta |v|\beta '\gamma '\rangle _{\cal A} ({n}_{\beta '} {n}_{\gamma '}  {\bar n}_{\beta}+{\bar n}_{\beta '} {\bar n}_{\gamma '} n_{\beta})$$ $$ - C_{\beta \gamma \alpha '\eta } \ \langle \alpha \eta |v|\beta '\gamma '\rangle _{\cal A} (n_{\beta'} {n}_{\gamma '} {\bar n}_{\alpha}+{\bar n}_{\beta '} {\bar n}_{\gamma '} n_{\alpha})
- C_{\alpha \beta \alpha '\eta } \ \langle \gamma \eta |v|\beta '\gamma '\rangle _{\cal A})  ({n}_{\beta '} {n}_{\gamma '} {\bar n}_{\gamma}+{\bar n}_{\beta '} {\bar n}_{\gamma '} n_{\gamma}))
$$
$$
+ (n_{\alpha} n_{\gamma} {\bar n}_{\gamma '}+ {\bar n}_{\alpha} {\bar n}_{\gamma} {n}_{\gamma '}) \langle \gamma \alpha |v|\gamma ' \eta \rangle _{\cal A} \   C_{\eta \beta \alpha '\beta '}
- (n_{\alpha} n_{\gamma} {\bar n}_{\beta '}+ {\bar n}_{\alpha} {\bar n}_{\gamma} {n}_{\beta '})\langle \gamma \alpha |v|\beta ' \eta \rangle _{\cal A} \  C_{\eta \beta \alpha '\gamma '}$$ $$ - (n_{\alpha} n_{\gamma} {\bar n}_{\alpha '}+ {\bar n}_{\alpha} {\bar n}_{\gamma} {n}_{\alpha '})\langle \gamma \alpha |v|\alpha ' \eta \rangle _{\cal A} \  C_{\eta \beta \gamma '\beta '})
 - (C_{\alpha \beta \eta \beta '} \ \langle \gamma \eta |v|\gamma '\alpha '\rangle _{\cal A}(n_{\alpha '} n_{\gamma '} {\bar n}_{\gamma}+ {\bar n}_{\alpha '} {\bar n}_{\gamma '} {n}_{\gamma}  )$$ $$ - C_{\gamma \beta \eta \beta '}\ \langle \alpha \eta |v|\gamma '\alpha '\rangle _{\cal A}(n_{\alpha '} n_{\gamma '} {\bar n}_{\alpha}+ {\bar n}_{\alpha '} {\bar n}_{\gamma '} {n}_{\alpha}  )
- C_{\alpha \gamma \eta \beta '} \  \langle \beta \eta |v|\gamma '\alpha '\rangle _{\cal A})  (n_{\alpha '} n_{\gamma '} {\bar n}_{\beta}+ {\bar n}_{\alpha '} {\bar n}_{\gamma '} {n}_{\beta}  )) \}
$$
where the explicit dependence of all terms on $t'$ has been suppressed. In the on-shell limit this gives
\beq \label{12.22f}
C^3_{\alpha \beta \gamma , \alpha ' \beta ' \gamma '}(t)\approx - \pi^2 \delta(\epsilon_{\alpha}+ \epsilon_{\beta}+ \epsilon_{\gamma} - \epsilon_{\alpha '} -\epsilon_{\beta '} - \epsilon_{\gamma '}) \sum_{\eta}
\eeq
$$
 \times
\{(n_\alpha n_\beta {\bar n}_{\alpha '}+ {\bar n}_\alpha {\bar n}_\beta {n}_{\alpha '})  \langle \alpha \beta |v|\alpha ' \eta \rangle _{\cal A} \  \langle \eta \gamma |{\tilde v}| \beta ' \gamma ' \rangle_{\cal A} (n_{\beta '} n_{\gamma '} {\bar n}_\eta {\bar n}_\gamma - {\bar n}_{\beta '} {\bar n}_{\gamma '} {n}_\eta {n}_\gamma )$$ $$
- (n_\alpha n_\beta {\bar n}_{\beta '}+ {\bar n}_\alpha {\bar n}_\beta {n}_{\beta '})\langle \alpha \beta |v|\beta ' \eta \rangle _{\cal A} \   \langle \eta \gamma |{\tilde v}|\alpha '\gamma '\rangle_{\cal A} (n_{\alpha '} n_{\gamma '} {\bar n}_\eta {\bar n}_\gamma - {\bar n}_{\alpha '} {\bar n}_{\gamma '} {n}_\eta {n}_\gamma )$$ $$
+  (n_\alpha n_\beta {\bar n}_{\gamma '}+ {\bar n}_\alpha {\bar n}_\beta {n}_{\gamma '})\langle \alpha \beta |v|\gamma ' \eta \rangle _{\cal A} \  \langle \eta \gamma |{\tilde v}|\alpha ' \beta '\rangle_{\cal A} (n_{\alpha '} n_{\beta '} {\bar n}_\eta {\bar n}_\gamma - {\bar n}_{\alpha '} {\bar n}_{\beta '} {n}_\eta {n}_\gamma )$$ $$
- \langle \beta \gamma|{\tilde v}| \eta \gamma '\rangle_{\cal A} (n_{\gamma '} n_{\eta } {\bar n}_\beta {\bar n}_\gamma - {\bar n}_{\gamma '} {\bar n}_{\eta } {n}_\beta {n}_\gamma )  \ \langle \alpha \eta |v|\alpha '\beta '\rangle _{\cal A}(n_{\alpha '}n_{\beta '} {\bar n}_\alpha + {\bar n}_{\alpha '} {\bar n}_{\beta '} {n}_\alpha)$$ $$
+ \langle \alpha \gamma|{\tilde v}| \eta \gamma '\rangle_{\cal A} (n_{\gamma '} n_{\eta } {\bar n}_\alpha {\bar n}_\gamma - {\bar n}_{\gamma '} {\bar n}_{\eta } {n}_\alpha {n}_\gamma )  \ \langle \beta \eta |v|\alpha '\beta '\rangle _{\cal A} (n_{\alpha '}n_{\beta '} {\bar n}_\beta +{\bar n}_{\alpha '} {\bar n}_{\beta '} {n}_\beta)$$ $$
- \langle  \alpha \beta|{\tilde v}|\eta \gamma '\rangle_{\cal A} (n_{\gamma '} n_{\eta } {\bar n}_\beta {\bar n}_\alpha - {\bar n}_{\gamma '} {\bar n}_{\eta } {n}_\beta {n}_\alpha )  \ \langle \gamma \eta |v|\alpha '\beta '\rangle _{\cal A} (n_{\alpha '} n_{\beta '} {\bar n}_\gamma + {\bar n}_{\alpha '} {\bar n}_{\beta '} {n}_\gamma)
$$
$$
+  (n_{\beta} n_{\gamma} {\bar n}_{\beta '} + {\bar n}_{\beta} {\bar n}_{\gamma} {n}_{\beta '}) \langle \beta \gamma |v|\beta ' \eta \rangle _{\cal A}\  \  \langle \alpha \eta |{\tilde v}|\alpha '\gamma '\rangle_{\cal A} (n_{\alpha '} n_{\gamma '} {\bar n}_\eta {\bar n}_\alpha - {\bar n}_{\alpha '} {\bar n}_{\gamma '} {n}_\eta {n}_\alpha )$$ $$
-  (n_{\beta} n_{\gamma} {\bar n}_{\alpha '} + {\bar n}_{\beta} {\bar n}_{\gamma} {n}_{\alpha '}) \langle \beta \gamma |v|\alpha ' \eta \rangle _{\cal A} \  \langle \alpha \eta |{\tilde v}|\beta '\gamma '\rangle_{\cal A} (n_{\beta '} n_{\gamma '} {\bar n}_\eta {\bar n}_\alpha - {\bar n}_{\beta '} {\bar n}_{\gamma '} {n}_\eta {n}_\alpha )$$ $$
- (n_{\beta} n_{\gamma} {\bar n}_{\gamma '} + {\bar n}_{\beta} {\bar n}_{\gamma} {n}_{\gamma '}) \langle \beta \gamma |v|\gamma ' \eta \rangle _{\cal A} \  \langle \alpha \eta |{\tilde v}|\alpha '\beta '\rangle_{\cal A} (n_{\alpha '} n_{\beta '} {\bar n}_\eta {\bar n}_\alpha - {\bar n}_{\alpha '} {\bar n}_{\beta '} {n}_\eta {n}_\alpha )$$ $$
 - \langle \alpha \gamma|{\tilde v}| \alpha '\eta \rangle_{\cal A} (n_{\alpha '} n_{\eta } {\bar n}_\gamma {\bar n}_\alpha - {\bar n}_{\alpha '} {\bar n}_{\eta } {n}_\gamma {n}_\alpha )  \ \langle \beta \eta |v|\beta '\gamma '\rangle _{\cal A} ({n}_{\beta '} {n}_{\gamma '}  {\bar n}_{\beta}+{\bar n}_{\beta '} {\bar n}_{\gamma '} n_{\beta})$$ $$
 + \langle \beta \gamma |{\tilde v}|\alpha '\eta \rangle_{\cal A} (n_{\alpha '} n_{\eta } {\bar n}_\beta {\bar n}_\gamma - {\bar n}_{\alpha '} {\bar n}_{\eta } {n}_\gamma {n}_\gamma )  \ \langle \alpha \eta |v|\beta '\gamma '\rangle _{\cal A} (n_{\beta'} {n}_{\gamma '} {\bar n}_{\alpha}+{\bar n}_{\beta '} {\bar n}_{\gamma '} n_{\alpha})$$ $$
+ \langle \alpha \beta |{\tilde v}|\alpha '\eta \rangle_{\cal A} (n_{\alpha '} n_{\eta } {\bar n}_\beta {\bar n}_\alpha - {\bar n}_{\alpha '} {\bar n}_{\eta } {n}_\gamma {n}_\beta )  \ \langle \gamma \eta |v|\beta '\gamma '\rangle _{\cal A})  ({n}_{\beta '} {n}_{\gamma '} {\bar n}_{\gamma}+{\bar n}_{\beta '} {\bar n}_{\gamma '} n_{\gamma})$$ $$
+ n_{\alpha} n_{\gamma} {\bar n}_{\gamma '}+ {\bar n}_{\alpha} {\bar n}_{\gamma} {n}_{\gamma '}) \langle \gamma \alpha |v|\gamma ' \eta \rangle _{\cal A} \   \langle \eta \beta |{\tilde v}|\alpha '\beta '\rangle_{\cal A} (n_{\alpha '} n_{\beta '} {\bar n}_\eta {\bar n}_\beta - {\bar n}_{\alpha '} {\bar n}_{\beta '} {n}_\eta {n}_\beta )$$ $$
- (n_{\alpha} n_{\gamma} {\bar n}_{\beta '}+ {\bar n}_{\alpha} {\bar n}_{\gamma} {n}_{\beta '})\langle \gamma \alpha |v|\beta ' \eta \rangle _{\cal A} \  \langle \eta \beta |{\tilde v}|\alpha '\gamma '\rangle_{\cal A} (n_{\alpha '} n_{\gamma '} {\bar n}_\eta {\bar n}_\beta - {\bar n}_{\alpha '} {\bar n}_{\gamma '} {n}_\eta {n}_\beta ) $$ $$
- (n_{\alpha} n_{\gamma} {\bar n}_{\alpha '}+ {\bar n}_{\alpha} {\bar n}_{\gamma} {n}_{\alpha '})\langle \gamma \alpha |v|\alpha ' \eta \rangle _{\cal A} \  \langle \eta \beta |{\tilde v}|\gamma '\beta '\rangle_{\cal A} (n_{\gamma '} n_{\beta '} {\bar n}_\eta {\bar n}_\beta - {\bar n}_{\gamma '} {\bar n}_{\beta '} {n}_\eta {n}_\beta )$$ $$
 - \langle \alpha \beta |{\tilde v}| \eta \beta '\rangle_{\cal A} (n_{\beta '} n_{\eta } {\bar n}_\beta {\bar n}_\alpha - {\bar n}_{\beta '} {\bar n}_{\eta } {n}_\alpha {n}_\beta )  \ \langle \gamma \eta |v|\gamma ' \alpha '\rangle _{\cal A}(n_{\alpha '} n_{\gamma '} {\bar n}_{\gamma}+ {\bar n}_{\alpha '} {\bar n}_{\gamma '} {n}_{\gamma}  )$$ $$
  +\langle \gamma \beta |{\tilde v}|\eta \beta '\rangle_{\cal A} (n_{\beta '} n_{\eta } {\bar n}_\beta {\bar n}_\gamma - {\bar n}_{\beta '} {\bar n}_{\eta } {n}_\gamma {n}_\beta ) \ \langle \alpha \eta |v|\gamma ' \alpha '\rangle _{\cal A}(n_{\alpha '} n_{\gamma '} {\bar n}_{\alpha}+ {\bar n}_{\alpha '} {\bar n}_{\gamma '} {n}_{\alpha}  )$$ $$
 +\langle  \alpha \gamma |{\tilde v}|\eta \beta '\rangle_{\cal A} (n_{\beta '} n_{\eta } {\bar n}_\gamma {\bar n}_\alpha - {\bar n}_{\beta '} {\bar n}_{\eta } {n}_\gamma {n}_\alpha )  \  \langle \beta \eta |v|\gamma ' \alpha '\rangle _{\cal A})  (n_{\alpha '} n_{\gamma '} {\bar n}_{\beta}+ {\bar n}_{\alpha '} {\bar n}_{\gamma '} {n}_{\beta}  ) \}
$$
using
$$
\langle  \alpha \gamma |{\tilde v}|\eta \lambda \rangle_{\cal A} = \delta(\epsilon_\alpha + \epsilon_\gamma - \epsilon_\eta - \epsilon_\lambda)\
\langle  \alpha \gamma |{v}|\eta \lambda \rangle_{\cal A} ,
$$
which is dimensionless.

Eq. (\ref{12.22f}) can be rewritten as:
\beq \label{12.22g}
C^3_{\alpha \beta \gamma , \alpha ' \beta ' \gamma '}(t)\approx - \pi^2  \sum_{\eta}
\{  \langle \alpha \beta |{\tilde v}|\alpha ' \eta \rangle_{\cal A}   \  \langle \eta \gamma |{\tilde v}| \beta '\gamma '\rangle_{\cal A} \ N(\alpha \beta \alpha '  \beta '\gamma '
\eta \gamma) \eeq $$
- \langle \alpha \beta |{\tilde v}|\beta ' \eta \rangle_{\cal A}   \   \langle \eta \gamma |{\tilde v}|\alpha '\gamma '\rangle_{\cal A} \ N(\alpha \beta \beta '  \alpha ' \gamma '
\eta \gamma)
+  \langle \alpha \beta |{\tilde v}|\gamma ' \eta \rangle_{\cal A}   \  \langle \eta \gamma |{\tilde v}|\alpha ' \beta '\rangle_{\cal A} \ N(\alpha \beta \gamma '  \alpha ' \beta ' \eta \gamma) $$ $$
+   \langle \beta \gamma |{\tilde v}|\beta ' \eta \rangle_{\cal A}  \  \  \langle \alpha \eta |{\tilde v}|\alpha '\gamma '\rangle_{\cal A} \ N( \beta \gamma  \beta '  \alpha ' \gamma ' \eta \alpha)
- \langle \beta \gamma |{\tilde v}|\alpha ' \eta \rangle_{\cal A}   \  \langle \alpha \eta |{\tilde v}|\beta '\gamma '\rangle_{\cal A} \ N( \beta \gamma  \alpha '  \beta ' \gamma ' \eta \alpha)$$ $$
-  \langle \beta \gamma |{\tilde v}|\gamma ' \eta \rangle_{\cal A}   \  \langle \alpha \eta |{\tilde v}|\alpha '\beta '\rangle_{\cal A} \ N( \beta \gamma  \gamma '  \alpha ' \beta ' \eta \alpha)
+  \langle \gamma \alpha |{\tilde v}|\gamma ' \eta \rangle_{\cal A}   \   \langle \eta \beta |{\tilde v}|\alpha '\beta '\rangle_{\cal A} \ N( \alpha \gamma  \gamma '  \alpha ' \beta ' \eta \beta)$$ $$
- \langle \gamma \alpha |{\tilde v}|\beta ' \eta \rangle_{\cal A}   \  \langle \eta \beta |{\tilde v}| \alpha '\gamma '\rangle_{\cal A} \ N( \alpha \gamma  \beta '  \alpha ' \gamma ' \eta \beta)
- \langle \gamma \alpha |{\tilde v}|\alpha ' \eta \rangle_{\cal A}   \  \langle \eta \beta |{\tilde v}|\gamma '\beta '\rangle_{\cal A} \ N( \alpha \gamma  \alpha '  \beta ' \gamma ' \eta \beta)
 \}
$$
with a function $N$ including the time-dependent occupation numbers and blocking factors
\beq  \label{Nalpha}
N( \alpha \gamma  \alpha '  \beta ' \gamma ' \eta \beta) = [n_\alpha n_\gamma {\bar n}_{\alpha '} + {\bar n}_\alpha {\bar n}_\gamma {n}_{\alpha '}] \cdot
[ n_{\beta '} n_{\gamma '} {\bar n}_\eta {\bar n}_\beta -  {\bar n}_{\beta '} {\bar n}_{\gamma '} {n}_\eta {n}_\beta ] .
\eeq
 Thus the matrix elements of $C^3$ are entirely described by antisymmetrized on-shell matrix elements of the interaction $v$ and associated occupation numbers. Due to the antisymmetry of the matrix elements we can rewrite (\ref{12.22g}) as
\beq \label{12.22i}
C^3_{\alpha \beta \gamma , \alpha ' \beta ' \gamma '}(t)\approx - \pi^2  \sum_{\eta}
\{  \langle \alpha \beta |{\tilde v}|\alpha ' \eta \rangle_{\cal A}   \  \langle \eta \gamma |{\tilde v}| \beta '\gamma '\rangle_{\cal A} \ N(\alpha \beta \alpha '  \beta '\gamma '
\eta \gamma) \eeq $$
- \langle \alpha \beta |{\tilde v}|\beta ' \eta \rangle_{\cal A}   \   \langle \eta \gamma |{\tilde v}|\alpha '\gamma '\rangle_{\cal A} \ N(\alpha \beta \beta '  \alpha ' \gamma '
\eta \gamma)
+  \langle \alpha \beta |{\tilde v}|\gamma ' \eta \rangle_{\cal A}   \  \langle \eta \gamma |{\tilde v}|\alpha ' \beta '\rangle_{\cal A} \ N(\alpha \beta \gamma '  \alpha ' \beta ' \eta \gamma) $$ $$
   -\langle \beta \gamma |{\tilde v}|\beta ' \eta \rangle_{\cal A}  \  \  \langle \eta \alpha |{\tilde v}|\alpha '\gamma '\rangle_{\cal A} \ N( \beta \gamma  \beta '  \alpha ' \gamma ' \eta \alpha)
 +\langle \beta \gamma |{\tilde v}|\alpha ' \eta \rangle_{\cal A}   \  \langle \eta \alpha |{\tilde v}|\beta '\gamma '\rangle_{\cal A} \ N( \beta \gamma  \alpha '  \beta ' \gamma ' \eta \alpha)$$ $$
  +\langle \beta \gamma |{\tilde v}|\gamma ' \eta \rangle_{\cal A}   \  \langle \eta \alpha |{\tilde v}|\alpha '\beta '\rangle_{\cal A} \ N( \beta \gamma  \gamma '  \alpha ' \beta ' \eta \alpha)
  -\langle \alpha \gamma |{\tilde v}|\gamma ' \eta \rangle_{\cal A}   \   \langle \eta \beta |{\tilde v}|\alpha '\beta '\rangle_{\cal A} \ N( \alpha \gamma  \gamma '  \alpha ' \beta ' \eta \beta)$$ $$
 +\langle \alpha \gamma |{\tilde v}|\beta ' \eta \rangle_{\cal A}   \  \langle \eta \beta |{\tilde v}| \alpha '\gamma '\rangle_{\cal A} \ N( \alpha \gamma  \beta '  \alpha ' \gamma ' \eta \beta)
 -\langle \alpha \gamma |{\tilde v}|\alpha ' \eta \rangle_{\cal A}   \  \langle \eta \beta |{\tilde v}|\beta ' \gamma '\rangle_{\cal A} \ N( \alpha \gamma  \alpha '  \beta ' \gamma ' \eta \beta)
 \} .
$$

\medskip
We, furthermore, have to insert the result (\ref{12.22i}) in
\beq \label{12.23b}
\langle \alpha \beta |V_C^3| \alpha ' \beta ' \rangle (t'): =  \sum_{\gamma=\gamma' } \sum_{\lambda  \mu} \{
\langle \alpha \gamma |v|\lambda \mu \rangle \ C^3_{\lambda \beta \mu ,\alpha '\beta '\gamma' }(t')
+ \langle \beta \gamma |v|\lambda \mu \rangle \ C^3_{\alpha \lambda \mu ,\alpha '\beta '\gamma' }(t') \} \eeq
$$
 - \sum_{\gamma=\gamma'} \sum_{\lambda  \mu} \ \{
C^{3}_{\alpha \beta \gamma ,\lambda \beta '\mu}(t') \ \langle \lambda  \mu|v|\alpha ' \gamma' \rangle
+ C^{3}_{\alpha \beta \gamma ,\alpha '\lambda \mu}(t')\ \langle \lambda  \mu |v|\beta ' \gamma' \rangle \} .
$$

 This gives
\beq \label{VB3}
\langle \alpha \beta |V^3_C(t)|\alpha ' \beta '\rangle = - \pi^2  \sum_{\gamma= \gamma '} \sum_{\lambda \mu} \sum_{\eta}\ \langle \alpha \gamma|v|\lambda \mu \rangle
\{  \langle \lambda \beta |{\tilde v}|\alpha ' \eta \rangle_{\cal A}   \  \langle \eta \mu |{\tilde v}| \beta '\gamma '\rangle_{\cal A} \ N(\lambda \beta \alpha '  \beta '\gamma ' \eta \mu) \eeq $$
- \langle \lambda \beta |{\tilde v}|\beta ' \eta \rangle_{\cal A}   \   \langle \eta \mu |{\tilde v}|\alpha '\gamma '\rangle_{\cal A} \ N(\lambda \beta \beta '  \alpha ' \gamma ' \eta \mu)
+  \langle \lambda \beta |{\tilde v}|\gamma ' \eta \rangle_{\cal A}   \  \langle \eta \mu |{\tilde v}|\alpha ' \beta '\rangle_{\cal A} \ N(\lambda \beta \gamma '  \alpha ' \beta ' \eta \mu) $$ $$
   -\langle \beta \mu |{\tilde v}|\beta ' \eta \rangle_{\cal A}  \  \  \langle \eta \lambda |{\tilde v}|\alpha '\gamma '\rangle_{\cal A} \ N( \beta \mu  \beta '  \alpha ' \gamma ' \eta \lambda)
 +\langle \beta \mu |{\tilde v}|\alpha ' \eta \rangle_{\cal A}   \  \langle \eta \lambda |{\tilde v}|\beta '\gamma '\rangle_{\cal A} \ N( \beta \mu  \alpha '  \beta ' \gamma ' \eta \lambda)$$ $$
  +\langle \beta \mu |{\tilde v}|\gamma ' \eta \rangle_{\cal A}   \  \langle \eta \lambda |{\tilde v}|\alpha '\beta '\rangle_{\cal A} \ N( \beta \mu  \gamma '  \alpha ' \beta ' \eta \lambda)
  -\langle \lambda \mu |{\tilde v}|\gamma ' \eta \rangle_{\cal A}   \   \langle \eta \beta |{\tilde v}|\alpha '\beta '\rangle_{\cal A} \ N( \lambda \mu  \gamma '  \alpha ' \beta ' \eta \beta)$$ $$
 +\langle \lambda \mu |{\tilde v}|\beta ' \eta \rangle_{\cal A}   \  \langle \eta \beta |{\tilde v}| \alpha '\gamma '\rangle_{\cal A} \ N( \lambda \mu  \beta '  \alpha ' \gamma ' \eta \beta)
 -\langle \lambda \mu |{\tilde v}|\alpha ' \eta \rangle_{\cal A}   \  \langle \eta \beta |{\tilde v}|\beta ' \gamma '\rangle_{\cal A} \ N( \lambda \mu  \alpha '  \beta ' \gamma ' \eta \beta)
 \}
$$
$$
- \pi^2  \sum_{\gamma= \gamma '} \sum_{\lambda \mu} \sum_{\eta}\ \langle \beta \gamma|v|\lambda \mu \rangle \{  \langle \alpha \lambda |{\tilde v}|\alpha ' \eta \rangle_{\cal A}   \  \langle \eta \mu |{\tilde v}| \beta '\gamma '\rangle_{\cal A} \ N(\alpha \lambda \alpha '  \beta '\gamma '
\eta \mu) $$ $$
- \langle \alpha \lambda |{\tilde v}|\beta ' \eta \rangle_{\cal A}   \   \langle \eta \mu |{\tilde v}|\alpha '\gamma '\rangle_{\cal A} \ N(\alpha \lambda \beta '  \alpha ' \gamma ' \eta \mu)
+  \langle \alpha \lambda |{\tilde v}|\gamma ' \eta \rangle_{\cal A}   \  \langle \eta \mu |{\tilde v}|\alpha ' \beta '\rangle_{\cal A} \ N(\alpha \lambda \gamma '  \alpha ' \beta ' \eta \mu) $$ $$
   -\langle \lambda \mu |{\tilde v}|\beta ' \eta \rangle_{\cal A}  \  \  \langle \eta \alpha |{\tilde v}|\alpha '\gamma '\rangle_{\cal A} \ N( \lambda \mu  \beta '  \alpha ' \gamma ' \eta \alpha)
 +\langle \lambda \mu |{\tilde v}|\alpha ' \eta \rangle_{\cal A}   \  \langle \eta \alpha |{\tilde v}|\beta '\gamma '\rangle_{\cal A} \ N( \lambda \mu  \alpha '  \beta ' \gamma ' \eta \alpha)$$ $$
  +\langle \lambda \mu |{\tilde v}|\gamma ' \eta \rangle_{\cal A}   \  \langle \eta \alpha |{\tilde v}|\alpha '\beta '\rangle_{\cal A} \ N( \lambda \mu  \gamma '  \alpha ' \beta ' \eta \alpha)
  -\langle \alpha \mu |{\tilde v}|\gamma ' \eta \rangle_{\cal A}   \   \langle \eta \lambda |{\tilde v}|\alpha '\beta '\rangle_{\cal A} \ N( \alpha \mu  \gamma '  \alpha ' \beta ' \eta \lambda)$$ $$
 +\langle \alpha \mu |{\tilde v}|\beta ' \eta \rangle_{\cal A}   \  \langle \eta \lambda |{\tilde v}| \alpha '\gamma '\rangle_{\cal A} \ N( \alpha \mu  \beta '  \alpha ' \gamma ' \eta \lambda)
 -\langle \alpha \mu |{\tilde v}|\alpha ' \eta \rangle_{\cal A}   \  \langle \eta \lambda |{\tilde v}|\beta ' \gamma '\rangle_{\cal A} \ N( \alpha \mu  \alpha '  \beta ' \gamma ' \eta \lambda)
 \}
$$
$$
 +\pi^2  \sum_{\gamma= \gamma '} \sum_{\lambda \mu} \sum_{\eta}\ \{  \langle \alpha \beta |{\tilde v}|\lambda \eta \rangle_{\cal A}   \  \langle \eta \gamma |{\tilde v}| \beta '\mu\rangle_{\cal A} \ N( \beta '\mu  \gamma \lambda \eta \alpha \beta ) $$ $$
- \langle \alpha \beta |{\tilde v}|\beta ' \eta \rangle_{\cal A}   \   \langle \eta \gamma |{\tilde v}|\lambda\mu \rangle_{\cal A} \
N(  \lambda \mu \gamma \beta ' \eta \alpha \beta )
+  \langle \alpha \beta |{\tilde v}|\mu \eta \rangle_{\cal A}   \  \langle \eta \gamma |{\tilde v}|\lambda \beta '\rangle_{\cal A} \
N(  \lambda \beta ' \gamma \mu \eta \alpha \beta ) $$ $$
   -\langle \beta \gamma |{\tilde v}|\beta ' \eta \rangle_{\cal A}  \  \  \langle \eta \alpha |{\tilde v}|\lambda \mu \rangle_{\cal A} \
   N(  \lambda \mu \alpha \beta ' \eta \beta \gamma)
 +\langle \beta \gamma |{\tilde v}|\lambda \eta \rangle_{\cal A}   \  \langle \eta \alpha |{\tilde v}|\beta '\mu\rangle_{\cal A} \
 N(  \beta ' \mu \alpha \lambda \eta \beta \gamma )$$ $$
  +\langle \beta \gamma |{\tilde v}|\mu \eta \rangle_{\cal A}   \  \langle \eta \alpha |{\tilde v}|\lambda \beta '\rangle_{\cal A} \
  N(  \lambda \beta ' \alpha \mu \eta \beta \gamma)
  -\langle \alpha \gamma |{\tilde v}|\mu \eta \rangle_{\cal A}   \   \langle \eta \beta |{\tilde v}|\lambda \beta '\rangle_{\cal A} \
  N( \lambda \beta ' \beta \mu \eta \alpha \gamma)$$ $$
 +\langle \alpha \gamma |{\tilde v}|\beta ' \eta \rangle_{\cal A}   \  \langle \eta \beta |{\tilde v}| \lambda \mu \rangle_{\cal A} \
 N(  \lambda \mu \beta \beta ' \eta \alpha \gamma)
 -\langle \alpha \gamma |{\tilde v}|\lambda \eta \rangle_{\cal A}   \  \langle \eta \beta |{\tilde v}|\beta ' \mu \rangle_{\cal A} \
 N(  \beta ' \mu \beta \lambda \eta \alpha \gamma)
 \} \langle \lambda \mu|v|\alpha ' \gamma ' \rangle
$$
$$
 +\pi^2  \sum_{\gamma= \gamma '} \sum_{\lambda \mu} \sum_{\eta}\ \{  \langle \alpha \beta |{\tilde v}|\alpha ' \eta \rangle_{\cal A}   \  \langle \eta \gamma |{\tilde v}| \lambda \mu \rangle_{\cal A} \ N(  \lambda \mu \gamma \alpha ' \eta \alpha \beta) $$ $$
- \langle \alpha \beta |{\tilde v}|\lambda \eta \rangle_{\cal A}   \   \langle \eta \gamma |{\tilde v}|\alpha '\mu \rangle_{\cal A} \
N( \alpha ' \mu \gamma \lambda \eta \alpha \beta)
+  \langle \alpha \beta |{\tilde v}|\mu \eta \rangle_{\cal A}   \  \langle \eta \gamma |{\tilde v}|\alpha ' \lambda\rangle_{\cal A} \
N(  \alpha ' \lambda \gamma \mu \eta \alpha \beta) $$ $$
 -\langle \beta \gamma |{\tilde v}|\lambda \eta \rangle_{\cal A}  \  \  \langle \eta \alpha |{\tilde v}|\alpha '\mu\rangle_{\cal A} \
 N( \alpha ' \mu \alpha \lambda \eta \beta \gamma)
 +\langle \beta \gamma |{\tilde v}|\alpha ' \eta \rangle_{\cal A}   \  \langle \eta \alpha |{\tilde v}|\lambda \mu \rangle_{\cal A} \
 N(  \lambda \mu \alpha \alpha ' \eta \beta \gamma)$$ $$
  +\langle \beta \gamma |{\tilde v}|\mu \eta \rangle_{\cal A}   \  \langle \eta \alpha |{\tilde v}|\alpha '\lambda \rangle_{\cal A} \
  N( \alpha ' \lambda \alpha \mu \eta \beta \gamma)
  -\langle \alpha \gamma |{\tilde v}|\mu \eta \rangle_{\cal A}   \   \langle \eta \beta |{\tilde v}|\alpha '\lambda\rangle_{\cal A} \
  N(  \alpha ' \lambda \beta \mu \eta \alpha \gamma)$$ $$
 +\langle \alpha \gamma |{\tilde v}|\lambda \eta \rangle_{\cal A}   \  \langle \eta \beta |{\tilde v}| \alpha '\mu\rangle_{\cal A} \
 N( \alpha ' \mu \beta \lambda \eta \alpha \gamma)
 -\langle \alpha \gamma |{\tilde v}|\alpha ' \eta \rangle_{\cal A}   \  \langle \eta \beta |{\tilde v}|\lambda \mu\rangle_{\cal A} \
 N(  \lambda \mu \beta \alpha ' \eta \alpha \gamma)
 \}  \langle \lambda \mu|v| \beta ' \gamma '\rangle.
$$
We recall that (\ref{VB3}) is the additional inhomogeneous term in the equation for ${\dot c}_2$ in (\ref{12.20}), i.e.
$$
  Tr_{(3=3')} [ v(13)+v(23), c_3] ,
$$
which has to be integrated in time. In the on-shell approximation this gives a factor $-i \pi \delta(\epsilon_\alpha +  \epsilon_\beta - \epsilon_\lambda - \epsilon_\mu ) $, which leads to an additional factor of $-i\pi$ and a replacement
$$
\langle \alpha \beta |v|\lambda \mu \rangle  \ \delta(\epsilon_\alpha +  \epsilon_\beta - \epsilon_\lambda - \epsilon_\mu ) = \langle \alpha \beta |{\tilde v}|\lambda \mu \rangle .
$$
For the 3-body part of the collision integral $I^3_{\alpha \alpha}$ we, furthermore, have to evaluate the commutator with the interaction $v$ and sum over the states $\beta$, i.e.
\beq \label{finalx}
I^3_{\alpha \alpha} (t) = - \pi \sum_\beta \sum_{\delta \nu} \{ \langle \alpha \beta|v|\delta \nu \rangle \langle \delta \nu|{\tilde V}_C^3(t)|\alpha \beta \rangle -
\langle \alpha \beta|{\tilde V}^{3 \dagger}_C(t)|\delta \nu \rangle \langle \delta \nu|v|\alpha \beta \rangle \} ,
\eeq
 	\end{widetext}
where in ${\tilde V}^3_C(t)$ all terms in $V^3_C$ with $v$ have to be replaced by ${\tilde v}$ as argued above (on-shell limit). This provides all interaction terms in a transparent way for numerical computations.  All terms involve an interaction $v$, the product of 3 on-shell interactions ${\tilde v}$ and the function of the occupation numbers $N$ (\ref{Nalpha}). All matrix elements of the interaction are constrained by energy-momentum conservation, which reduces the sum over the intermediate states drastically and provides total energy-momentum conservation.

{The equation for the time evolution of occupation numbers finally is given by 
\beq \label{finaleq}
\frac{\partial}{\partial t} n_\alpha(t) = I^2_{\alpha \alpha}(t)+I^3_{\alpha \alpha}(t)
\eeq
with $I^2_{\alpha \alpha}(t)$ given by (125) and $I^3_{\alpha \alpha}(t)$ given by (\ref{finalx}).}
 
Furthermore, summing $I^3_{\alpha \alpha}(t)$ additionally over $\alpha I^2_{\alpha \alpha}(t)$ we obtain
\beq \label{energyf}
\sum_\alpha\frac{\partial}{\partial t} n_\alpha(t) = \sum_\alpha I_{\alpha \alpha}(t) = 0 ,
\eeq
which implies particle number conservation and provides a further control for numerical realizations.

\bibliography{references}

\end{document}